\documentclass{elsart}
\def\CRRC2{$CR-RC^2$}
\newcommand{\promille}{%
  \relax\ifmmode\promillezeichen
        \else\leavevmode\(\mathsurround=0pt\promillezeichen\)\fi}
\newcommand{\promillezeichen}{%
  \kern-.05em%
  \raise.5ex\hbox{\the\scriptfont0 0}%
  \kern-.15em/\kern-.15em%
  \lower.25ex\hbox{\the\scriptfont0 00}}
\usepackage{epsfig}

\usepackage{amssymb}

\begin{document}

\begin{frontmatter}
\vspace*{-1.8cm}
\begin{flushright}
{\bf\small LAL 04-05}\\
\vspace*{0.1cm}
{\small February 2004}
\end{flushright}

\title{\Large Technical design and performance\\ of the NEMO~3 detector}

\begin{center}{\small
\author{R. Arnold~$^a$}, \author{\bf C. Augier~$^b$}, \author{A.M. Bakalyarov~$^c$}, \author{J. Baker~$^d$}, \author{A. Barabash~$^e$}, \author{Ph. Bernaudin~$^b$},  
\author{M. Bouchel~$^b$}, \author{V. Brudanin~$^f$}, \author{A.J. Caffrey~$^d$}, \author{J. Cailleret~$^a$}, \author{J.E. Campagne~$^b$}, 
\author{D. Dassi\'e~$^g$}, 
\author{V. Egorov~$^f$}, \author{K. Errahmane~$^b$}, \author{A.I. Etienvre~$^b$}, 
\author{T. Filipova~$^f$}, 
\author{J. Forget~$^b$}, \author{A. Guiral~$^g$}, \author{P. Guiral~$^g$}, \author{J.L. Guyonnet~$^a$}, 
\author{F. Hubert~$^g$}, \author{Ph. Hubert~$^g$}, \author{B. Humbert~$^a$}, \author{R. Igersheim~$^a$}, \author{P. Imbert~$^b$}, \author{C. Jollet~$^a$}, \author{S. Jullian~$^b$}, \author{I. Kisel~$^f$}, \author{A. Klimenko~$^g$}, \author{O. Kochetov~$^f$}, \author{V. Kovalenko~$^f$}, \author{D. Lalanne~$^b$}, 
\author{F. Laplanche~$^b$}, \author{B. Lavigne~$^b$}, \author{V.I. Lebedev~$^c$}, \author{J. Lebris~$^i$},
\author{F. Leccia~$^h$}, \author{A. Leconte~$^i$},
\author{I. Linck~$^a$}, 
\author{C. Longuemare~$^j$}, \author{Ch. Marquet~$^h$}, \author{G. Martin-Chassard~$^b$}, \author{F. Mauger~$^j$}, \author{I. Nemchenok~$^f$}, 
\author{I. Nikolic-Audit~$^h$}, 
\author{H. Ohsumi~$^k$}, \author{S. P\'ecourt~$^b$}, \author{F. Piquemal~$^h$}, \author{J.L. Reyss~$^l$}, \author{A. Richard~$^i$}, \author{C.L. Riddle~$^d$}, \author{J. Rypko~$^b$}, \author{X. Sarazin~$^b$}, \author{L. Simard~$^b$}, 
\author{F. Scheibling~$^a$}, 
\author{Yu. Shitov~$^f$}, \author{A. Smolnikov~$^g$}, \author{I. \v{S}tekl~$^m$}, 
\author{C.S. Sutton~$^n$}, \author{G. Szklarz~$^b$}, \author{V. Timkin~$^f$}, \author{V. Tretyak~$^f$}, \author{V. Umatov~$^e$}, \author{L. V\'ala~$^{b,m}$}, \author{I. Vanushin~$^e$}, \author{S. Vasiliev~$^g$}, \author{V. Vasilyev~$^e$}, \author{V. Vorobel~$^o$}, \author{Ts. Vylov~$^f$}, \author{J. Wurtz~$^a$}, \author{S.V. Zhukov~$^c$}
}
\end{center}
\vspace{3mm}
\small
\address[$^a$]{IReS, IN2P3-CNRS et Universit\'e Louis Pasteur, 67037 Strasbourg, France}
\address[$^b$]{LAL, IN2P3-CNRS et Universit\'e de Paris-Sud, 91405 Orsay, France}
\address[$^c$]{RRC "Kurchatov Insitute", 123182 Moscow, Russia}
\address[$^d$]{INEEL, Idaho Falls, ID 83415, U.S.A.}
\address[$^e$]{ITEP, 117259 Moscow, Russia}
\address[$^f$]{JINR, 141980 Dubna, Russia}
\address[$^g$]{INR RAS, 117312 Moscow, Russia}
\address[$^h$]{CENBG, IN2P3-CNRS et Universit\'e de Bordeaux I, 33170 Gradignan, France}
\address[$^i$]{IPN, IN2P3-CNRS et Universit\'e de Paris-Sud, 91405 Orsay, France} 
\address[$^j$]{LPC, IN2P3-CNRS et Universit\'e de Caen, 14032 Caen, France}
\address[$^k$]{Saga University, Saga 840-8502, Japan}
\address[$^l$]{LSCE, CNRS, 91190 Gif-sur-Yvette, France}
\address[$^m$]{CTU, Prague, Czech Republic}
\address[$^n$]{MHC, South Hadley, Massachusetts 01075, U.S.A.}
\address[$^o$]{Charles University, Prague, Czech Republic}
\vspace*{0.5cm}
\begin{abstract}[h]
The development of the NEMO~3 detector, which is now running in the Fr\'ejus Underground Laboratory (L.S.M. Laboratoire Souterrain de Modane), was begun more than ten years ago. The NEMO~3 detector uses a tracking-calorimeter technique in order to investigate double beta decay processes for several isotopes. The technical description of the detector is followed by the presentation of its performance. 
\end{abstract}
\vspace*{1cm}
\begin{center}
{\large\it Preprint submitted to Nucl. Instrum. Methods A}
\end{center}

\end{frontmatter}
\clearpage
\tableofcontents
\clearpage
\section{Introduction}
\subsection{Objective of the experiment}

The primary objective of the NEMO~3 experiment is to search for neutrinoless double beta decay  for several isotopes. This research is one of the most pressing topics in neutrino physics, for which there is the fundamental problem of whether or not neutrinos are massless. If double beta decay without neutrino emission, $\beta\beta 0 \nu$, is observed the neutrino can be a massive Majorana particle, which is its own antiparticle.

It was proposed years ago~\cite{Furry} that there could be an exchange of neutrinos between two neutrons in the same nucleus leading to the emission of two electrons and no neutrinos. The Majorana mass term enables such a transition through a $V-A$ interaction. The observation of the $\beta\beta 0 \nu$ process would then prove the Majorana nature of the neutrino.

It is also possible to investigate $\beta\beta 0 \nu$ transitions to the 2$^+$ excited state, which involves a Majorana mass term through the $V+A$ interaction. Other mechanisms may contribute to the $\beta\beta 0 \nu$ process, in particular the emission of a Majoron $M^0$, the boson associated with spontaneous symmetry breaking of lepton number. The search for the $\beta \beta M^0$ process involves a three body decay spectrum with the Majoron avoiding detection, which imposes additional constraints on the design of the detector. The objective of the NEMO~3 experiment (Neutrino Ettore Majorana Observatory) is to investigate these 
three decay modes to further the understanding of double beta decay.

In all double beta decays, which are second order weak interactions, nuclei decay into daughter nuclei by emitting two electrons accompanied by two undetected neutrinos. This is the $\beta\beta 2 \nu$ process which has already been observed for 10 isotopes: $^{48}$Ca, $^{76}$Ge, $^{82}$Se, $^{96}$Zr, $^{100}$Mo, $^{116}$Cd, $^{128}$Te, $^{130}$Te, $^{150}$Nd and $^{238}$U (see~\cite{Barabash} for a review article).

The NEMO~3 experiment aims to search for the effective Majorana neutrino 
down to a mass $\langle m_{\nu} \rangle$ at the level of 0.1~eV. If only a limit is reached on the $\beta\beta 0\nu$ half-life, an upper limit on $\langle m_{\nu} \rangle$ can be inferred from the relation
\begin{equation} 
(T_{1/2}^{0\nu})^{-1} = \left(\langle m_{\nu} \rangle \, / \, m_e \right)^2 \times |M_{0\nu}|^2 \times G_{0\nu}
\end{equation} 
where $G_{0\nu}$ is the phase-space factor that is analytically calculated and proportional to the transition energy to the fifth power, $Q_{\beta\beta}^5$. $M_{0\nu}$ is the nuclear matrix element of the relevant isotope for which calculations have large theoretical uncertainties. Given the uncertainty in $M_{0\nu}$, a mass limit of 0.1~eV corresponds to a neutrinoless double beta decay with a half-life limit of the order of 10$^{25}$ years for $^{100}$Mo. To improve the sensitivity of a double beta decay experiment it is preferable to study an isotope with a large $Q_{\beta\beta}$, not only to get a larger $G_{0\nu}$, but also to reduce the background in the search for a  $\beta\beta 0\nu $ signal.

Measurements of a half-life as long as 10$^{25}$ years is challenging. It requires a solid understanding of natural and cosmogenic radioactive backgrounds in the materials from which the detector is made and backgrounds induced by the radioactivity from the walls and atmosphere in the underground laboratory. Two prototypes, NEMO~1~\cite{NEMO1_bib} and NEMO~2~\cite{NEMO2-NIM}, were constructed and run to demonstrate the feasibility of the experimental technique. The development of NEMO~3 was begun more than ten years ago~\cite{NEMO3}. It reflects a more than 10-fold enhancement on the NEMO~2 sensitivity in order to measure the double beta decay half-life limits of 10$^{25}$~yr for the $\beta\beta 0\nu$ process, 10$^{23}$~yr for the $\beta \beta M^0$ process, and 10$^{22}$~yr for the $\beta\beta2 \nu$ process.

\subsection{General description of the NEMO~3 detector}

\begin{figure}[h]
\begin{center}
\includegraphics*[height=12cm]{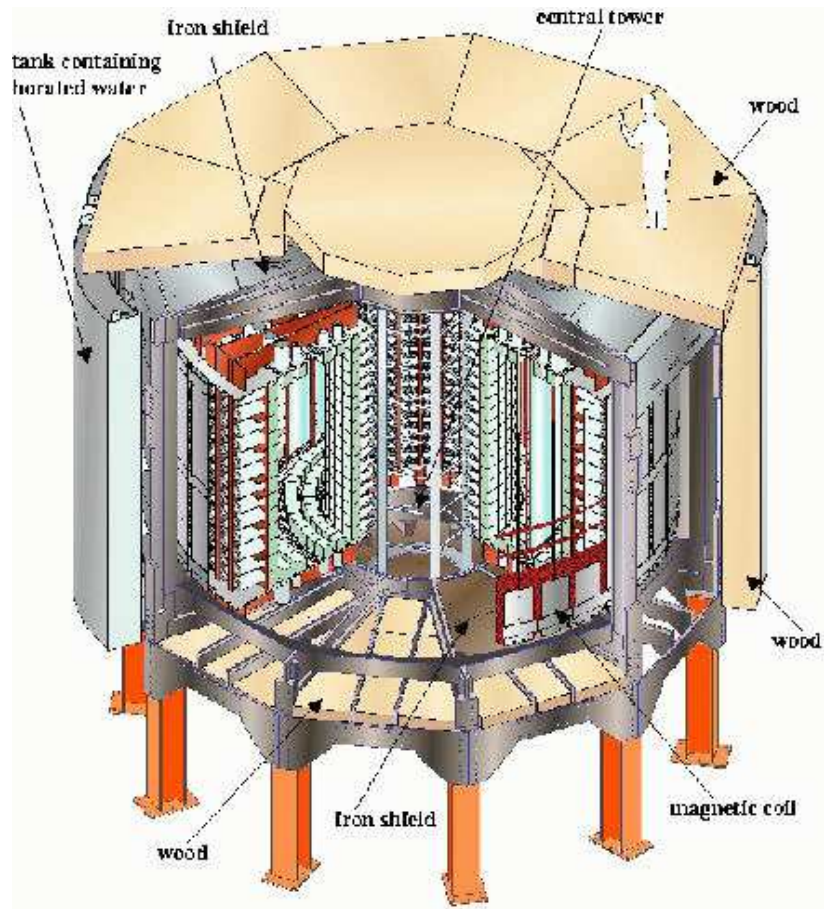}
\caption{\label{nemo3_last} An exploded view of the NEMO~3 detector. Note the  coil, iron $\gamma$-ray shield, and the two different types of neutron shields, composed of water tanks and wood. The paraffin shield under the central tower is not shown on the picture.}
\end{center}
\end{figure}

The philosophy behind NEMO~3 is the direct detection of the two electrons from $\beta\beta$ decay by a tracking device  and a calorimeter. 
The NEMO~3 detector, for which the general layout is shown in Fig.~\ref{nemo3_last}, is similar in function to the earlier prototype detector,
 NEMO~2, but has lower radioactivity and is able to accommodate up to 10 kg of double beta decay isotopes.

\begin{figure}[h]
\begin{center}
\includegraphics*[width=9cm]{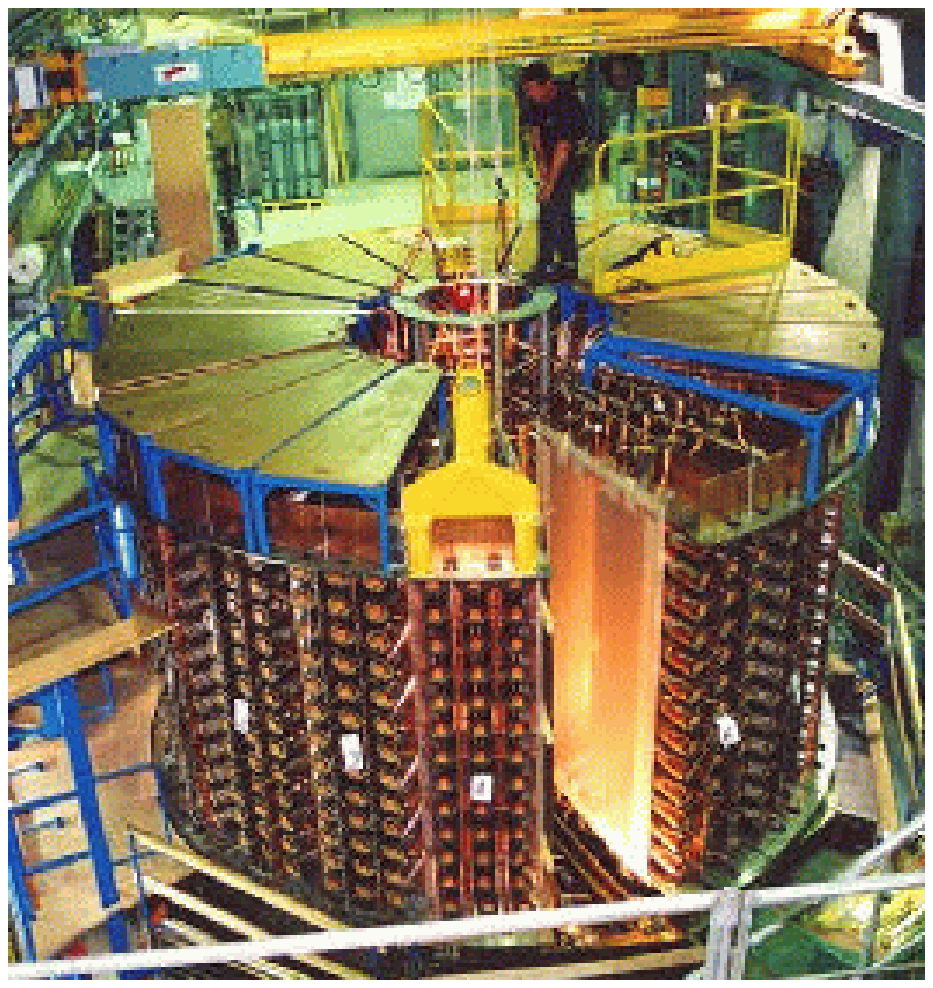}
\caption{\label{nemo3_photo} View of the NEMO~3 detector in the LSM before the installation of the last sector.}
\end{center}

\vspace*{0.5cm}

\begin{center}
\includegraphics[width=12cm,scale=1]{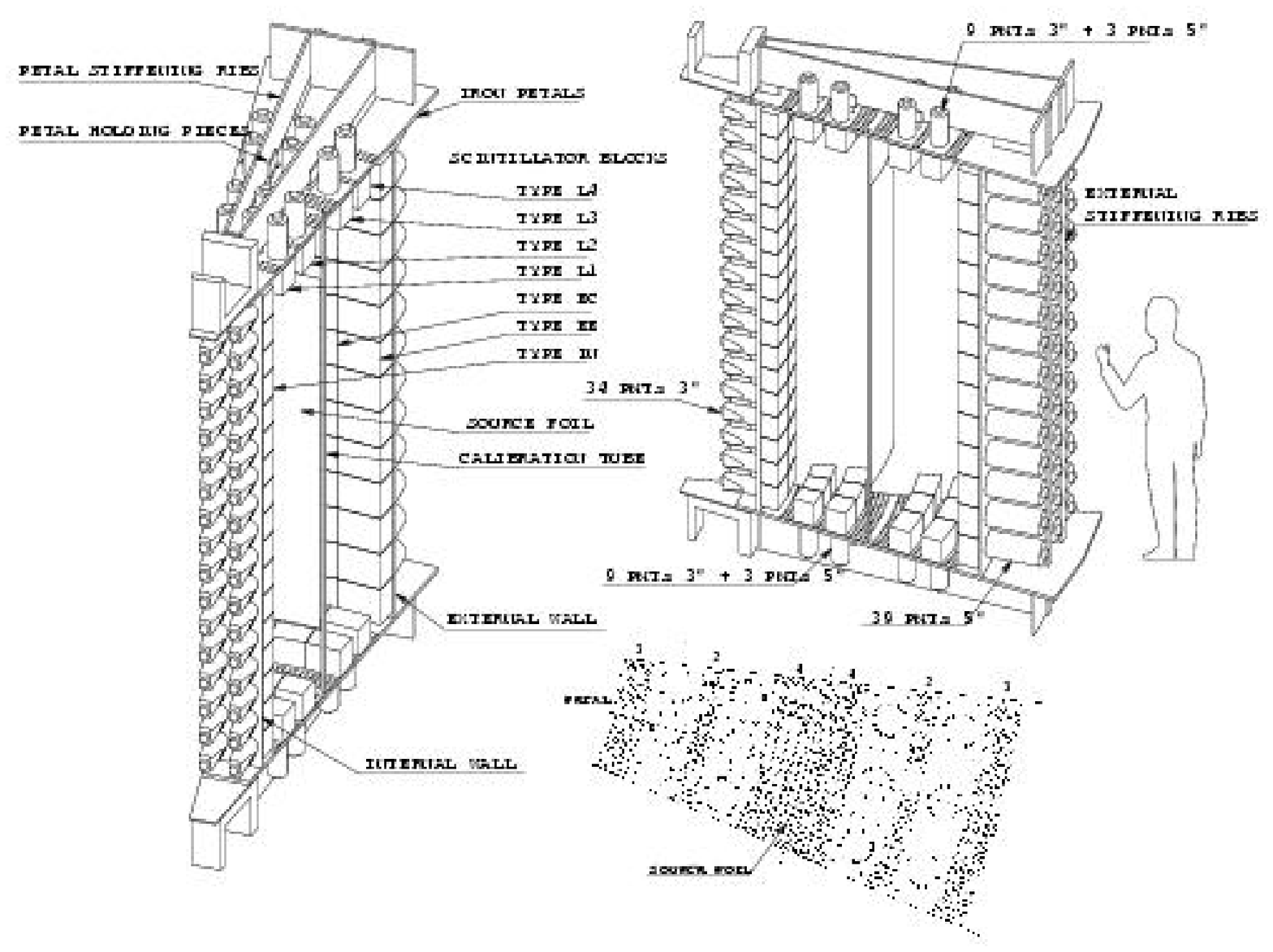}
\vspace*{-0.5cm}
\caption{\label{nemo3_secteur_PM} One sector of NEMO~3 with details on the source foil, scintillator blocks and photomultipliers. The Geiger cells are located between the internal and external walls. L4, L3,... IN identify different shapes of scintillator blocks. A petal (end-cap) is also shown with the ``4-2-3'' layer configuration for the Geiger cells.}
\end{center}
\end{figure}

The NEMO~3 detector is now installed in the Fr\'ejus Underground Laboratory (LSM\footnote{Laboratoire Souterrain de Modane}) in France. It is cylindrical in design and divided into 20 equal sectors, as shown in Fig.~\ref{nemo3_photo} and Fig.~\ref{nemo3_secteur_PM}. The segmentation permits easy access to a patchwork of source foils of
the different isotopes. This patchwork is also cylindrical in form. It is 3.1~m in diameter, 2.5~m in height and $30-60$~mg/cm$^2$ thick.
\clearpage

\clearpage

The source foils are fixed vertically between two concentric cylindrical tracking volumes composed of 6180 open octagonal drift cells. The drift cells are 270~cm long, operating in Geiger mode at 7 mbar above atmospheric pressure, with a partial pressure of 40~mbar of ethyl alcohol in a mixture with helium gas. The cells run vertically and three-dimensional tracking is accomplished with the arrival time of the signals on the anode wires and the plasma propagation times to the ends of the drift cells.

\begin{figure}[h]
\begin{center}
\includegraphics*[height=12cm]{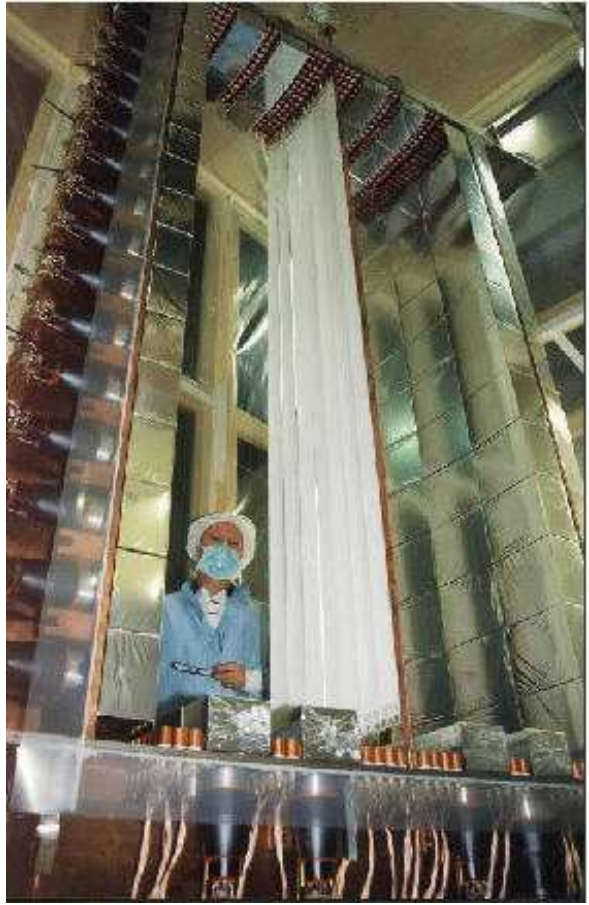}
\caption{\label{nemo3_secteur_photo} View of the third sector in the source mounting room just after the installation of the tellurium source.}
\end{center}
\end{figure}

Energy and time-of-flight measurements are acquired from plastic scintillators covering the two vertical surfaces of the active tracking volume. To further enhance the acceptance efficiency, the end-caps (the top and bottom of the detector) are also equipped with scintillators in the spaces between the drift cell layers. This calorimeter is made of 1940 large blocks of scintillators coupled to very low radioactivity 3'' or 5'' photomultiplier tubes (PMTs). The 10~cm thick blocks of scintillator yield a high photon detection efficiency. Fig.~\ref{nemo3_secteur_photo} shows a picture of one sector of the NEMO~3 detector.

A solenoid surrounding the detector produces a 25 Gauss magnetic field parallel to the foil axis, in order to identify the particle charge.  Pairs ($e^+e^-$) are produced  in the source foils in the 1 to 10~MeV region by high-energy $\gamma$-rays  from neutron capture. The curvature measurements also permit an efficient rejection of incoming electrons.

Finally, an external shield, in the form of 20 cm of low radioactivity iron, covers the detector to reduce $\gamma$-rays and thermal neutrons. Outside of this iron there is a borated water shield to thermalize fast neutrons and capture thermal neutrons. 

In the NEMO~3 detector, electrons, positrons, photons and $\alpha$-particles can be identified.
Thus, the detector is able to detect multi-particle events in the low energy domain of natural radioactivity.

\subsection{Background of the experiment\label{bkg_explain}}

The most significant concern in this double beta decay experiment is the background. The calorimeter measures the energy of the two electrons emitted from a common vertex in the source foil. The energy region of interest for the $\beta \beta 0 \nu$ signal is around 3~MeV ($Q_{\beta\beta}(^{100}$Mo) = 3.034~MeV and $Q_{\beta\beta}(^{82}$Se) = 2.995~MeV~\cite{Audi}). This energy region is shared by some energetic natural radioactivity, which can produce two electrons in the source which mimic  $\beta \beta 0 \nu$ decays. The key to the success of the experiment is to be able to positively identify these background events with high efficiency.

\subsubsection{Natural radioactivity decay chains and other radioactive isotopes}

In general, natural radioactivity coming from very long half-life isotopes such as potassium (K), uranium (U) and thorium (Th) needs to be carefully monitored. 
The decay chain for $^{235}$U is not taken into account, because even though the half-life is $7.04\times 10^8$~yr its natural isotopic abundance is only 0.7\% and its decay chain daughter nuclei do not generate enough energy to mimic the $\beta \beta 0 \nu$ signal. Concerning $^{40}$K, 
the energy range of these decays is again not a background concern for the $\beta \beta 0 \nu$'s signal.

From the natural decay chains of $^{238}$U and $^{232}$Th, only $^{214}$Bi and $^{208}$Tl are $\beta$-decay isotopes with $Q_{\beta}$ greater than 3~MeV (with respective $Q_{\beta}$ values of 3.270 and 4.992~MeV, and respective half-lives of 19.9 and 3.05~minutes~\cite{Lederer}). 
Thus, $^{214}$Bi and $^{208}$Tl produce $\gamma$-rays and electrons that are energetic enough to simulate $\beta\beta 0 \nu$ events at 3~MeV (energies and intensities of $\gamma$-rays from natural radioactivity decay chains of $^{238}$U and $^{232}$Th can be found in Ref.~\cite{Browne}).  The most energetic $\gamma$-rays are from $^{208}$Tl (2.615~MeV) for which the branching ratio is 36\%. 

Radon ($^{222}$Rn, $T_{1/2} = 3.824$~days) and thoron ($^{220}$Rn, $T_{1/2} = 55.6$~s) are $\alpha$-decay isotopes, which have  $^{214}$Bi and $^{208}$Tl as daughter isotopes respectively. Coming mainly from the rocks and present in the air, the $^{222}$Rn and $^{220}$Rn are very diffusion prone rare gases which can enter the detector. Subsequent $\alpha$-decays of these gases give $^{218}$Po and $^{216}$Po respectively, which can contaminate the interior of the detector.

\subsubsection{External and internal backgrounds}

When describing the background, it is convenient to distinguish between the ``internal''  and ``external'' sources. The internal backgrounds come from radioactive contaminants inside the $\beta\beta$ source foil, while external backgrounds come from radioactive contaminants outside the $\beta\beta$ source foils, which interact with the detector. 

The internal background for the $\beta\beta 0\nu$ signal in the 3~MeV region has two origins. The first is the tail of the $\beta\beta 2\nu$ decay distribution of the source, which cannot be separated from the $\beta\beta 0\nu$ signal and the level of overlap depends on the energy resolution of the detector. Thus, this ultimately defines the half-life limits to which $\beta\beta 0\nu$ can be searched for.
The second background comes from the $\beta$-decays of  $^{214}$Bi and $^{208}$Tl, which are present in the source at some level. They can mimic $\beta\beta$ events by three mechanisms. These are $\beta$-decay accompanied by an electron conversion process, M\"oller scattering of $\beta$-decay electrons in the source foil and $\beta$-decay emission to an excited state followed by a Compton scattered $\gamma$-ray. The last mechanism can be detected as two electron events if the $\gamma$-ray is not detected.
Thus, the experiment requires ultra-pure source foils. The maximum levels of impurities in the source have been calculated so as to produce fewer events than the tail of the $\beta\beta 2\nu$ decay gives in the region of interest for $\beta\beta 0\nu$.

The external background is defined as events produced by $\gamma$-ray sources located outside the source foils and interacting with them. The interaction of $\gamma$-rays in the foils can lead to two electron-like events by $e^+e^-$ pair creation, double Compton scattering or Compton followed by M\"oller scattering.
One of the main sources of this external radioactivity comes from the PMTs, but there are other sources too, such as cosmic rays, radon and neutrons.
To decrease the background for the NEMO~3 detector it was placed at a depth of 4800~m water equivalent, where cosmic ray fluxes have been found 
to be negligible. 
Two shields and a magnetic field suppress the backgrounds from $\gamma$-rays and neutrons. 
The vigorous air ventilation system in the laboratory reduces radon levels down to 10-20~Bq/m$^3$.
All the materials used in the detector have been selected for their radiopurity properties and  in particular, a substantial effort has been made to reduce the contamination of the PMTs from $^{40}$K, $^{214}$Bi  and $^{208}$Tl.

Consequently, in the construction of the NEMO~3 experiment every attempt has been made to minimize internal and external backgrounds by purification of the enriched isotope samples and by carefully selecting all the detector materials. As it was shown with the NEMO~2 prototype, the NEMO~3 detector will be able to characterize and measure its own background.

\section{Technical design of the NEMO~3 detector}

\subsection{The NEMO~3 sources\label{sources_mesures}}

\subsubsection{Introduction}

The primary design feature of the NEMO~3 experiment was to have the detector and the source of the double beta decay independent, unlike
the case of the $^{76}$Ge experiments. This permits one to study several double beta decay isotopes, a critical point is to be able to confirm an excess of $\beta \beta 0 \nu$ events from one isotope with another isotope. It also reduces the dependence of the interpretation of the result 
on the nuclear matrix elements. Furthermore, a rich study of the backgrounds and systematic effects is possible. 

The choice of which nuclei to study was affected by several parameters. These include the transition energy ($Q_{\beta \beta}$), the nuclear matrix elements ($M_{0\nu}$ and $M_{2\nu}$) of the transitions for $\beta\beta 0\nu$ and $\beta\beta 2\nu$ decays, the background in the energy region surrounding the $Q_{\beta \beta}$ value, the possibility of reducing the radioactivity of the isotope studied to acceptable levels, and finally the natural isotopic abundance of the candidate. Basing the choice singularly on $M_{0\nu}$ is not advisable  because the calculations are too uncertain. A good criterion for isotope selection is the $Q_{\beta \beta}$ value with respect to backgrounds. As explained in Section~\ref{bkg_explain}, the 2.615~MeV $\gamma$-ray produced in the decay of $^{208}$Tl is consistently a troublesome source of background and it is important to select $\beta\beta$ candidates with a $Q_{\beta \beta}$ value above this transition. The natural isotopic abundance is another useful criterion because in general the higher the abundance the easier the enrichment process. Typically only isotopic abundances greater than 2\% were considered.  
Five nuclei satisfy these two criteria: $^{116}$Cd , $^{82}$Se, $^{100}$Mo, $^{96}$Zr and $^{150}$Nd (with respective $Q_{\beta \beta}$ values of 2804.7, 2995.2, 3034.8, 3350.0 and 3367.1~keV and respective isotope abundance values of 7.5, 9.2, 9.6, 2.8 and 5.6\%~\cite{Audi}). Given this list and the availability of $^{100}$Mo, much effort has been focused by the NEMO collaboration on this isotope (it had already been studied by the NEMO~2 prototype~\cite{NEMO2-MO}). However the focus is not exclusively on $^{100}$Mo, in view of the fact that the detector can house several different sources.

There have been improvements in the isotopic enrichment processing in Russia, where all the double beta decay sources were produced. Thus the $^{48}$Ca isotope has been added to the list of interesting sources. Note that $^{48}$Ca fails to meet the abundance selection criterion but has an impressive $Q_{\beta \beta}$ value ($Q_{\beta \beta} = 4272.0$~keV and isotope abundance of 0.187\%~\cite{Audi}). Finally, $^{130}$Te (with $Q_{\beta \beta}$ value of 2528.9~keV and isotope abundance of 33.8\%~\cite{Audi}) has been added for $\beta\beta 2\nu$ studies. Historically, $^{130}$Te has had two different geochemical half-life measurements~\cite{Te,Tebis}, which are inconsistent with each other and a reliable one is sought here. 

A description of the current population of the 20 sectors of NEMO~3 follows. 
To study the $\beta\beta 0\nu$ processes 6914~g of $^{100}$Mo and 932~g of $^{82}$Se are housed in twelve and two sectors respectively. The objective here is to reach a sensitivity for the effective neutrino mass on the order of 0.1~eV. Several other $\beta\beta$ decay isotopes in smaller quantities have been introduced to study the $\beta\beta 2\nu$ processes which will complement the very detailed studies provided by the $^{100}$Mo and $^{82}$Se on angular distributions and single electron energy spectra. This second tier of isotope samples consists of  454~g of $^{130}$Te (2 sectors), 405~g of $^{116}$Cd (1 sector), 37~g of $^{150}$Nd, 9~g of $^{96}$Zr and 7~g of $^{48}$Ca. Finally, source foils with high levels of radiopurity, so they are effectively void of internal backgrounds, will measure the external background in the experiment. This not only accounts for the presence of 621~g of copper but also a very pure oxide of natural tellurium, which permits one to study the background near 3~MeV. This natural tellurium also provides an investigation of the $\beta\beta 2\nu$ because the natural abundance of isotope 130 for tellurium is 33.8\%, which gives 166~g of $^{130}$Te. 

\begin{figure}[h]
\begin{center}
\includegraphics{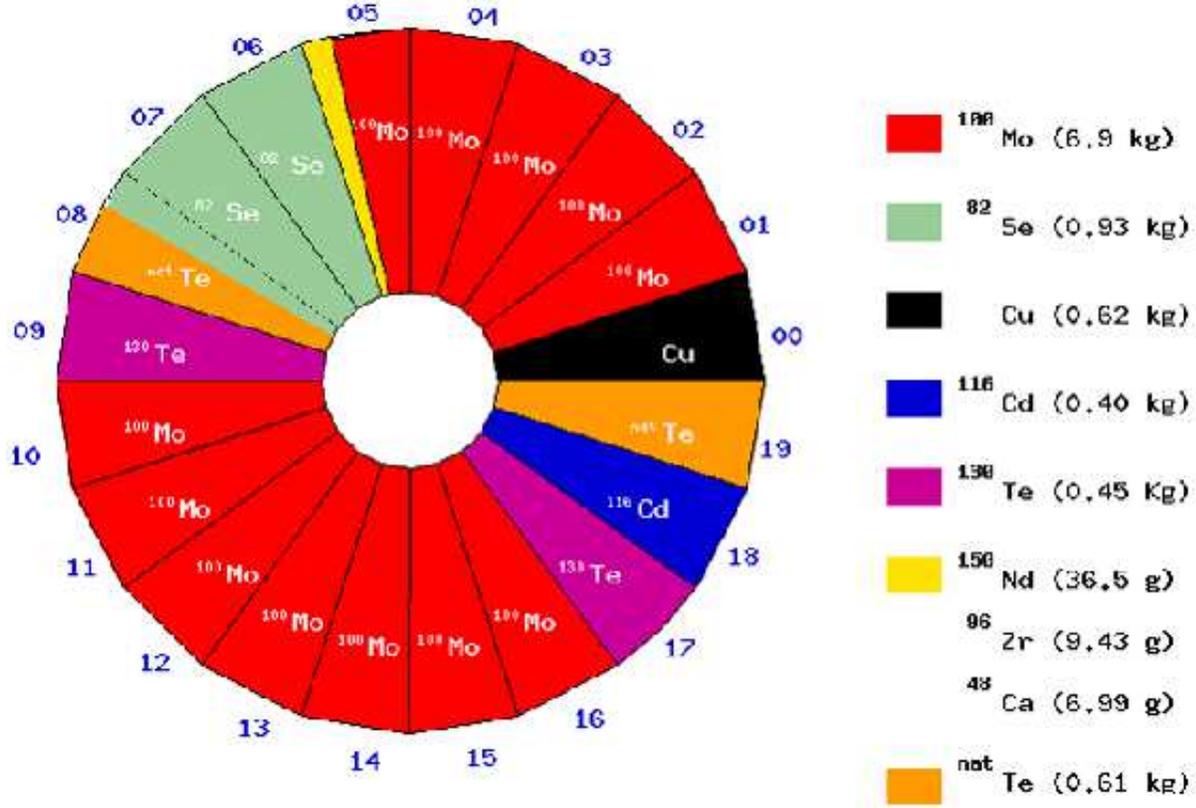}
\caption{\label{camembert} The source distribution in the 20 sectors of NEMO~3.}
\end{center}
\end{figure}

For each sector, a source frame was constructed on which were placed seven strips. The mean length of the strips is 2480~mm with a width of 63~mm if they are on the edges of the frame or 65~mm for the five strips in the middle of the frame.
All the strips are attached to the frames in a clean room at the LSM where they are then introduced into the sectors. The so-called ``NEMO~3 camembert'' depicts the distribution of the sources in the 20 sectors, Fig.~\ref{camembert}.

The thickness of the source foils was chosen to take into account the energy resolution, which is fixed by the calorimeter design. The detector efficiency for the $\beta \beta 0 \nu$ process is not compromised  as long as the surface densities of the foils do not exceed  60~mg/cm$^2$. As a consequence the source foils have surface densities between 30 and 60~mg/cm$^2$, which means a thickness lower than 60~$\mu$m for the metallic foils (density of $\sim 10$~g/cm$^3$) and lower than 300~$\mu$m for composite foils (density of $\sim 2$~g/cm$^3$). 

As indicated above, NEMO~3 sources are metallic or composite. Cadmium, copper and a fraction of the molybdenum foils are  metallic sources. Composite foils are a mixture of source powder and organic glue. For the $^{100}$Mo 64\% are in the form of composite strips. The selenium, tellurium, zirconium, neodymium and calcium foils are all composite foils.

For composite foils, the glue is made from water and some percentage of PVA (polyvinyl alcohol). This mixture is laid down on a Mylar sheet and then covered by another sheet forming a sandwich-like structure. These sheets are often referred to as backing films, which provide mechanical rigidity. The Mylar sheets have undergone a special processing in which a large number of microscopic holes (around 0.4~$\mu$m in diameter) have been created to insure a good bond with the glue. The holes are made first by irradiating the Mylar at JINR with a $^{84}$Kr ion beam of 3~MeV/nucleon and a luminosity of around $5\times 10^{11}$~ions/s. Nearly 30\% of the backing film surface is affected by the ion tracks. The next step in preparing the film is chemically etching it with NaOH (5 M) at 70$^o$~C, then the film is washed with water and 1\% of CH$_3$COOH (acetic acid) and finally dried with hot air. There are three types of backing film used in the experiment.
Type~1 has a thickness of 18~$\mu$m and around $7\times 10^7$~holes/cm$^2$. Type~2 has a thickness of 19~$\mu$m and around $2 \times10^7$~holes/cm$^2$. Type~3, which is 23~$\mu$m thick, has around $7 \times 10^7$~holes/cm$^2$. All the products (Mylar, water, acid...) used to process the backing film have been selected for their radiopurity with High Purity Germanium (HPGe) detector measurements at the LSM (the HPGe detectors are from Eurisys Mesures Company). 
The characteristics of all the source foils' strips in the 20 sectors are summarized in Tables~\ref{sources_1} and \ref{sources_2}.

\subsubsection{Radiopurity of the sources with respect to $^{214}$Bi and $^{208}$Tl}

\begin{table}[h]
\centering
\begin{tabular}{|c|c|c|c|c|c|} \hline
\multicolumn{1}{|c|}{Sector} & \multicolumn{1}{|c|}{Source strips} & \multicolumn{1}{|c|}{$\eta$ (\%) } &
\multicolumn{1}{|c|}{M$_1$ (g)} & \multicolumn{1}{|c|}{M$_2$ (g)} &
\multicolumn{1}{|c|}{M$_3$ (g)} \\
\hline 
 00 & 7 of $^{nat}$Cu {\bf (M)} & / & 620.8 & 620.8 & 620.8 {\bf $^{\rm nat}$Cu} \\
\hline 
 01 & 5 of $^{enr}$Mo {\bf (M)} & 95.14 & 424.21 & 423.22 &  401.76 {\bf $^{100}$Mo} \\ \cline{2-6}
 &    2 of $^{enr}$Mo {\bf (C)} &  95.14 & 176.22 & 145.08   & 137.72 {\bf $^{100}$Mo}  \\ 
\hline 
 02 & 7 of $^{enr}$ Mo {\bf (M)} &  1 and 2 : 96.81 & 186.44 & 186.06 & 179.76 {\bf $^{100}$Mo}\\ \cline{3-6}
 &  &  3 to 7 : 98.51    & 434.88 & 434.40   &  426.94 {\bf $^{100}$Mo} \\
\hline 
 03 & 7 of $^{enr}$Mo {\bf (M)} & 98.90 & 697.32 & 696.47 &   686.29 {\bf $^{100}$Mo}\\
\hline 
 04 & 7 of $^{enr}$Mo {\bf (M)} & 97.90 & 614.63 & 614.14 &   600.05 {\bf $^{100}$Mo}\\
\hline 
 05 & 2 of $^{enr}$Mo {\bf (M)} &  1 and 2 : 98.20 & 188.27& 187.89 & 184.14 {\bf $^{100}$Mo} \\ \cline{2-6}
 &    3 of $^{enr}$Mo {\bf (C)} &  3 : 96.66  & 109.22 & 90.07    & 86.89 {\bf $^{100}$Mo} \\ \cline{3-6}
 &   &  4 : 98.20  & 108.76  & 90.16   & 88.34 {\bf $^{100}$Mo}  \\ \cline{3-6}
 &    & 5 : 95.80  & 87.00 & 70.85   &  67.73 {\bf $^{100}$Mo} \\ \cline{2-6}
 &    1 of $^{enr}$Nd$_2$O$_3$ {\bf (C)}& 6 : 91.0 & 56.68 & 40.18  &  36.55 {\bf $^{150}$Nd} \\  \cline{2-6}
 & 1/2 of $^{enr}$ZrO$_2$ {\bf (C)}& 57.3 & 11.57 & 7.15  & ITEP : 4.10 {\bf $^{96}$Zr}\\
 & (2 parts) & 57.3 & 14.94 & 9.27&  INR : 5.31 {\bf $^{96}$Zr} \\\cline{2-6}
 &    1/4 of $^{enr}$CaF$_2$ {\bf (C)}&  73.1 & 18.516 & 9.572  &  6.997 {\bf $^{48}$Ca} \\  \cline{2-6}
 &    1/4 of back. film & &  & &  \\
\hline 
 06 &  7 of $^{\rm enr}$Se {\bf (C)}& 97.02 & 455.67 & 385.31  &  373.80 {\bf $^{82}$Se} \\
\hline 
 07 &  7 of $^{\rm enr}$Se {\bf (C)}& 96.82 & 535.04 & 460.65  &  446.03 {\bf $^{82}$Se} \\
\hline
 08 &  2 of $^{\rm enr}$Se {\bf (C)} &  1 : 96.95  & 73.58 & 63.24    & 61.31 {\bf $^{82}$Se} \\ \cline{3-6}
 & &  2 : 97.02 & 62.78 & 52.82  & 51.25 {\bf $^{82}$Se}  \\\cline{2-6}
 &    5 of $^{\rm nat}$TeO$_2$ {\bf (C)} &  3 to  7 : 33.8  & 346.44 & 189.19  & 63.94 {\bf $^{130}$Te} \\
\hline
09 &  7 of $^{\rm enr}$TeO$_2$  {\bf (C)}& 89.4 & 380.86 & 255.77  &  228.61 {\bf $^{130}$Te} \\
\hline
\end{tabular}
\vspace*{1cm}
\caption{\label{sources_1}Characteristics of the source strips for each of the NEMO~3 sectors 00-09: 
$\eta$ is the percentage of $\beta\beta$ decay isotope in the enriched sample; $M_1$, $M_2$ and $M_3$ are respectively the {\textbf total} mass of material in the foils, the mass of the investigated {\textbf element} in the foils, and the mass of the relevant $\beta\beta$ decay {\textbf isotope} in the foils. In this table, {\bf (M)} and {\bf (C)} identify the metallic and composite foil.}
\vspace*{0.5cm}
\end{table}

As presented in Section~\ref{bkg_explain}, the presence of impurities in the source foils may give rise to two-electron events which mimic $\beta\beta$ decay and produce background events in the region of the  $\beta\beta 0 \nu $ signal.  These impurities have been sufficiently reduced that given the energy resolution of the calorimeter, the ultimate background for the $\beta\beta 0 \nu$ signal is the tail of the $\beta\beta 2 \nu $ decay distribution. This is why acceptable levels of $^{214}$Bi and $^{208}$Tl in the foils  depends on the number of $\beta\beta 2 \nu $ events in the region 2.8 to 3.2 MeV. For 10~kg of $^{100}$Mo ($Q_{\beta\beta} = 3.035$~MeV), one background event/yr is expected from the $\beta \beta 2 \nu$ process above 2.8~MeV. As a consequence, the maximum levels of $^{214}$Bi and $^{208}$Tl contamination in the Mo source have been calculated to ensure that $\beta\beta 2 \nu $ is the limiting background, that means that $^{214}$Bi and $^{208}$Tl should yield less than 0.4 background events/yr above 2.8~MeV for 10~kg of $^{100}$Mo. The associated limits are thus:
\begin{equation}
\label{lim_max_bdf_Bi_Mo}
A_{(^{100}{\rm Mo})} (^{214}{\rm Bi})  \, < \, 0.3 \,\, {\rm mBq/kg}
\end{equation}
\begin{equation}
\label{lim_max_bdf_Tl_Mo}
A_{(^{100}{\rm Mo})} (^{208}{\rm Tl}) \, < \, 0.02 \,\,  {\rm mBq/kg}
\end{equation}

\clearpage
\begin{table}[h]
\centering
\begin{tabular}{|c|c|c|c|c|c|} \hline
\multicolumn{1}{|c|}{Sector} & \multicolumn{1}{|c|}{Source} & \multicolumn{1}{|c|}{$\eta$ (\%) } &
\multicolumn{1}{|c|}{M$_1$ (g)} & \multicolumn{1}{|c|}{M$_2$ (g)} &
\multicolumn{1}{|c|}{M$_3$ (g)} \\
\hline 
10 & 7 of $^{\rm enr}$Mo  {\bf (C)} &  1 and 2 : 95.14 & 205.9 & 170.14  & 161.51 {\bf $^{100}$Mo} \\\cline{3-6}
 & & 3 to 6 : 96.66 & 414.68 & 339.94 &   327.92 {\bf $^{100}$Mo} \\ \cline{3-6}
 & &  7 : 96.32 & 102.91 &  84.73 & 81.45 {\bf $^{100}$Mo}\\
\hline
11 & 7 of $^{\rm enr}$Mo  {\bf (C)} &  5 : 95.14 & 107.88 & 89.44  & 84.92 {\bf $^{100}$Mo} \\ \cline{3-6}
 & & others : 96.66 & 614.12 & 503.73 &  485.93 {\bf $^{100}$Mo}\\
\hline
12 & 7 of $^{\rm enr}$Mo  {\bf (C)} & 95.14 & 728.25 & 601.59  &  571.89 {\bf $^{100}$Mo} \\
\hline
13 & 7 of $^{\rm enr}$Mo  {\bf (C)} & 2 and  4 : 98.95 & 213.73 & 177.74  &  175.46 {\bf $^{100}$Mo} \\ \cline{3-6}
 & &  others : 96.20 & 508.93 & 420.9 & 404.1 {\bf $^{100}$Mo}\\
\hline
14 & 7 of $^{\rm enr}$Mo  {\bf (C)} & 98.95 & 735.11 & 608.07  &  601.00 {\bf $^{100}$Mo} \\
 \hline
15 & 7 of $^{\rm enr}$Mo  {\bf (C)} & 96.20 & 753.85 & 627.59  &  602.62 {\bf $^{100}$Mo} \\
\hline
16 & 7 of $^{\rm enr}$Mo  {\bf (C)} & 1, 2, 4, 7 : 95.14 & 391.64 & 318.97  &  302.79 {\bf $^{100}$Mo} \\ \cline{3-6}
 & & 3 and 5 : 96.20 & 217.74 & 181.23  &  174.0 {\bf $^{100}$Mo}\\ \cline{3-6}
 & &  6 : 95.30 & 102.35 & 84.49  &  80.34  {\bf $^{100}$Mo}\\
\hline
17 &  7 of $^{\rm enr}$TeO$_2$ {\bf (C)} & 89.4 & 375.52 & 252.01  &  225.29 {\bf $^{130}$Te} \\
\hline
18 &  7 of $^{\rm enr}$Cd {\bf (M)}& 93.2 & 491.18 & 434.42  &  404.89 {\bf $^{116}$Cd} \\
\hline
19 &  7 of $^{\rm nat}$TeO$_2$  {\bf (C)}& 33.8 & 547.18& 301.89  & 102.04 {\bf $^{130}$Te} \\
\hline
\end{tabular}
\vspace*{0.5cm}
\caption{\label{sources_2}Characteristics of the source strips
 for each of the NEMO~3 sectors 10-19: 
$\eta$ is the percentage of $\beta\beta$ decay isotope in the enriched sample; $M_1$, $M_2$ and $M_3$ are respectively the {\textbf total} mass of material in the foils, the mass of the investigated {\textbf element} in the foils, and the mass of the $\beta\beta$ decay {\textbf isotope} in the foils. In this table, {\bf (M)} and {\bf (C)} are written respectively for metallic and composite foil.}
\vspace*{0.6cm}
\end{table}

For $^{82}$Se the $T_{1/2}^{2\nu}$ is 10 times longer than for $^{100}$Mo, however the available mass of this isotope is only 1~kg. Similarly, simulations have given the maximum levels of contamination for $^{214}$Bi and $^{208}$Tl in a 1 kg Se source foil:
\begin{equation}
\label{lim_max_bdf_Bi_Se}
A_{(^{82}{\rm Se})} (^{214}{\rm Bi}) \, < \, 0.7 \,\,  {\rm mBq/kg}
\end{equation}
\begin{equation}
\label{lim_max_bdf_Tl_Se}
A_{(^{82}{\rm Se})} (^{208}{\rm Tl}) \, < \, 0.05 \,\, {\rm mBq/kg}
\end{equation}
No specific limits for activities of these contaminants were required for the other isotopes.
Given the low mass of these isotopes, the limits obtained are not expected to be as competitive with the Mo and Se sources.

\subsubsection{Production and enrichment of molybdenum}

The isotopic abundance of $^{100}$Mo is 9.6\% in $^{nat}$Mo. Using enrichment processes in Russia under the control of ITEP, Mo samples with levels of $95.14\pm 0.05$\% to $98.95\pm 0.05$\%  $^{100}$Mo were produced having a total mass of 10~kg.

The enrichment process involves the production of MoF$_6$ gas from natural Mo. This gas is then centrifuged to isolate the heavier Mo isotope such as $^{100}$Mo. The next step is an oxidation-reduction reaction on the enriched $^{100}$MoF$_6$ gas which yields $^{100}$MoO$_3$ and finally
$^{100}$Mo metallic powder.

Radioactivity measurements of this enriched Mo powder have shown that the enrichment process must be complemented with a purification process, more specifically thorium extraction. However the best measurements obtained with the HPGe spectrometer ($\sim (1-8)$~mBq/kg for $^{214}$Bi and $\sim (0.4-2)$~mBq/kg for $^{208}$Tl) did not satisfy the specific requirements for NEMO~3 given in Eq.~\ref{lim_max_bdf_Bi_Mo} and Eq.~\ref{lim_max_bdf_Tl_Mo}. To reach these levels, the collaboration decided to investigate two different purification methods in parallel: a physical process and a chemical process. The methods were refined using samples of natural molybdenum. HPGe measurements were made before and after processing to identify improvements in the purification processes.

\subsubsection{Physical purification of the enriched Mo powder and metallic strip fabrication}

Enriched Mo powder is used directly to both purify and produce metallic foils. This purification process, developed by ITEP, involves transforming the powder into an ultrapure monocrystal with a  mass of around 1~kg.

The powder is first pressed to obtain a solid Mo sample. Then the Mo is locally melted in a vacuum with an electron beam and a monocrystal is drawn from the liquid portion. Impurities coming from natural radioactivity decay chains make a migration towards the crystal extremities, because these are more soluble in the melting zone than molybdenum. Finally, cutting the skin of impurities off the crystal and repeating the process, one obtains a very pure sample, from which an enriched purified Mo monocrystal can be grown, with a 20~mm diameter.

``Short'' metallic strips, which are between 44 and 63~$\mu$m thick and between 64 and 1445 mm long, are fabricated from the cut monocrystal by heating and rolling it in a vacuum to avoid pollution. The next step is to trim the edges to obtain short strips 63 to 65~mm wide. Wastes from each step can be recycled, either by the physical or chemical method.

After the radioactivity measurements (see Table~\ref{activ_autres_sources}
 for the ``best'' limits), three to five short strips are attached end-to-end to
 create a NEMO~3 strip with a length of around 2500~mm. 

Metallic Mo strips were placed in sectors 02, 03 and 04. There are also five additional strips in sector 01 and two strips in sector 05, which give a combined  mass of $(2479 \pm 5)$~g of $^{100}$Mo.

\subsubsection{Chemical purification of the enriched Mo powder and fabrication of composite strips}

The chemical purification process also starts with the metallic powder. The focus of this method is to remove long lived radioactive isotopes of the $^{238}$U and $^{232}$Th decay chains while filling Ra sites with Ba by spiking the sample during the processing. 
The process takes advantage of an equilibrium break in the $^{238}$U and $^{232}$Th decay chains, which can selectively transform these chains to non-equilibrium states in which only short lifetime daughters exist. 
The purification process was carried out in a class 100 clean room at INEEL. It is described in Ref.~\cite{Sutton}.
\clearpage

\begin{table}[h]
\centering
\begin{tabular}{|l|l|l|l|l|l|l|l|l|} \hline
\multicolumn{1}{|c|}{Source sample} & \multicolumn{1}{|c|}{Meas.} & \multicolumn{1}{|c|}{Exp.} & \multicolumn{1}{|c|}{$^{40}$K} & \multicolumn{1}{|c|}{$^{235}$U}  & \multicolumn{2}{|c|}{ $^{238}$U chain } & \multicolumn{2}{|c|}{$^{232}$Th chain} \\ \cline{6-9}
\multicolumn{1}{|c|}{Activity} & mass & & & & \multicolumn{1}{|c|}{$^{234}$Th} & \multicolumn{1}{|c|}{$^{214}$Pb} & \multicolumn{1}{|c|}{$^{228}$Ac} & \multicolumn{1}{|c|}{$^{208}$Tl}  \\
\multicolumn{1}{|c|}{(mBq/kg)} & (g)& (h)& & &  & \multicolumn{1}{|c|}{$^{214}$Bi} & &  \\
\hline 
$^{100}$Mo {\bf (M)} &  &  &  & $1.5$& & & & \\
{\bf 2479 g} &  733  & 840  & $< \, 5 $ & $\pm 0.3 $ & $< \, 15$ & $< \, 0.39 $ & $< \, 0.5 $  & $< \, 0.11 $\\ \hline
$^{100}$Mo {\bf (C)} &  &  &  & & & & & \\
{\bf 4435 g} & 735  & 648  & $< \, 6 $ & $< \, 0.3 $ & $< \, 15$ & $< \, 0.34 $ & $< \, 0.3 $  & $< \, 0.10 $\\ \hline
 &   &  &$55$ &$20.0$ &  & $1.2 $ &  & $0.4 $\\
$^{82}$Se {\bf (C)} & 800  &  628 &$\pm 5$ &$\pm 0.7$ & $< \, 18$ & $\pm 0.5 $ & $< \, 1$ & $\pm 0.1 $\\ \cline{2-9}
{\bf 932 g} &  &  & $200$ & $8.5$ & &  &   & \\
 & 292 & 500 & $\pm20$ & $\pm0.9$ &$< \, 25$ & $< \, 4.2$ & $< \, 4$  & $< \, 0.70$\\ \hline
$^{130}$TeO$_2$ {\bf (C)}&  & & & & & &$1.7$ & \\
{\bf 454 g of $^{130}$Te} & 633& 666 & $< \, 8 $ & $< \, 0.5$ & $< \, 20$ & $< \, 0.67$ & $\pm 0.7$ & $< \, 0.46$ \\ \hline
$^{116}$Cd & 257 & 778 & $< \, 13 $ &$< \, 0.5 $ &$< \, 12 $ &$< \, 1.5 $ &$< \, 2 $ & $< \, 0.5 $\\
 ({\bf (M)} + mylar)&  & & & & & & & \\
{\bf 405 g of $^{116}$Cd} & 299& 368 & $< \, 20 $ & $< \, 1$ & $< \, 56$ & $< \, 1.7$ & $< \, 4$ & $< \, 0.83$  \\ \hline
$^{150}$Nd$_2$O$_3$ {\bf (C)}& &  & & & & & $20$ &$10$ \\
{\bf 37.0 g of $^{150}$Nd} & 58.2& 458& $< \, 70 $ & $< \, 1$ & $< \, 66$ & $< \, 3.0$ & $\pm 7$ & $\pm 2$ \\ \hline
$^{96}$ZrO$_2$ ITEP {\bf (C)} & & & & & & & & \\
{\bf 4.1 g of $^{96}$Zr} & 13.7& 624& $< \, 217 $ & $< \, 7$ & $< \, 222$ & $< \, 16$ & $< \, 23$ & $< \, 10$ \\ \hline
$^{96}$ZrO$_2$ INR {\bf (C)}&  && $583$ & & & & & \\
{\bf 5.3 g of $^{96}$Zr} & 16.6 & 456 & $\pm 167$ & $< \, 10$ & $< \, 211$ & $< \, 14$ & $< \, 27$ & $< \, 5.5$ \\ \hline
$^{48}$CaF$_2$ {\bf (C)}&& &  & & & & & \\
{\bf 6.99 g of $^{48}$Ca} & 24.56 & 1590 & $< \, 50$ & $< \, 2$ & $< \, 15$ & $< \, 4$ & $< \, 6$ & $< \, 2$ \\ \hline
$^{nat}$TeO$_2$ {\bf (C)} & & &$8$ & & & & & \\
{\bf 166 g of $^{130}$Te} & 620 & 700  & $\pm 3 $ & $< \, 0.3$ & $< \, 17$ & $< \, 0.17$ & $< \, 0.9$ & $< \, 0.090$ \\ \hline
Cu~{\bf (M)} {\bf 621 g} & 1656& 853 & $< \, 8$ & $< \, 0.2$ & $< \, 5$ & $< \, 0.12$ & $< \, 0.4$ & $< \, 0.040$ \\
\hline 
\end{tabular}
\vspace*{0.7cm}
\caption{\label{activ_autres_sources}Radioactivity measurements for the NEMO~3 source foils (in mBq/kg). The total enriched mass of each isotope is given in bold characters. The error bars are statistical uncertainties at the 1$\sigma$ level while the limits are at the 2$\sigma$ level. A systematic uncertainty of about 10\% is associated with the Monte Carlo calculations for the HPGe detector efficiencies. Only the lower limits obtained for $^{100}$Mo are presented, for both metallic and composite strips. In the case of $^{48}$CaF$_2$ the results are for the powder.}
\vspace*{0.7cm}
\end{table}

Radioactivity measurements of the purified enriched Mo powder samples were made with HPGe spectroscopy in the LSM. The limits, $A(^{214}{\rm Bi}) < 0.2$~mBq/kg and $A(^{208}{\rm Tl}) < 0.05$~mBq/kg, are the achievable levels for the HPGe detectors. The required limit on $^{214}$Bi (Eq.~\ref{lim_max_bdf_Bi_Mo}) is directly measurable. The task of measuring the required limit for $^{208}$Tl (Eq.~\ref{lim_max_bdf_Tl_Mo}) is beyond the practical measuring limits of the HPGe detectors in the LSM. However, the chemical extraction factors defined as the ratio of contamination before and after purification were measured~\cite{Sutton} for natural and enriched Mo. This study implied by Ra extraction limits and indirectly inferred by measurable quantities of $^{235}$U in the enriched Mo samples, showed that there is strong evidence that the $^{208}$Tl contamination will be below the NEMO~3 design criteria. Ultimately, the NEMO~3 detector will measure this activity.

The chemically purified $^{100}$Mo is used to make composite foils in the method previously discussed. There are two types of backing film used, Type~1 for sector 01 and Type~2 for sectors 05 and 10 through 16.

To produce the composite strips, the first step involves sieving the powder to keep only grains with diameters smaller than 45~$\mu$m. Then, the residual is ground up and several additional sieving processes are undertaken so the grains are small enough to ensure a good bond to the backing foil. Next the powder is mixed with the glue (water and PVA). The mixture is introduced into a syringe, which is heated with ultra-sound to obtain a paste. This paste of desired thickness is uniformly spread onto one of the two Mylar foils (backing film). After 10 hours of drying, the composite strip is cut to length with a surface density lower than 60~mg/cm$^2$.
The total mass of $^{100}$Mo in composite foils is $(4435 \pm 22)$~g.

\subsubsection{Production, enrichment and purification processes for other isotopes}

\noindent {\bf $\bullet$ $^{82}$Se source}

There is sufficient mass of $^{82}$Se to study the $\beta\beta 0\nu$ process. One can use a similar enrichment process to produce $^{82}$SeF$_6$ gas as that used for the Mo. The next step is an electrical discharge in the gas to obtain the enriched Se powder. 

Two different production runs of 500~g for $^{82}$Se powder were carried out. They had an enrichment factor of $97.02\pm 0.05$\% for run~1 and $96.82\pm 0.05$\% for run~2. No subsequent purification process was carried out. A portion of run~1 was already used in the NEMO~2 prototype and a value for the $^{214}$Bi contamination was measured, but the contaminants were found to be concentrated in small "hot spots" and rejected in the analysis via identification of the vertex of the candidate events~\cite{Se}. The $^{82}$Se used in NEMO~2 foils was recovered and 
used to produce composite strips for NEMO~3. The sample of material from run~2 plus the remaining part of run~1 were also used to produce composite strips. Low activities in $^{214}$Bi ($1.2 \pm 0.5$~mBq/kg) and $^{208}$Tl ($0.4 \pm 0.1$~mBq/kg) were measured for 0.8~kg of $^{82}$Se strips with the HPGe detector, as shown in Table~\ref{activ_autres_sources}.
These correspond to an expected background of 0.2~events/yr/kg from $^{214}$Bi and 1~event/yr/kg from $^{208}$Tl, but it is expected that the measured contamination in these Se foils may again be localized and will be suppressed through data analysis. In the mean time a purification process is being developed at INEEL for potential future runs with several kilograms of Se.

Se enriched powder was used to make composite strips at ITEP, with the Type~3 backing film  in sectors 06 and 07, and Type~1 in sector 08. 
The total mass of $^{82}$Se is $(932 \pm 5)$~g.

{\bf $\bullet$ $^{130}$Te source}

The Te was enriched ($89.4\pm 0.5$\% of isotope 130) by the production of $^{130}$TeF$_6$ gas, followed by oxidation and reduction to obtain enriched TeO$_2$ powder. The reason for $^{130}$TeO$_2$ versus $^{130}$Te is that it is easier to work. The Kurchatov Institute (Moscow, Russia) provided this powder to the NEMO collaboration after three separate purifications. For this sample the radioactivity limits for $^{214}$Bi and $^{208}$Tl were measured and a small contamination of $^{228}$Ac ($^{232}$Th decay chain) was detected suggesting that the limit on $^{208}$Tl is close to a value which NEMO~3 should measure (see Table~\ref{activ_autres_sources}). Composite strips were made with the Type~1 backing film. A total mass of $454 \pm 2$~g of $^{130}$Te was placed in sectors 09 and 17.

{\bf $\bullet$ $^{116}$Cd source}

Metallic enriched cadmium ($93.2\pm 0.2~\%$ of isotope 116) was obtained again by the centrifuged separation method. Part of the sample had been measured with the NEMO~2 prototype~\cite{Cd}. Another part was purified by a  distillation technique. 

Despite the metallic quality of the cadmium source, strips were glued between Mylar foils to provide mechanical strength in the vertical position. A total mass of $(405 \pm 1)$~g of $^{116}$Cd was placed in sector 18.

{\bf $\bullet$ $^{150}$Nd source}

The $^{150}$Nd$_2$O$_3$ powder was provided by INR (Moscow, Russia), after enrichment ($91.0\pm 0.5$\% of isotope 150) by electromagnetic separation and chemical purification. Radioactivity measurements (see Table~\ref{activ_autres_sources}) showed $A(^{214}{\rm Bi})< 3.0$~mBq/kg (the maximum level of contamination required for NEMO~3 is $83$~mBq/kg) but there was a small contamination of $^{208}$Tl ($(10 \pm 2)$~mBq/kg instead of $A(^{208}{\rm Tl})< 5.5$~mBq/kg for NEMO 3). As a consequence, this source will be used to check the ability of NEMO~3 to measure internal backgrounds.

The one neodymium composite strip (number 6 of sector 05) is 
made with 40.2~g of enriched Nd$_2$O$_3$ powder  
and backing films of Type~1. This gives a total mass of $37.0\pm0.1$~g of $^{150}$Nd.

{\bf $\bullet$ $^{96}$Zr source}

Enriched zirconium was obtained by an electromagnetic separation technique, with the samples averaging $57.3\pm 1.4\%$ $^{96}$Zr. The samples were a powder of $^{96}$ZrO$_2$, from two different origins.

The first sample came from ITEP and was measured in the NEMO~2 prototype. Some contamination of $^{40}$K, $^{228}$Ac and $^{208}$Tl was measured. Similar to  the Se contaminants they were concentrated in ``hot-spots'' and removed in data analysis~\cite{Zr}. The $^{96}$ZrO$_2$ powder was recovered from NEMO~2 foils and purified using a chemical process. It represents 9.6~g of ZrO$_2$ or $4.1\pm 0.1$~g of $^{96}$Zr. The second sample comes from INR (Moscow, Russia) and is 12.4~g of ZrO$_2$ or $5.3\pm 0.1$~g of $^{96}$Zr.

The zirconium composite strips were made with enriched ZrO$_2$ powder and   
 backing films of Type~2. The strip is the 7th  in sector 05. The total mass of $^{96}$Zr is $9.4\pm 0.2$~g.

{\bf $\bullet$ $^{48}$Ca source}

A CaCO$_3$ sample is enriched in the isotope calcium 48 ($73.2\pm 1.6\%$). It was produced by electromagnetic separation methods in Russia. Additionally, a purification process was developed collaboratively by JINR and the Kurchatov Institute. It  removes $^{226}$Ra, $^{228}$Ra, $^{60}$Co and $^{152}$Eu, as well as elements from the uranium and thorium decay chains. The measured purification factors for $^{226(228)}$Ra, $^{60}$Co and $^{152}$Eu are greater than 1300, 3300 and 250 respectively. After the purification process, with 64~g of enriched CaCO$_3$, JINR had a yield of 42.1~g of enriched CaF$_2$ powder.

The first portion (24.6~g of enriched CaF$_2$) of this powder was used for radioactivity measurements with HPGe studies in the LSM. Only limits were obtained (see Table~\ref{activ_autres_sources}).

The second portion of this powder (17.5~g) was used to make nine 40~mm diameter disks. 
Mylar was again used and cut in the shape of the disks. To create the calcium portion of strip 25\% of the 7th foil was populated with this material in  sector 05. The disks were glued between two Type~2 backing films. In total there is  $6.99\pm 0.05$~g of $^{48}$Ca.

{\bf $\bullet$ $^{nat}$Te and copper sources}

The foils of $^{nat}$TeO$_2$ placed in the detector allows the NEMO collaboration to measure the external background for $^{100}$Mo. The effective $Z$ of these foils ($Z(^{nat}$TeO$_2) = 43.2$) is nearly the same as that of the molybdenum foils ($Z$(Mo)$ = 42$). This is useful because the external $\gamma$-ray background can give rise to contamination processes, which are all proportional to $Z^2$. Thus, the background for $^{100}$Mo and $^{nat}$TeO$_2$ foils should give rise to similar event rates. 
In addition,  $^{nat}$TeO$_2$ has 33.8\% $^{130}$Te and the $\beta\beta2 \nu$ process expected has $Q_{\beta\beta} = 2.53$~MeV. Thus these events would not enter the $\beta\beta 0 \nu$  region of interest for $^{100}$Mo. Consequently, a background subtraction is possible for the $^{100}$Mo foils using an analysis of the $^{nat}$TeO$_2$ spectrum.

It is also useful to study $\beta\beta$ processes for the $^{130}$Te part of $^{nat}$TeO$_2$ (33.8\%) compared to enriched TeO$_2$. Foils of $^{nat}$TeO$_2$ were not purified, but radioactivity measurements showed limits lower than 0.17 and 0.09~mBq/kg for $^{214}$Bi and $^{208}$Tl respectively, as shown in Table~\ref{activ_autres_sources} (semi-conductor purity levels).

The $^{nat}$TeO$_2$ composite strips were made with Type~1 backing films for 5~strips in sector 08 and Type~3 for the 7~strips in sector 19. This gives a total mass of 614~g of $^{nat}$TeO$_2$ or 166~g of $^{130}$Te.

The copper foil provides a similar study of external backgrounds for a smaller value of $Z$. The metallic copper source is very pure ($A(^{214}{\rm Bi})< 0.12$~mBq/kg and $A(^{208}{\rm Tl})< 0.04$~mBq/kg) with a mass of 621~g and was placed in sector 00.

\subsection{Design of the calorimeter\label{calorimetre}}

\subsubsection{Description}

The three functions of the NEMO~3 calorimeter are to measure the particle energy, make time-of-flight measurements and give a fast trigger signal. 
The calorimeter is constructed with 1940 counters, each of which is made with a plastic scintillator, light guide and PMT (3'' or 5''). 
The gains of the PMTs have been adjusted to cover energies up to 12~MeV. The plastic scintillators were chosen to minimize backscattering 
and for their radiopurity. The scintillators completely cover the two cylindrical walls which surround the tracking volume. There is also partial coverage of the top and bottom end-caps (also called petals).

The scintillator blocks are inside the helium-alcohol gas mixture of the tracking detector in order to minimize energy loss in the detection of electrons. The blocks are supported by a rigid frame, which allows the PMTs to be outside the helium environment. This configuration
prevents rapid aging of the PMTs due to the helium.

The nomenclature for the calorimeter scintillator arrays is shown in Fig.~\ref{nemo3_secteur_PM}. Note that the arrays of scintillator for the calorimeter's petals are identified as L1 through L4 as one goes out radially. For the cylindrical walls of a sector the internal  wall uses the designation IN for the array and for the external wall there are EC and EE arrays to distinguish the blocks in the center (EC) versus the edge (EE) of the wall. These seven types of scintillator are distinguished by their different shapes, which have been designed to fit the circular geometry of the NEMO~3 detector, except for the scintillator thickness (10~cm for all blocks). This thickness has been chosen in order to obtain a high efficiency (50\% at 500~keV) for $\gamma$-ray  detection so as to measure the residual radioactivity of the source foil and also to reject background events.

\subsubsection{Scintillator and light guide characteristics}

The INR Kiev-Kharkov collaboration (Ukraine) was given the charge of producing the 480 end-cap scintillators and JINR was assigned the 1460 wall scintillators. The chemical nature of the material using for scintillator production is the solid solution of a scintillating agent p-Terphenyl (PTP) and a wavelength shifter 1.4-di-(5-phenyl-2-oxazoly)benzene (POPOP) in polystyrene.
After studies in both production laboratories, the mass fractions of polystyrene, PTP and POPOP were chosen. These are respectively 98.75, 1.2 and 0.05\% for the end-caps scintillators and  98.49, 1.5 and 0.01\% for the blocks of the walls.

\begin{figure}[h]
\begin{center}
\includegraphics{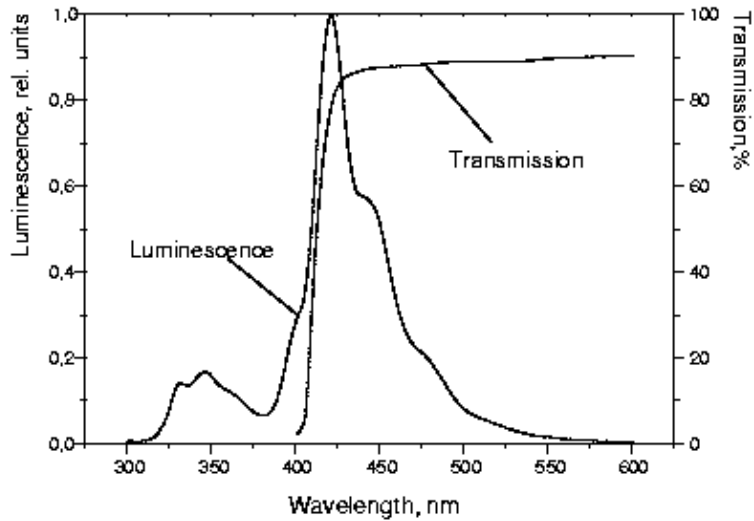}
\caption{\label{transmission} Luminescence and transmission spectra of polystyrene-based plastic scintillators, such as that used in the wall blocks, but not a NEMO~3 block, rather a cylinder 10~cm long and 3~cm in diameter. }
\end{center}
\vspace*{0.7cm}
\end{figure}

As performance objectives, the energy resolution $\sigma(E)/E$ for 1~MeV electrons had to be better than 6.2\%. This resolution was checked during production using 482 and 976~keV conversion electrons produced by a $^{207}$Bi source. After etching the blocks under water to obtain diffusive reflection at the surfaces, there was an improvement in the resolution by about 1\%. The average values of the energy resolution were respectively 5.1\% for IN blocks and 5.5\% for EE and EC blocks. To ensure the use of scintillators with the best resolution, a greater number of scintillator blocks were produced than necessary: 1093 IN blocks for 680 used (62\%), 994 EE blocks for 520 used (52\%) and 428 EC blocks for 260 used (61\%). 
In order to compare the performance of the different types of scintillator several tests were made such as optical transmission. For a 10 cm thick samples from Dubna, the light transmission was on average 75\% and always greater than 70\% for the wavelength $420$~nm (see Fig.~\ref{transmission}).
The radiopurity of the scintillator was measured and found to be respectively 430 and 60 times lower in  $^{214}$Bi and $^{208}$Tl activity than the  PMTs used to read it out, which are also low radioactivity PMTs (see Table~\ref{activ_PM_tot}). The scintillator blocks were then sent to 
CENBG and IReS to mount the blocks with the best resolution on 
the walls and petals of the detector.

For the light guides optical PMMA (polymethylmethacrylate) was manufactured for the experiment  
and used for the scintillator and PMT interface. This also protects the PMTs from helium.
The light transmission through the guides is 98\% in the wavelength range 380-420~nm.  To ensure rigidity the light guide is glued to an iron ring, which provides a pressfit between the guide and the petal or wall.

\begin{table}[h]
\centering
\begin{tabular}{|c|c|c|c|} \hline
\multicolumn{1}{|c|}{Total activity in Bq} & \multicolumn{1}{|c|}{$^{40}$K} & \multicolumn{1}{|c|}{$^{214}$Bi} & \multicolumn{1}{|c|}{$^{208}$Tl} \\
\hline
3'' PMTs R6091 - 1040 pieces   & & & \\
(230~g/PMT) & 354 &  86 &  5.2 \\
\hline
5'' PMTs R6594 - 900 pieces & & & \\ 
(385~g/PMT) & 477 & 216 & 12.6 \\
\hline
$\Sigma$ PMTs & 831 & 302 & 17.8 \\
\hline
\end{tabular}
\vspace*{0.5cm}
\caption{\label{activ_PM_tot}Total radioactivity, in Bq, for all NEMO~3 Hamamatsu photomultipliers.}
\vspace*{0.3cm}
\end{table}

\subsubsection{Photomultiplier tube characteristics\label{PMT}}

Development of low background PMTs was begun in 1992, in a collaboration between different manufacturers and physicists studying dark matter, double beta decay and neutrino oscillations. The selection criterion for the low radioactivity glass was for the contamination in $^{40}$K, $^{214}$Bi and $^{208}$Tl to be lower than 1.7, 0.83 and 0.17~Bq/kg respectively.  
The Hamamatsu company was chosen to produce the PMTs for NEMO~3, with the radiopurity of their glass being 100 to 1000 times better than standard glass (see Table~\ref{activ_PM_tot}). The other parts of their PMTs also have very low contamination (see Section~\ref{radiopurity} for the radioactivity measurements of the PMTs).

The IN, L1, L2 and L3 scintillator blocks  were coupled to R6091 3'' PMTs (230~g, 1040 pieces). These tubes have 12 dynodes and a flat photocathode. The EE, EC and L4 scintillator blocks were coupled to R6594 5'' PMTs (385~g, 900 pieces). These 5'' tubes have 10 dynodes and a hemispherical photocathode for structural integrity and thus need a second interface guide to match the design between the PMT and the light guide.

The Hamamatsu PMTs were chosen not only for their low background but also for their performance. A dedicated test station using a H$_2$ lamp was developed. 
The energy and time resolutions were measured at 1~MeV, the average values were respectively 4\% and 250~ps. The linearity of the response of the PMTs was studied as a function of the energies between 0 and 12~MeV. Also, the symmetry and the uniformity of the photocathode was investigated and finally the noise at the minimal threshold, 10~mV ($\sim 33$~keV).

\subsubsection{Preparation and installation of the scintillator blocks\label{scintillator}}
 
A  visual check on the color of the block was made. Then five layers of 70~$\mu$m thick Teflon ribbon were wrapped around the four lateral faces of the scintillator block to diffusely scatter the scintillation light for improved light collection.

The energy resolution and the peak position at 1~MeV were checked for several locations on the entrance face of the block with an electron spectrometer using a $^{90}$Sr source. The spectrometer had an intrinsic energy resolution of 0.6\%. This test identified and rejected blocks with inhomogeneities.

All six faces of the blocks, with the exception of the contact region for the light guide, were covered with sheets of aluminized Mylar. The Mylar not only protects the scintillators from ambient light and from light produced by Geiger propagation plasma in the tracking region, but also enhances the light reflection inside the scintillator, while minimizing energy lost by the electrons at the entrance face.

After gluing\footnote{A systematic check of the radiopurity for all the glues used in the NEMO~3 detector was carried out in the LSM~\cite{Busto_colle}} the scintillator block to the petals and the walls, 
the peak position and energy resolution at 1~MeV were measured at the center of each block, using the electron spectrometer.
This information was used to identify particular scintillator block and PMT combinations  so as to obtain an energy resolution for the calorimeter which was as uniform as possible. The details of the performance of the calorimeter are given in Section~\ref{cluster_def}.

\subsubsection{Assembly of the PMTs in the LSM}

Once a sector was transported to the LSM,
the 3'' PMTs were glued directly to the light guides, while the 5'' PMTs first had the interface guide glued to the light guide and then the interface/light guide combination glued to the 5'' PMTs. 
After a check to see that the system was light-tight, a $\mu$-metal shield was placed around each PMT. 

\subsection{The tracking detector\label{geigerdesc}}

The tracking volume of the NEMO~3 detector is made of layers of vertical drift cells working in Geiger mode. After a long period of research and 
development at LAL, which had the responsibility for the tracking portion of the detector, the optimal parameters were identified for the best resolution and 
efficiency while minimizing multiple scattering effects. This optimization involved a balance between two parameters which contribute to aging effects of the wires. The first is the diameter of the cell's wires which needs to be as low as possible for better transparency in the tracking device. The second is the proportion of helium and alcohol in the gas mixture.

\subsubsection{Elementary Geiger cell\label{nemo3_cell_descrip}}

The cross section of each cell is ``octagonal'' in design with a diameter of 
3~cm that is outlined by 8 wires. The basic cell consists of a central anode wire surrounded by the 8 ground wires, which belong to four adjacent cells to minimize the number of wires. However, between layers an extra ground wire was added to each cell to avoid electrostatic cross talk.
All the wires are stainless steel, 50~$\mu$m in diameter and 2.7~m long.
The wires are strung between the two iron petals of each sector which form the top and bottom of the detector (see  Fig.~\ref{nemo3_secteur_PM}). On each end of the cell, there is a cylindrical cathode probe, which will be referred to as the cathode ring. The cathode ring is 3~cm in length and 2.3~cm in diameter. 
The anode wire runs through the center of this ring while the ground wires are supported just outside the ring. If one compares NEMO~3 to the previous prototypes NEMO~1 and NEMO~2, there is a new mechanism for securing the wires in the petals. The advantages of this new mechanism are that it allows easy wire installation, avoids the radioactivity of solder, and provides simple serviceability if replacement is necessary. 

The cells work in the Geiger mode with initially a mixture of helium 
gas with ethyl alcohol. The characteristic operating voltage for the anode wires is 1800~V. When a charged particle crosses a cell the ionized gas yields around six electrons per centimeter. These electrons drift towards the anode wire at a speed of about 2.3~cm/$\mu$s when the electrons are close to the anode. When they are far away from the anode wire the mean drift velocity is around 1~cm/$\mu$s, because the layout of the wires (field and ground) establishes a varying electric field within each cell.

Measurements of these drift times are used to reconstruct the transverse position of the particle in the cells. The Geiger regime has a fast rise time for the anode pulse (around 10~ns) which can be used with a fixed threshold to provide a good time reference for the TDC measurements (see Section~\ref{elec}). In the Geiger regime, the avalanche near the anode wire develops into a Geiger plasma which propagates along the wire at a speed of 6 to 7~cm/$\mu$s, depending on the working point (Geiger plateau) which is a function of the gas mixture and operating voltage.  
The arrival of the  plasma at the ends of the wire is detected with the cathode rings mentioned above.  
The propagation times are used to determine the longitudinal position of the particle as it passes through the cell.

\subsubsection{NEMO~3 tracking device}

Tracking simulations in a 25 Gauss magnetic field were investigated. The study revealed the optimum configuration for the layers of cells in the sectors of NEMO~3. Taking into account multiple scattering in the track reconstruction the result was four layers near the source foil followed by a gap, then two layers and another gap followed by three layers near the scintillator wall (the ``4-2-3'' layer configuration, see Fig.~\ref{nemo3_secteur_PM}). 
Thus, there are nine layers on each side of the source foil to reconstruct tracks. The gaps between the groups of cell layers is due to the position of the plastic scintillators on the petals. The four layers near the source foil are sufficient to provide a precise vertex position. Two layers in the middle and three layers close to the plastic scintillator walls provide good trajectory curvature measurements. 

\subsubsection{Assembly and wiring of a sector}

To suspend the wire cells the iron petals have an intricate pattern of holes drilled into them in order to support the 4-2-3 layer configuration. 
The sectors were wired in a class 10,000 clean room.
Studies with prototypes confirmed the need for very high quality wire for proper plasma propagation. To fulfill this requirement a special production run of a 200~km long stainless steel wire, that was 50~$\mu$m in diameter,  was contracted\footnote{Trakus factory, Germany}. Precision in the diameter is better than 1\%.  Wires were strung layer by layer, alternating ground and anode wires.
The wiring of one sector took about four weeks for the 1991 wires.

Cells are of high quality if two conditions are satisfied simultaneously: at  operating voltage the counting rate is free of secondary triggers and electrical discharges while having the Geiger plasma propagated to both extremities of the cell. A special container measuring 6.4~m$^3$ was used to test each of the 20 sectors. 
The measured counting rates in the cells using cosmic rays is around 60~Hz compared to 0.2~Hz in the LSM before the coil and shields were added.
Thus a test for one week at LAL corresponds to several years of operation in the LSM. As a result of the anode tests, it was found that on average 10 of the 340 anode wires were replaced per sector. Fewer than 10 ground wires per sector needed to be replaced. This exercise also demonstrated the efficiency of cell repair. 

The final step, to ensure a tight seal to contain the helium, used silicon RTV to outline the regions to be sealed, and then Araldite.
A gas seal was also formed between the petals and walls, at the exterior of the sectors.

\subsection{Electronics, trigger and data acquisition systems\label{elec}}

The NEMO~3 detector has calorimeter and tracking detector independent electronics and data acquisition systems with a trigger system that can be inter-dependent.
The electronics, trigger and data acquisition are separated into modules which share a VME bus. This design permits not only $\beta\beta$ runs, but also different tests to adjust and calibrate the detector.

\subsubsection{Calorimeter electronics\label{elec_calo}}

The PMT bases are designed with a progressive voltage divider to improve the linearity under conditions of high current.
Tests carried out at the IReS laboratory identified fixed sets of resistors for the PMT bases which control the voltage between the photocathode and the first dynode so as to optimize the time resolution. 

Three large power supplies from C.A.E.N.\footnote{C.A.E.N. HV power supply type SY527 10 boards A938 AP, 24 channels each, with 
AMP multicontact connectors} are used to supply  the HV for the 1940 PMTs. Each HV channel is shared by three PMTs via a distribution board. The three PMTs must be similar in gain and fine tuning is done with two carefully selected fixed resistors on the distribution board for each PMT. 
 Data taken with a $^{207}$Bi source was used to set the PMTs' gains. There are nine distribution boards per sector and each board has four HV channels that in turn supply 12 PMTs.

The 97 PMTs of each sector are divided by the source foil into the interior and exterior regions.  A total of 46 PMTs are used for the interior 
region of which 12 PMTs are on the top and bottom petals. The exterior region is covered with 51 PMTs, again with 12 PMTs on the top and 
bottom petals.
Thus, there is a total of 40 half-sectors for which front-end electronics boards are designed. 
The corresponding 40 mother boards are housed
 in three VMEbus crates  
and each mother board supports 46 or 51 analog-NEMO (ANEMO) daughter boards.

The ANEMO boards have both a low and a high threshold discriminator. If the PMT signal exceeds the lower level threshold it starts a TDC measurement and opens a charge integration gate for 80~ns. The high threshold discriminator is adjustable up to 1~V but generally runs at 48~mV corresponding to 150~keV. The high threshold discriminator works as a one shot that delivers an event signal to the mother board.

Each mother board in turn provides an analog signal to the trigger logic which reflects the number of channels that have exceeded the upper threshold. The signal strength is 1~mA per channel. This level is used to trigger the system (1st level trigger) if the desired multiplicity of active PMTs is achieved.  
The trigger logic then produces a signal called {\it STOP-PMT}, which is sent to all the calorimeter electronic channels, to save their data. So the TDCs are stopped and the integrated charge is stored.  Then digital conversions begin.
At the same time, a signal  
is sent to the calorimeter acquisition processor, which  
permits the read out of the digitized times and charges  for the active channels. 

The analog-to-digital conversions of the charge  and the timing  signal are made with two 12-bit ADCs.
The energy resolution is 0.36~pC/channel ($\sim~3$ keV/channel) and the time resolution is 53~ps/channel. 
If any PMT signal exceeds the high level threshold then the TDC measurement and charge integration are aborted and the
system resets after 200~ns.

\subsubsection{Tracking detector electronics}

The Geiger electronics are  
divided into two types of boards. The first is for secondary voltage distribution, which provides 1800~V to the anode wires. 
Included on the secondary distribution boards are the analog signals from the anode wire and the two cathode rings. These boards receive high voltage from two of the C.A.E.N. power supplies\footnote{C.A.E.N. HV power supply type SY527 with A734P boards, 16 channels each}.
The second type of board is for the tracking electronics and is an acquisition board which is connected to the distribution board. It uses ASICs\footnote{Application Specific Integrated Circuit} and has an interface with a 50~MHz clock (20~ns per channel).

The functions of the acquisition board are first amplification and then discrimination of analog signals coming from the 
distribution board. Time measurements for each cell are acquired for the anode wire and the two cathode rings. Note that the two cathode times are identified as $t_{LC}$ and $t_{HC}$ where $LC$ and $HC$ stand for low and high cathode times. The low one corresponds to the cathode ring on
the bottom of the detector and the high  for the top.
 
Each of the 20 sectors needs the following electronics. Eight secondary distribution boards receive a total of 15 daughter boards, of which five for the anode, five for $LC$ and five for $HC$ signals. The five sets of three different daughter boards services eight cells per set, so that there are 40 cells per distribution board.
Then each sector also needs eight acquisition boards, which receive 10 analog ASIC and 10 digital ASIC circuits, with each ASIC handling four cells, so 40 cells per board.

Each of the four channels of the analog ASIC\footnote{1.20 micron technology from AMS inside PLCC 44 boxes} is used to amplify the anode and two cathode signals by a factor of 60, and to compare them to anode and cathode thresholds generated by a software programable  8-bit DAC. 
For signals exceeding the thresholds, a comparator provides a TTL signal which is sent to the TDC scalers of the digital ASIC.
There are four TDCs for each of the four channels of the digital ASIC\footnote{1.00 micron technology from ES2 inside PLCC 68 boxes}. The first three are for the anode ($tdc_A$), low cathode ($tdc_{LC}$) and high cathode ($tdc_{HC}$) contents which are measured with a 12-bit TDC and give times between 0 and 82~$\mu$s. The alpha TDC ($tdc_{\alpha}$) is 17-bits, which provides time measurements between 0 and 2.6~ms.

The anode signal starts the TDCs and creates an OR signal (called {\it HIT}) for all the cells of a layer of a given sector; so 360 TTL signals are sent to the T2 trigger (see Section~\ref{Trigg}).

The propagation of the Geiger plasma is detected by the cathode rings. 
These signals stop their respective cathode TDCs and give $tdc_{LC}$ and $tdc_{HC}$ values. Physical propagation times are proportional to these values:
\begin{equation}
t_{LC} = [tdc_{LC}  \times 20] \, {\rm ns}
\label{time_LC}
\end{equation}
\begin{equation}
t_{HC} = [(tdc_{HC} \times 20) - 17.5] \, {\rm ns}
\label{time_HC}
\end{equation}
The time constant, 17.5~ns, is removed from $t_{HC}$ to take into account the difference in cable lengths: the low cathode cables are 6~m and high cathode cables 9.5~m.

Concerning the anode signal, there are two cases which have to be distinguished.
The anode signal can exceed its threshold before or after the arrival of the {\it STOP-A} signal which comes from the trigger (see Section~\ref{Trigg}).
In the first case, for $\beta$-type events, the {\it STOP-A} signal is used to stop the $tdc_{A}$ channels which have received an earlier start signal. Anode times $t_{A}$, correspond to the transverse drift time given by:
\begin{equation}
t_{A} = [(tdc_{max} - tdc_{A}) \times 20 ] \, {\rm ns}
\label{anode_time}
\end{equation}
where $tdc_{max}$ corresponds to 6.14~$\mu$s. 

In the second case, for $\alpha$-type events, all Geiger cells not already triggered can register delayed hits which occur after the {\it STOP-A} has
been
received for up to 704~$\mu$s. Anode signals exceeding their threshold start not only their anode and cathode TDCs but also their alpha TDCs ($tdc_{\alpha}$).

Cathode signals stop the cathode TDCs and give $tdc_{LC}$ and $tdc_{HC}$ values, but the anode and alpha TDCs are stopped by the {\it STOP-$\alpha$} signal coming from the trigger. As a consequence, in\linebreak
\clearpage
\noindent
 this case the $tdc_{A}$ and $tdc_{\alpha}$ have the same value modulo 4096 for these cells. The corresponding alpha time $t_{\alpha}$ is then given by:
\begin{equation}
t_{\alpha} = [(tdc_{\alpha \_ max} - tdc_{\alpha}) \times 20 ] \, {\rm ns}
\label{alpha_time}
\end{equation}
where $tdc_{\alpha \_ max}$ corresponds to $\sim 704~\mu$s. 

\subsubsection{The NEMO~3 trigger system\label{Trigg}}

The trigger system
was developed by LPC. 
It receives one analog signal from each of the 40 half-sectors that is proportional to the number of PMTs that have exceeded their high threshold in that sector. The 40 signals are summed resulting in an analog signal. The trigger then goes onto the second level which involves the Geiger layers.
 In this case 360 channels are read out by treating each Geiger layer in each sector as a bit, which is on if the layer is hit. This information allows the use of a rough track recognition program to be run on the available Geiger cell information. It is then possible to refine the identified tracks by spatially connecting the Geiger cells and triggered PMTs.

\begin{figure}[h]
\vspace*{0.5cm}
\begin{center}
\includegraphics*[width=13cm]{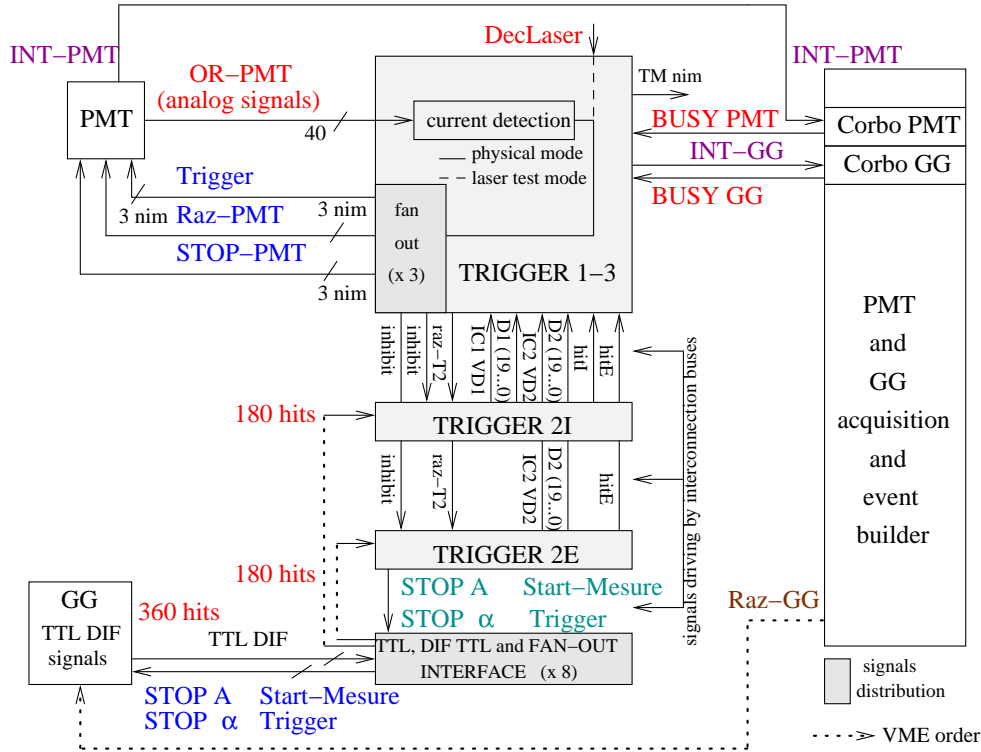}
\caption{\label{synop_trigg} Overview of the NEMO~3 trigger system.}
\end{center}
\end{figure}

Timing constraints and trigger strategy lead to a two level trigger system (see Fig.~\ref{synop_trigg}) for normal running with a third trigger for calibration runs.

The {\bf first level (T1)} of the trigger, is embedded on the T1-3 board, and is based on the number of PMTs required to initiate a readout. It is used to identify the number of active scintillators by using the summed
current from the 40 analog signals 
to define the multiplicity of the event {\it (MULT)}. 

If the trigger logic encounters a multiplicity higher than that set by the {\it MULT} threshold up to 20~ns after the first triggered PMT, T1 generates the {\it STOP-PMT} signal, 
which is the timing reference of the experiment, with an electronic accuracy better than 150~ps.

The {\bf second level (T2)} consists of track recognition in the Geiger layers ({\it GG}), using the 40 half-sectors. The track recognition is first performed on a per half-sector basis. Nevertheless, the probability for an electron to cross more than one sector is high, so tracks that cross two adjacent half-sectors are searched for in a second step. Thus, there are two secondary levels.

The first level search is between four different possible tracking patterns, which correspond respectively to a long track, 
a short track near the foil, 
a short track near the scintillator block 
or no track; the second level consists of making special associations between adjacent half-sectors. For example 
a long track in a half-sector and any kind of track or no track in the adjacent half-sector. One possibility is to use two short tracks in two adjacent half-sectors, one near the foil and the other near the calorimeter walls,  
which allows the trigger to select a full track contained in more than one sector.

This level is embedded on the T2I (internal) and T2E (external) boards, which receive 180 {\it HIT} signals each  and provide nine logical signals from a logical OR  which lasts for 1.5~$\mu$s on the cells of each layer. 
If the trigger logic is satisfied for the chosen track condition, a second level local trigger is generated. The T2 level conditions on the T2I/E boards are set in programmable memories.

The {\bf third level (T3)}, which is only used for calibration runs, is embedded on the T1-3 board and checks on the possible coincidence between pre-tracks from T2's second stage and fired {\it PMT} half-sectors. This level selects electron tracks coming from radioactive sources installed in the calibration tubes. It  is implemented in hardware without the possibility of changing the algorithm. 

For the case of an active PMT and Geiger cell trigger condition ({\it PMT+GG}), if the second level trigger is detected, the {\it STOP-A} signal is sent to Geiger acquisition boards with the programmable delay. This delay is set to 6.14~$\mu$s after the {\it STOP-PMT} signal. Then two trigger signals are sent to the Geiger and calorimeter electronics with a  programmable delay set to 6.14~$\mu$s  after the {\it STOP-PMT} signal. The first signal stops the automatic time-out which occurs 102~$\mu$s after the {\it STOP-A} signal. The second permits the digitization of the analog signals of the activated PMTs. In case these signals are not received there is an automatic reset of each of the PMT channels which have started measurements. Finally, the {\it STOP-$\alpha$}  
is sent to the Geiger acquisition boards with a fixed delay of 710~$\mu$s after the {\it STOP-PMT} signal.

\subsubsection{The NEMO~3 data acquisition system}

The control and readout of the calorimeter and Geiger cell crates is performed with the inter-crate VICbus and two CES RIO 8062 computer modules equipped with PowerPC 604E-300MHz CPU chips. The data acquisition system is based on {\it Cascade}~\cite{Cascade} operating under the Lynx-OS software package developed at CERN. It uses two boards: Corbo PMT for the calorimeter readout trigger and Corbo GG for the track detector readout trigger. The two independent acquisition processors collect  information in parallel. After the processors have read out the event and de-activated their ``busy'' signal, the trigger system reinitializes its logic electronics for the next event. Data buffers for the calorimeter and Geiger cells are then sent to the event builder processor (EVB) via the PVIC bus (multidrop PCI-to-PCI high speed link), as described in Section~\ref{acqui_result}.

\subsubsection{Survey of the experiment\label{slow-control}}

Monitoring and control of the experiment from remote sites is possible with two dedicated PCs in the LSM. The two PCs also operate locally. The tasks of these two PCs are not overlapping. The first controls the gas system of the tracking chamber, the current in the magnetic coil and the high voltage on the Geiger and PMT boards. The second controls the on/off power for the acquisition boards and crates, the high voltage crates and the uninterruptible power supply.

\subsubsection{The NEMO~3 database: {\it NEMO DB}}

The MySQL database management system is used for NEMO~3. Data synchronization in the {\it NEMO DB} server network is based on the replication concept of the  MySQL package. Here any number of servers can be replicate and transfer data from one primary server. The structure of the NEMO MySQL servers
includes the primary server at the LSM  for information stored in the on-line database, the main server at Lyon CCIN2P3\footnote{Centre de calcul de l'IN2P3: computer center of the institute for nuclear and particle physics} which contains all the data, and a set of local servers mirroring the main server.

The {\it NEMO DB} contains the electronic logbook of the runs, the ca\-li\-bra\-tion parameters of scintillator counters and Geiger cells, as well as other information about the run conditions.

\subsection{Energy and time calibration of the counters\label{calib}}

\subsubsection{Introduction}

In order to measure accuratly the absolute energy released in a double beta decay $(Q_{\beta\beta})$, a calibration procedure was established. 
The solution for NEMO~3 is to use radioactive sources that can be introduced into the detector and present only during runs dedicated to calibration.
These absolute energy measurements run for extended periods and take place only two or three times a year. Thus, daily studies of the stability of the counters are done with a laser survey system. 

Timing information is used to discriminate between external and internal events for background studies (see Section~\ref{bkg_explain}). The relative timing offsets for each of the 1940 counters has to be determined,  
using particles emitted in coincidence from $^{60}$Co radioactive sources. Particle times-of-flight are also corrected for several effects specifically the amplitude corrections due to leading edge discriminators (called time-energy corrections) and TDC slope corrections. These corrections are also checked with the laser survey system.

\subsubsection{Mechanics of the calibration tubes}

Each of the 20 sectors of the detector is equipped with a vertical tube made of flattened copper that is 
located along the edge of the source foils (see Fig.~\ref{nemo3_secteur_PM}) 
and three pairs of kapton windows: 
one window of the pair is oriented towards the internal wall and the other towards the external one. The size of the windows ($\sim 500$~mm$^2$) and their vertical positions ($z = -90,~0$ and +90~cm with an accuracy better than 1~mm) have been chosen to obtain an approximately uniform illumination of the scintillator blocks by three radioactive sources placed inside the tube. 
The source carrier is a long narrow delrin rod 
supporting three source holders which are introduced into the copper tube from the top of the detector, after the removal of some shields on the top. 

\subsubsection{Radioactive sources}

For energy calibration of the counters, one needs radioactive sources which emit electrons. The choice was made to use $^{207}$Bi and $^{90}$Sr sources.
Decay of the first one provides conversion electrons of 482 and 976~keV energy (K lines) suited for an energy calibration up to 1.5~MeV.  The products in $^{207}$Bi decay are essentially $\gamma$-rays, thus the tracking chamber must be in operation to select electrons originating directly from the sources. 
To measure energies up to 3~MeV or more, one needs at least one additional calibration point, which is obtained using electrons from $^{90}$Y (daughter of $^{90}$Sr) and measuring the end-point of the $\beta$ spectrum at 2.283~MeV. This calibration does not require pattern recognition because the events of interest are located in the tail of the spectrum, which can only contain electrons coming directly from the sources. Relatively intense sources are used here for short runs.

For timing calibration, the relative offsets for each channel are determined with a $^{60}$Co source, which emits two coincident $\gamma$-rays with energies of 1332 and 1173~keV. Spectra of arrival time differences are collected to establish time delays between the 1940 channels. This time calibration does not require the use of the tracking chamber and allows the use of relatively intense sources.

\subsubsection{Laser calibration system}

\begin{figure}[h]
\begin{center}
\includegraphics[width=14cm]{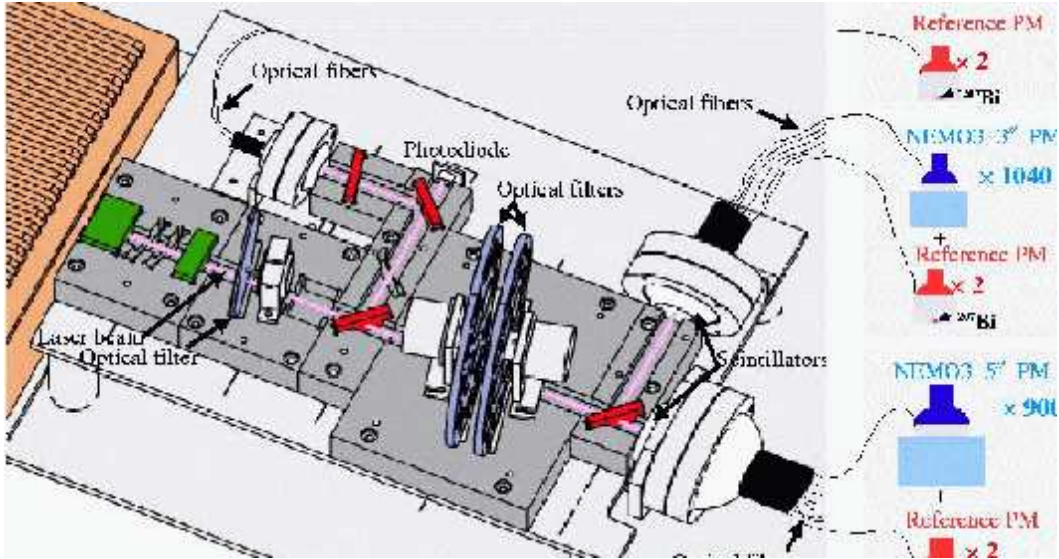}
\caption{\label{laser_system} The laser calibration system.}
\end{center}
\end{figure}

In an experiment such as NEMO~3, which requires stability for many years over a large number of PMTs, frequent tests of the detector's stability are of paramount importance. 
The objectives of the chosen laser calibration system\footnote{MNL 202 laser from Lasertechnik Berlin; 200~kW, 10~Hz}, as shown in Fig.~\ref{laser_system}, are first a daily check of the absolute energy and time calibration, which permits the calculation of the corrections according to the emission peak. Next measurements of the PMTs linearity between 0 and 12~MeV are made and finally is determined the time-energy relation. To accomplish this, the shape of the laser signal has to be very similar to the one produced by an electron. Thus the laser light must be known with very high accuracy ($< 1\%$) and must be stable during the measurement. One needs also a precise calibration of the optical filter transmission for the energy range 0 to 12~MeV and a good common time reference (STOP).

The N$_2$ laser wavelength of $337\pm 15$~nm was selected. Then the light beam is split into two parts. The first is sent to a photodiode, which monitors the laser light intensity. A variable optical attenuator is used to adjust the flux. The second beam goes through  two optical filter wheels to simulate the full range of energy. This beam is then wavelength shifted to 420~nm and sent to the NEMO~3 PMTs by means of optical fibers. The shifter is a sphere of scintillator, wrapped in Teflon and aluminum to increase diffusive reflections, used to mimic the signal of electrons in the scintillators. Each optical fiber\footnote{Toray PFU-CD501-10-E} is divided into two strands and the optical distance between these two is adjustable using individual attenuators, which allow the distribution of the same flux of light to all the counters. Six reference counters (independent of the NEMO~3 calorimeter) equipped with $^{207}$Bi sources allow the monitoring of the laser by measuring energies of both the laser and the 976~keV conversion electrons emitted from $^{207}$Bi. 

The daily laser procedure is carried out in two steps. The first one is during the run under standard conditions for the acquisition of $\beta\beta$ events ($\beta\beta$ run). It consists in stabilizing the laser and  checking parameters, such as photodiode pedestal and laser light. During this step the flux can be corrected if necessary. The second step is made at the end of the $\beta\beta$ run. It consists of a pedestal measurement for the PMTs\footnote{1940 PMTs from the NEMO~3 calorimeter and six PMTs for the reference counters}, then a rotation of two filter wheels to obtain a 1~MeV equivalent flux, this is followed by the acquisition of the laser and $^{207}$Bi events. Finally the laser is turned off and the filter wheels moved to opaque settings.
For ``complete'' calibration run up to 12~MeV, the procedure is only repeated few times during the year with different fluxes and changing the rotation of the two optical attenuator disks for each run. 

\subsubsection{Absolute energy calibration method}

Over the full 12~MeV range of energies measured by NEMO~3, the relation between the charge signal and energy deposited in the counter is not necessarily linear. The relation is, however, linear up to 4~MeV, where the greatest accuracy is required: 
\begin{equation}
E \, = \, a \, (C - P) \,  + \, b
\label{energy}
\end{equation}
here $C$ is the ADC value of the scintillator and $P$ is the pedestal.
The energy calibration constants $a$ and $b$ are determined using at least two points from measurements with radioactive sources, while the 
 relation for the energies greater than 4~MeV is determined using a ``complete'' laser calibration run.

The energy resolution is assumed to have essentially two contributions. The principal component originates from the statistical fluctuations of the scintillation photons and from the number of photo-electrons at the PMT anode. It increases as the square root of the energy. The second component is due to the instrumental effects which are energy independent. These two terms contribute to the resolution in quadrature in the form:
\begin{equation}
\sigma(E) \, = \, A \sqrt{E} \, \oplus \, B
\label{resol}
\end{equation}
Using 60 $^{207}$Bi sources ($\sim 220$~Bq each) with the high threshold set at 48~mV ($\sim 150$~keV), a 24 hour run yields about 5000 events for each PMT with identifiable tracks. Then the positions of the two peaks, corresponding to the 482 and 976~keV conversion electrons, are extracted and the resolution is obtained as a result of a fit to the 976~keV line.  

Using four $^{90}$Sr sources simultaneously ($\sim 6$~kBq each) eight runs are taken with the sources in different positions and a threshold of 48~mV ($\sim 150$~keV). For each PMT about $5 \times 10^4$ events are used to form a spectrum (see top of Fig.~\ref{fit_90sr}).  A fit to the high-energy tail of the spectrum is made with a function describing the shape of a single $\beta$ spectrum of $^{90}$Y, convolved with the energy resolution function $\sigma(E)$ and taking into account the mean energy loss 
of the electrons in the gas of the wire chamber.

Finally, these three results (two peaks and one end-point) from the calibration runs with $^{207}$Bi and $^{90}$Sr are combined to extract the parameters $a$ and $b$ of a linear calibration valid for the energy region up to 4~MeV. The results of energy calibration runs are presented in Section~\ref{cluster_def}.

\subsubsection{Energy corrections using the laser system}
\label{laser_description}

The laser procedure is carried out as a reference just after the calibration runs and gives a fiducial reference energy, $E_0$, for each counter (one assumes there is no correction to be applied to the $b$ offset). At a later time $t$ a new value of the energy, $E_t$, is measured:
$$E_0 = a_0 (C_0 - P_0) +b~~~~~{\rm and}~~~~~E_t = a_t (C_t - P_t)+b = e_{corr} \times a_0 (C_t - P_t) + b$$
where $e_{corr}$ is the correction factor to be applied. It represents the variation of the calibration slope of the counter as a function of time between $t_0$ and $t$.
This variation of the laser is determined by comparing the laser peak position ($C_{laser}$) and the 976~keV peak position from the $^{207}$Bi ($C_{Bi}$) between $t_0$ and $t$ for the six reference counters. Results are then transferred to the database. 

The ``real'' energy value $E^{(i)}_{(t)}$ for counter number $i$ at instant $t$ is finally given by:
\begin{equation}
E^{(i)}_{(t)} = [ (C^{(i)}_{(t)} - P^{(i)}_{(t)})  \times a^{(i)} \times e_{corr_{(t)}}^{(i)}] \, + \, b^{(i)}
\label{final_energy}
\end{equation}

\subsubsection{Time calibration}

The timing response of two counters detecting two particles emitted in coincidence depends not only on the time-of-flight of each particle, but also on several effects which have to be corrected for.

$\bullet$ {\bf Time alignment of the counters}

All PMTs must be aligned in time. The time taken for each PMT to respond to a signal depends on the individual characteristics of the counter. In order to use only one time scale, an alignment procedure has been developed to obtain the individual time shifts $\varepsilon^{(i)}$ ($i = 1$ to 1940). 
The procedure to find $\varepsilon^{(i)}$ uses $\gamma$-rays emitted in coincidence by a decay in a $^{60}$Co radioactive source and detected by a pair of scintillators. 
There is a common {\it START} in the electronics for both counters and the electronics uses a common {\it STOP-PMT} for all the counters, which allows the individual delays to be extracted.

In order to calculate the time-of-flight correctly, only one $^{60}$Co source of 15.5~kBq is used per run. Ten runs with different source positions are performed in order to cover all possible combinations of PMTs.
A threshold of 150~mV ($\sim 500$~keV) is set for the arrival time difference spectra and the relative timing offset $\varepsilon^{(i)}$ for each counter $i$  (see Section~\ref{cluster_def}).
	
$\bullet$ {\bf Other corrections}

The effect of leading edge discriminators is to induce a time-vs-energy dependence, which can be described with a formula using four parameters
\begin{equation}
t(C) = p_1 - \frac{p_2}{p_3 \sqrt{C} + p_4}
\label{time_vs_energy}
\end{equation}
and taking into account the pulse shape. Determination of the parameters $p_k^{(i)}$ ($k = 1$ to 4 for counter $i$) is accomplished through a ``complete'' laser run producing equivalent energies between 0 and 12~MeV. The relative timing offset $\varepsilon^{(i)}$ for counter $i$ is then included in the asymptotic value $p_1^{(i)}$.

The laser survey system is also used to obtain daily time response corrections for each counter, which correspond to TDC slope variations:
$t_{corr} = tdc_{t} - tdc_{0}$.

Finally, the ``real'' time, $T^{(i)}_t$, used for a time-of-flight calculation for counter number $i$ at instant $t$ is:
\begin{equation}
T^{(i)}_{(t)} = tdc^{(i)}_{(t)} - t_{corr_{(t)}}^{(i)} - t(C_{(t)}^{(i)})
\label{final_time}
\end{equation}

\subsection{Coil and shields}

\subsubsection{Introduction}

To reach the desired sensitivity for the effective neutrino mass, there must be no events ($<0.1$) from external backgrounds in the energy range [2.8 - 3.2]~MeV during five years of data collection. 
The external background contribution coming from neutrons is due to $(\alpha,{\rm n})$ reactions, spontaneous fission of uranium and the interaction of cosmic ray muons in the rocks. The other external background contribution is the $\gamma$-ray flux produced in the LSM, which has been studied using a NaI detector surrounded by different shields~\cite{Ohsumi}. The origins of these $\gamma$-rays are natural radioactivity, radiative neutron captures and the bremsstrahlung of muons.

As shown in Section~\ref{radiopurity}, the detector has been designed  with stringent radiopurity  for its construction materials. For the external background  coming from $\gamma$-rays and neutrons, several studies were made with the NEMO~2 detector as well as the NEMO~3 simulations~\cite{Marquet2}. These have shown that there is a large reduction in these backgrounds given the following conditions to the experiment. A solenoid capable of producing a field of 25~G is surrounded by two external shields, the first one to reduce $\gamma$-ray and thermal neutron fluxes, the second to suppress the contribution of slow and fast neutrons. The design of the coil and  shields allows for partial dismantling of the detector to access each sector.

\subsubsection{The magnetic coil}

The simulation of the fast neutrons coming from the laboratory into NEMO~3 was carried out with 20~cm of iron shield. The contribution to the $\beta\beta 0\nu$ background by the $\gamma$-rays created from neutron captures leads mainly to $(e^+ e^-)$ events and also to a few $(e^- e^-)$ events produced in the source foils. The detection by the calorimeter of the $\gamma$-rays associated with these events provides a high rejection efficiency (80\%). A 25~G magnetic field, which provides the charge recognition rejects 95\% of the $(e^+ e^-)$ pair events.

The coil surrounds the entire detector and access to the detector was a necessary design consideration. Thus, the coil is made of ten sections 
with 203 copper rings connecting every other sector to form one loop of the helix  (see Fig.~\ref{nemo3_last}). 
The finished coil  is cylindrical, 5320~mm in diameter, 2713~mm in height and has a mass of 5~tons ($\sim 3.1$~tons of high radiopurity copper).

\subsubsection{Iron shield}

The iron shield is also made in ten sections (165~tons) with two end-caps (6~tons each). The lower end-cap is fixed and the upper one is removable. The iron shield is 20~cm thick, except for a few places where it is 18~cm on account of mechanical supports. The iron was selected for its radiopurity, as recorded in Table~\ref{bilan_activ}.

\subsubsection{Neutron shield}

The remaining $(e^+ e^-)$ pair events (5\% not rejected by curvature measurements) and the $(e^- e^-)$ events due to the $\gamma$-rays created by neutron capture can be suppressed only if the flux of neutrons inside the detector is decreased. NEMO~3's neutron shield is optimized to stop fast neutrons with an energy of a few MeV, it also suppresses thermal and epithermal neutrons.

The neutron shield is made of three parts, as shown in Fig.~\ref{nemo3_last}. The first one 
is situated below the central tower of the detector (not shown in the figure) and consists of paraffin 20~cm thick. The second part covers the end-caps below and above the detector, and consists of 28~cm of wood. The cylindrical external walls are covered with ten large tanks  
containing borated water which are 35~cm thick and separated by wood 28~cm thick.

\subsection{Mounting and assembly of the detector in the LSM\label{realisation}}

\subsubsection{Supporting structure}

The steel framework, which supports the NEMO~3 sectors, is made of two parts. 
It was installed in the LSM at the end of 1998 and supports the 20 sectors of the detector, the magnetic coil and the various shields. All the components of the framework were selected for their radiopurity.
To properly enclose the active detector and have access to readout electronics in the narrow experimental hall, the detector was raised off the ground two meters. Thus the control and readout electronics can be housed under the detector. This structure preserves proximity and serviceability of the electronics and detector.

\subsubsection{Placing of the sectors on the supporting structure}

Once the sectors were made helium tight, the anode and cathode cables were connected at LAL before being transported to the LSM, where PMTs were attached. After gluing and testing the PMTs, the sectors were carried inside the source mounting room and cleaned using alcohol and nitrogen gas.

The source frame, which supports the seven foils of each sector, was prepared simultaneously in the LSM clean room.  
This source frame was then mounted in the sector. Finally, the calibration tube was set in position and the sector was cleaned once again with nitrogen and closed with sheets of Mylar.

The source mounting room was then opened to move the sectors to the supporting structure. The calorimeter cables (HV and signal) were soldered to the PMT bases and to the distribution daughter boards. These operations were followed by the introduction of the plastic optical fibers into brass tubes which are present in each light guide of the calorimeter.
Finally there was a through test of the connections for the calorimeter, the tracking detector and the laser system (both electrical and optical). During these procedures the sectors were filled with nitrogen to avoid  deposition of impurities on the wires.

These tasks were previously checked in the year 2000 for three sectors (00, 18 and 19).
The 17 remaining sectors of NEMO~3 were gradually installed and tested until August 2001. The 20 sectors were then placed in their final positions
and the helium seals made using RTV glue.

The total fiducial volume inside the tracking detector is 28~m$^3$ (excluding the scintillator volume). The detector was filled with gas in December 2001, using a fast flushing of helium gas (5 volumes), followed by introduction of ethyl alcohol in the gas regulation system. The whole process took approximately 24~h. A constant pressure of gas was maintained with a flux of the order of 130~l/h until December 2002. The magnetic coil was installed in February 2002 and $\gamma$-shield in April 2002. Finally, the installation of the neutron shield was completed in February 2003. Between each step, $\beta\beta$ and calibration runs were taken to monitor stability and the effectiveness of the shields for comparisons with simulations. 

Data collection was temporarily stopped in December 2002 to make several improvements. These involved changes in HV power supplies for the calorimeter PMTs' overvoltage protection, while noise and gain adjustments were made on several PMTs. Helium leaks were sealed and there was an addition of argon into the gas mixture of the tracking detector in order to improve the Geiger cell stability and efficiency of the plasma's longitudinal propagation as well as limiting the aging problems. At the same time, the proportion of quencher in the mixture was reduced from 40 to 39~mbars partial pressure (ethyl alcohol at 14$^o$C instead of 15$^o$C).
The gas mixture used in the day-to-day running is now 
He with Ar (with $V(Ar)/V(He)=1\%$) and ethyl alcohol as the quencher  at a flux of  300 l/h. An overpressure of $\sim 10$~mbars is maintained to flush contaminants from the surrounding gas volume. 

In mid-February 2003, new $\beta\beta$ runs commenced under improved conditions, with the coil and both shields.

\subsection{Radiopurity of the detector\label{radiopurity}}




$\gamma$-ray spectra were acquired using low background HPGe detectors located at CENBG and in the LSM. These HPGe spectrometers typically use the ``Marinelli'' geometry to measure the abundance of selected isotopes in the decay chains via the strength of selected gamma lines. 

$\bullet$ {\bf Radioactivity measurements of the source foils}

A 400~cm$^3$ HPGe spectrometer in the LSM has a sensitivity capable of reaching 0.2~mBq/kg of $^{214}$Bi and 0.06~mBq/kg of $^{208}$Tl in a month with a 1~kg sample. 
The maximum allowed levels of $^{214}$Bi and  $^{208}$Tl contamination in the Mo sources are given in Eq.~\ref{lim_max_bdf_Bi_Mo} and \ref{lim_max_bdf_Tl_Mo}. The level of sensitivity necessary for  $^{214}$Bi was achieved, but not for  $^{208}$Tl. A solution could be reached by counting for more than one month but the collaboration has decided to make a compromise between the source measurement and the detector's construction. 
Radioactivity measurements for all the NEMO~3 sources are summarized in Table~\ref{activ_autres_sources}. Only the lower limits obtained for $^{100}$Mo are presented, for both the metallic and composite strips. In this table it can be seen that for $^{208}$Tl only limits have been obtained except in $^{82}$Se and $^{150}$Nd$_2$O$_3$ strips. In the case of $^{82}$Se, the presence of ``hot-spots'', which can be suppressed by analysis, is anticipated. For $^{150}$Nd, it is not a problem because the $Q_{\beta\beta}$ value (3367.1~keV) is well above the 2615~keV $\gamma$-ray produced in the decay of $^{208}$Tl.

$\bullet$ {\bf Radioactivity measurements for the other components of NEMO~3\label{Choix_PM}}

As explained in Section~\ref{PMT}, the Hamamatsu company was chosen to produce the low background PMTs, with very low contamination in all the components (glass, insulator, ceramics, etc).  
In NEMO~3 the total mass of the PMTs is 239.2~kg for the 3'' tubes and 346.5~kg for the 5'' tubes, which yield the total activities presented in Table~\ref{activ_PM_tot}.

After selection of the PMTs, the other materials (except shields) were chosen if their total activities in $^{40}$K, $^{214}$Bi and $^{208}$Tl were at most one tenth of the total activities for the PMTs. It was not always possible to satisfy this criterion for all three contaminants simultaneously. If $^{40}$K exceeds the threshold, the material may have been accepted but not for $^{214}$Bi and $^{208}$Tl which produce background events in the $Q_{\beta\beta}$ region. In Table~\ref{activ_compo_PM}  the measurements for all selected components of a 5'' PMT are summarized. The two last lines of this table provide a comparison between the activity of the selected calorimeter materials and the total activity of the PMTs, the first one being at most of the same order as the PMTs. 

\begin{table}[h]
\centering
\begin{tabular}{|c|c|c|c|c|} \hline
\multicolumn{1}{|c|}{Sample} & \multicolumn{1}{|c|}{Weight } &\multicolumn{1}{|c|}{$^{40}$K } &
\multicolumn{1}{|c|}{$^{214}$Bi } & \multicolumn{1}{|c|}{$^{208}$Tl} \\ 
& {\bf (kg)} & {\bf (mBq/kg)} & {\bf (mBq/kg)} & {\bf (mBq/kg)} \\ \hline \hline
Magnetic shield & & & & \\
(1.5~mm thick) & 1.385 & $< \, 20$ & $< \, 2$ & $< \, 2$\\ \hline
PMMA (light guide + interface) & 1.500 & $< \, 50$ & $< \, 5$ & $< \, 3$\\ \hline
Iron ring for light guide & 0.555 & $< \, 35$ & $< \, 3$ & $< \, 3$\\ \hline
Light-tightness sleeve   & 0.191 & $< \, 80$ & $< \, 17$ & $< \, 12$\\ \hline
PMT-Guide glue RTV 615  & 0.072 & $380\pm 100$ & $< \, 40$& $< \, 7$\\ \hline
Light-protection glue for PMT & & & & \\
RTV 106   & 0.025 & $250\pm 70$ & $< \, 20$ & $< \, 7$\\ \hline
Teflon ribbon (5 layers) & 0.023 & $< \, 170$ & $< \, 20$ & $< \, 5$\\ \hline
Guide-Iron ring & & & & \\
glue Epotek 310  & 0.017 & $< \, 110$ & $< \, 15$ &$< \, 3$ \\ \hline
Isolated circuit FR2 & & & &  \\
for PMT divider & 0.0105 & $320\pm 100$ & $57 \pm 8$ & $10 \pm 3$ \\ \hline
Light-protection glue RTV 116  & 0.006 & $< \, 830$ & $< \, 45$ &$< \, 4$ \\ \hline
Scintillator-Guide  & & & & \\
glue BC 600 & 0.004 & $< \, 235$ & $< \, 42$ &$< \, 7$ \\ \hline
Aluminized mylar  & 0.004 & $< \, 700$ & $< \, 35$ &$< \, 20$ \\ \hline
White RTV 160 & 0.003 & $370\pm 130$ & $35\pm 10$ &$< \, 4$ \\ \hline
Capacitor 3.3 nF - 2 kV (1 piece) & 0.003 & $< \, 230$ & $< \, 17$ & $48\pm 5$ \\ \hline
Capacitor 22 nF - 250 V (4 pieces) & 0.002 & $< \, 1300$ & $< \, 80$ &$< \, 55$ \\ \hline
{\it ``Radiall''} solder & & & & \\
Sn(63\%) Pb(37\%) & 0.002 & $< \, 400$ & $< \, 50$ &$< \, 30$ \\ \hline
CMS resistors (23 pieces) & 21~10$^{-5}$ & $< \, 9200$ & $1400\pm 500$ & $< \, 600$ \\ \hline  \hline
Total activity of these  & & & & \\
components {\bf in Bq/PMT}  & & $< \, 0.25$ & $< \, 0.028$ & $< \, 0.016$ \\ \hline
Total activity for one & & & & \\
Hamamatsu 5'' PMT {\bf in Bq/PMT} & & 0.53 & 0.24 & 0.014 \\ \hline
\end{tabular}
\vspace*{1.cm}
\caption{\label{activ_compo_PM}Radioactivity measurements for 5'' PMT components (in mBq/kg). Total radioactivity of these components is compared to the radioactivity of one PMT (in Bq/PMT).}
\end{table}

Finally, to conclude the NEMO~3 radiopurity report, a summary of the radioactivity measurements is given in Table~\ref{bilan_activ} for the majority of the components of the detector.

\begin{table}[h]
\centering
\begin{tabular}{|c|c|c|c|c|c|} \hline
\multicolumn{1}{|c|}{Components} & \multicolumn{1}{|c|}{Weight} &
\multicolumn{4}{|c|}{Total radioactivity (Bq)} \\ \cline{3-6}
\multicolumn{1}{|c|}{of NEMO~3} &\multicolumn{1}{|c|}{in kg} & \multicolumn{1}{c|}{$^{40}$K} &
\multicolumn{1}{c|}{$^{214}$Bi} & \multicolumn{1}{c|}{$^{208}$Tl}& \multicolumn{1}{c|}{$^{60}$Co}\\
\hline \hline
{\bf Photomultiplier} & & & & & \\
{\bf Tubes}  & ~600 & 831 & 302 & 17.8 & $//$ \\
\hline \hline
Scintillator blocks & ~6400 & $< \, 102$ & $< \, 1.2$ & $< \, 0.6 $ & $< \, 3 $\\ \hline
Copper & ~25000 & $< \, 125$ & $< \, 25$ & $< \, 10 $ & $< \, 6 $ \\ 
\hline
Iron petals  & ~10000 & $< \, 50$ & $< \, 6$ & $< \, 8 $ & $17 \pm 4$ \\ \hline
$\mu$-metal PMT  & & & & & \\
shield & ~2000 & $< \, 40$ & $< \, 4$ & $< \, 4$ & $< \, 4$ \\ \hline
Tracking detector & & & & & \\
wires & ~1.7 & $< \, 8 \, 10^{-3}$ & $< \, 10^{-3}$ & $< \, 6 \, 10^{-4}$ &  $< \, 10^{-2}$\\ \hline\hline
Iron shield & ~180000 & $< \, 3000$ & $< \, 400$ & $< \, 300$ &  $< \, 600$\\ \hline
\end{tabular}
\vspace*{0.5cm}
\caption{\label{bilan_activ}Total radioactivity for the components. The PMT activities are lower than the design specification. Moreover, with the exception of the iron shield, all activities are clearly lower than the PMT's.}
\end{table}

\section{Performance of the detector}

\subsection{The simulation program}

The NEMO~3 simulation program~\cite{Roger_crn97} has been developed in the framework of GEANT~\cite{GEANT},  with the EUCLID industrial software and the EUCLID-GEANT interface~\cite{EUCLID-GEANT}. The description of the NEMO~3 device (geometric information, description of more than 60 materials and tracking media) is given with 20 identical sectors, except for the source foils which are placed in their exact positions inside the detector. 

The event generator of the program, called {\it GENBB}, provides the possibility of generating different double beta decays (0$\nu$, 2$\nu$, Majorons).  Also internal and external background events due to decay of radioactive nuclei may be generated. This event generator also generates the kinematics of special events, such as Compton scattering of external $\gamma$-rays or M\"oller scattering of external electrons, thereby speeding up the computations compared to the simulated events with GEANT.

\begin{figure}[h]
\begin{center}
\includegraphics[width=15cm,scale=1]{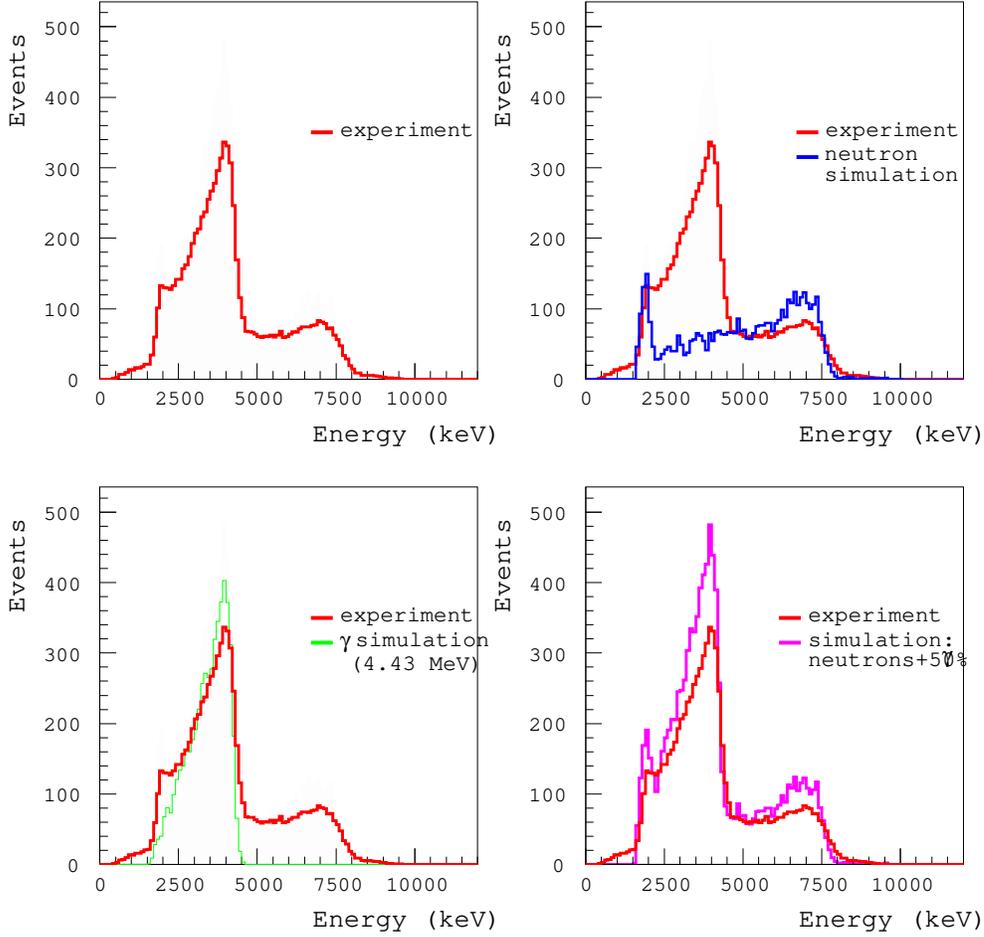}
\caption{\label{neutron_Cecile} Comparison between simulation and experimental data with a magnetic field and no shield, for one-crossing-electron events (with and without $\gamma$-rays) obtained with a neutron Am-Be source. See text for more details.}
\end{center}
\end{figure}

Important work has been done with the simulation of neutron interactions via the GEANT/MICAP code~\cite{GEANT/MICAP}. This code tracks neutrons from 20~MeV to 10$^{-5}$~eV. It also takes into account $\gamma$-ray emission from (n,$\gamma$) captures and (n,n'$\gamma$) scatterings. Using the results of several tests done with a Germanium detector and the NEMO~2 detector~\cite{Marquet2}, the $\gamma$-ray generation subroutines were improved by including additional spectroscopic information related to nuclei. Also a new library, GAMLIB, was developed taking into account branching ratios down to as low as 0.1\%. This library takes into account the possibility of emission of conversion electrons, which is particularly important for neutron captures with the large internal conversion coefficients observed in many nuclei. A comparison between the simulation and the data with a Am-Be neutron source, using the magnetic field and no shield, is presented in Fig.~\ref{neutron_Cecile} and shows the excellent agreement for high energy crossing electron events $e_{cross}$ (with $E_e >3.2$~MeV)~\cite{Cecile_these}. 

In Table~\ref{compar_neutron}, the experimental number of ($e_{cross}N\gamma$) events (with $N \ge 0$) observed per hour without a shield is compared to the expected number of events from simulations. These simulations take into account the $\gamma$-ray flux in the laboratory as well as thermal, epithermal and fast neutron fluxes, for which the respective proportions are given in this table. It was assumed that the epithermal neutron flux was equal to the thermal one. The resulting number of ($2.19 \pm 0.17$) observed events is  compared to ($2.05 \pm 0.25$) simulated events per hour. The agreement is still very good with the $\gamma$-ray shield, which provides an attenuation factor of $3.2 \pm 0.4$ with a $1200$ rejection factor for $\gamma$-rays and a $100$ one for thermal neutrons. The number of associated two-electron background events due to the neutron flux, with the $\gamma$-ray shield, is $13.6 \pm 4.4$ for the summed electron energies greater than 2.75~MeV~\cite{Cecile_these}. Recollect that the experiment requires that there be no external background. So there was the addition of the neutron shield, which provides an attenuation factor of $70$, which according to the simulation will suppress this background.

\begin{table}[h]
\centering
\begin{tabular}{|l|l|l|r|} \hline \hline
\multicolumn{1}{|c|}{Type of events} & \multicolumn{1}{|l|}{Simulated} & \multicolumn{1}{|c|}{$e_{cross}N\gamma$ events} & \multicolumn{1}{|c|}{Attenuation} \\
 & flux & in 1~h ($N \ge 0$)& factor with  \\
 & (x 10$^{-6}$~s$^{-1}$cm$^{-2}$) & ($E_e > 3.2$~MeV) & $\gamma$ shield \\
 & & without shield & \\ \hline \hline
$\gamma$-rays from laboratory & $3.2 \pm 0.9$ & $0.20 \pm 0.05$ & 1176  \\
Thermal neutrons  & $1.6 \pm 0.1$ & $0.50 \pm 0.05$ & 98  \\
($E~\le~0.1$~eV) & & & \\
Epithermal neutrons & $1.6 \pm 0.1$ & $0.8 \pm 0.2$ & 3.2  \\
(0.1~eV~$<~E~<$~1~MeV) & & & \\
Fast neutrons & $4 \pm 1$ & $0.55 \pm 0.15$ & 1.1 \\ 
($E~\ge~1$~MeV) & & & \\ \hline
Simulation total & & $2.05 \pm 0.25$ & 2.7  \\ \hline \hline
Experiment total & & $2.19 \pm 0.17$ & 3.2  \\ \hline \hline
\end{tabular}
\vspace*{0.5cm}
\caption{\label{compar_neutron}Comparison between simulated data and experimental data that studies the neutron contamination, without shield (3rd column) and with the $\gamma$-ray shield (4th column).}
\vspace*{0.5cm}
\end{table}

\subsection{Trigger and data acquisition\label{acqui_result}}
\vspace*{-0.5cm}
\subsubsection{Acquisition and data file building}

Data transfer is achieved via a dedicated Ethernet. The EVB (400~MHz, with a Powerful Ethernet controller) sends data event by event to a PC running Linux.  In the transfer, Linux processes the data by decoding it and re-organizing it into Random Access ZEBRA files. The initial format of {\it Cascade} events uses a 12-byte header, followed by calorimeter and Geiger cell information with variable lengths (two bytes per triggered PMT and three bytes per fired cell), and finally a 3-byte trailer. The final format is an ``ntuple'' format for data analysis. The ntuple building is done on the local acquisition disc and the Zebra file is written every 5000 events.

\subsubsection{Trigger type properties\label{trigger_perform}}

$\bullet$ {\bf The $\beta\beta$ trigger configuration}

During the $\beta\beta$ runs, the trigger configuration (see Section~\ref{Trigg}) is the two level one, which enables the readout  of events with at least one electron or electron-gamma event. The associated counting rate is $(7.4\pm 0.1)$~Hz. This configuration has the following properties.  
At least one triggered scintillator is required with energy deposited greater than 150~keV. Next, a track is reconstructed in a half-sector corresponding to the activation of Geiger cells in at least three of the nine layers and at least two fired cells in neighbouring layers (using at least two out of the four layers near the source foil, the two intermediate layers, or at least two out of the three layers near the scintillator wall). 

The $\beta\beta$ trigger efficiency  
was estimated by applying trigger criteria to simulated signal and background events with at least one fired Geiger plane and at least one active PMT. The trigger efficiency check was made with $\beta\beta 0\nu$ or $\beta\beta 2\nu$ events from the $^{100}$Mo source foils, $^{214}$Bi and $^{208}$Tl background events from the $^{100}$Mo source foils, and $^{214}$Bi decays in the gas. The number of generated events for each type was 10000. The results are summarized in Table~\ref{trigger_eff}, for events with at least one  PMT or two PMTs. 
The proportion of accepted simulated events are given in the first and third columns for at least one and two PMT(s) respectively. The lost events are mainly due to geometrical cuts. The proportion of accepted events after applying the trigger criteria are given in the second and fourth columns. It represents the trigger efficiency (the error bars are statistical uncertainties).  The efficiency relative to the number of events with exactly two fired PMTs is the number between squared brackets in the last column.
This study shows that trigger criteria do not provide additional cuts and allows one to keep data for further analysis of all interesting events. 
The trigger efficiency relative to the number of events with exactly two fired PMTs (criterium for $\beta\beta$ analysis) are 100\% for $\beta\beta$ events (both $0\nu$ and $2\nu$), 96.7\% and 94\% for $^{214}$Bi and $^{208}$Tl in the sources and 74.5\% for $^{214}$Bi in the gas.

\begin{table}[h!]
\vspace*{0.2cm}
\centering
\begin{tabular}{|c|c|c|c|c|} \hline \hline
\multicolumn{1}{|c|}{Type} & \multicolumn{1}{|c|}{Percentage of events} & \multicolumn{1}{|c|}{Associated}  & \multicolumn{1}{|c|}{Percentage of events} & \multicolumn{1}{|c|}{Associated}  \\
of  & with at least & trigger & with at least & trigger \\
events & one PMT & efficiency & two PMTs & efficiency  \\
 & (simulation & (\%)& (simulation & (\%)  \\
 & output) & & output) & [see caption] \\
\hline
$\beta \beta 0 \nu$ & & & & \\
 (in Mo foils) & 96.1 & $96.1\pm 1.0$ & 56.8 & $56.8 \pm 0.7$ [100] \\
$\beta \beta 2 \nu$ & & & & \\ 
 (in Mo foils) & 73.6 & $73.0\pm 0.9$ & 22.7 & $22.7 \pm 0.5$ [100] \\
$^{208}$Tl & & & & \\ 
 (in Mo foils) & 65.7 & $63.5\pm 0.8$ & 16.9 & $16.3 \pm 0.4$ [96.7] \\
$^{214}$Bi & & & & \\ 
 (in Mo foils) & 58.7 & $56.0\pm 0.8$ & 13.8 & $12.9 \pm 0.4$ [94.0] \\
$^{214}$Bi & & & & \\
 (in gas) & 79.4 & $60.0\pm 0.8$ & 20.7 & $15.4 \pm 0.4$ [74.5] \\
 \hline
\end{tabular}
\vspace*{0.3cm}
\caption{\label{trigger_eff}Results of simulations giving trigger efficiencies relative to the number of events coming from $\beta\beta 0 \nu$, $\beta\beta 2 \nu$, $^{208}$Tl and $^{214}$Bi impurities from Molybdenum source foils and finally $^{214}$Bi impurities from gas. The number of generated events for each case is 10000.}
\vspace*{0.1cm}
\end{table}

In the usual trigger conditions, dead time is due both to the 710~$\mu$s delay devoted to the search for alpha particles and to the Geiger readout dead time, which is at least 587.5~$\mu$s for three triggered Geiger cells. For each event the Geiger processor reads the 160 Geiger acquisition cards' status bits and five counters for each fired cell. This is accomplished through a VICbus single read cycle which needs 2.5~$\mu$s. Finally the interrupt handling and Cascade overhead take around 150~$\mu$s. Thus a $\beta\beta$ event with two triggered PMTs and an average of 16 triggered Geiger cells has a dead time of 1.5~ms (710+750~$\mu$s). 

$\bullet$ {\bf Other acquisition types}

With the same trigger configuration, one can have higher counting rates and lower dead times using acquisition processes without looking for delayed events. It is however impossible to remove the 710~$\mu$s fixed delay, but the Geiger readout can begin with the calorimeter interrupt. In this case the busy Geiger and the busy calorimeter are superimposed and the overlap is reduced to about $200~\mu$s, which gives a dead time of $\sim 1$~ms.

It is possible to take data with just the calorimeter or just the tracking detector. For laser runs, the counting rate is 250~Hz, with 10~Hz/PMT of that being the laser data, and the remainder coming from $^{207}$Bi in the six reference counters.

Calibration runs with 60 $^{207}$Bi sources use a $\beta\beta$ trigger, with an acquisition rate of $240$~Hz. Other calibration runs use only calorimeter acquisition, with a special one developed for $^{90}$Sr to determine the beta end-point. The acquisition of only the energy spectra is realized without ntuple building for the whole calorimeter, which permits very high counting rates (up to 30~kHz) with the smaller output files of 32 Mbytes. All of the 32~Mbytes of the calorimeter processor are used to construct histograms.

Counting rates associated with different triggers and acquisition types are presented in Table~\ref{trigg_types}.

\begin{table}[h]
\centering
\begin{tabular}{|l|l|l|} \hline \hline
\multicolumn{1}{|c|}{Trigger conditions} & \multicolumn{1}{|c|}{Event type} & \multicolumn{1}{|c|}{Counting}  \\
 & & rate (Hz) \\
\hline
$\ge 1$~triggered PMT & PMT singles & \\ 
with $E_{PMT} > 150$~keV & & 580 \\
$\ge 1$~track & GG singles & 65 \\
$\ge (1$~triggered PMT + 1~track) & (e), (e, e), (e, $N\gamma$), (e, e, $N\gamma$), & \\
with $E_{PMT} > 150$~keV & with $N\ge 1$ & 7.4 \\ 
$\ge (2$~triggered PMTs + 1 track) & (e, e), (e, $N\gamma$), (e, e, $N\gamma$), &  \\
with $E_{PMTs} > 150$~keV & with $N\ge 1$ & 1.15 \\ \hline
\end{tabular}
\vspace*{0.5cm}
\caption{\label{trigg_types}Different counting rates for the $\beta\beta$ trigger.}
\end{table}

\subsubsection{Proportion of events in a $\beta\beta$ run}

The combination of a tracking volume, delayed tracking electronics, calorimeter and a magnetic field allows NEMO~3 to identify electrons, positrons, $\gamma$-rays and $\alpha$-particles. The characteristics of events with these particles are outlined below.

\begin{itemize}

\item {electron (positron):} one $e^-$ ($e^+$) is reconstructed as a track, defined by active Geiger cells in time, with negative (positive) curvature starting from the source foil, passing through the wire chamber and being detected by only one scintillator given that electrons and positrons have a low probability of emitting bremsstrahlung $\gamma$-rays, which could trigger neighboring scintillators.

\item {alpha particle from the source foil:} an $\alpha$-particle is reconstructed as at least one delayed Geiger cell near an electron or a positron vertex, or as a short straight track defined with delayed hits within 1.5~$\mu$s of each other and passing through a fraction of the wire chamber. 

\item {gamma:} a $\gamma$-ray corresponds to one or two adjacent scintillators being triggered,  without any associated track and without a single Geiger cell hit in front of the scintillator. If there are two active scintillators, they must have been simultaneously triggered, or equivalently their time differences must be lower than the sum of their temporal resolutions. 
\end{itemize}

\begin{figure}[h]
\vspace*{0.2cm}
\begin{center}
\includegraphics[width=9.2cm,scale=1]{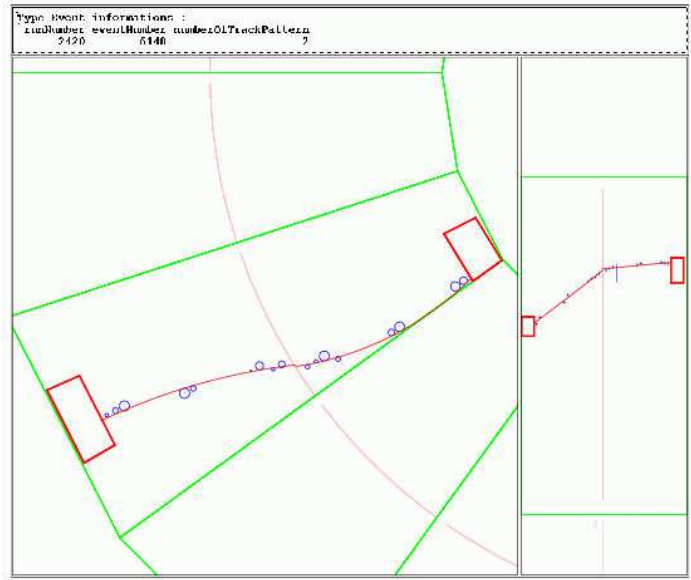}
\caption{\label{2e} Two-electron event produced in molybdenum source foil (sector 11). The left portion of this figure shows the transverse top view of NEMO~3, while the right part presents the associated longitudinal view. The circles radii correspond to the transverse distance from the anode wire for each fired cell, they are not error bars.  The electrons have energies of 1029~keV (internal wall) and 750~keV (external wall).}
\end{center}
\end{figure}

\begin{figure}[h]
\begin{center}
\includegraphics[width=9.2cm,scale=1]{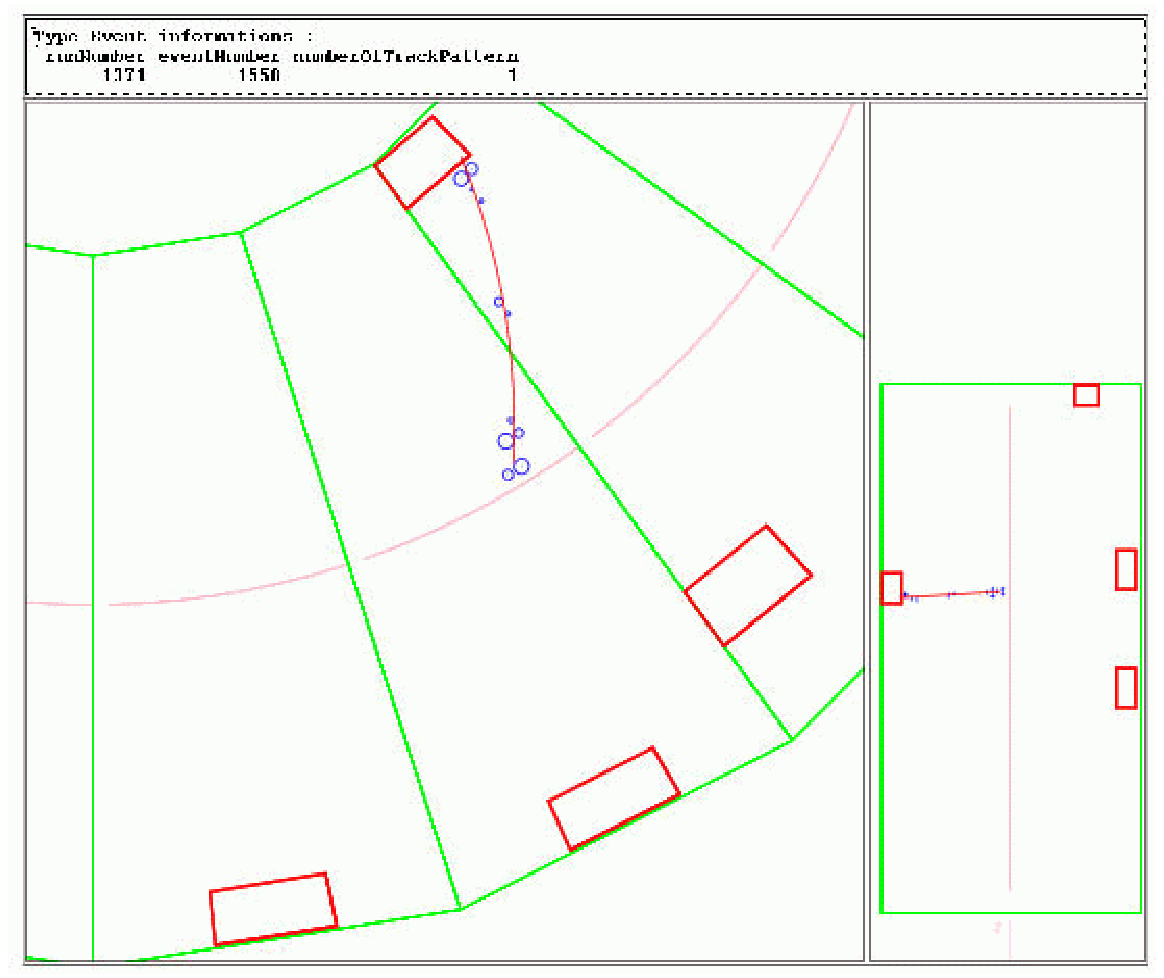}
\caption{\label{e_gamma} A $e^-\gamma\gamma\gamma$ background event produced in a molybdenum foil (sector 16).}
\end{center}
\end{figure}

\begin{figure}[h]
\vspace*{1cm}
\begin{center}
\includegraphics[width=9.5cm,scale=1]{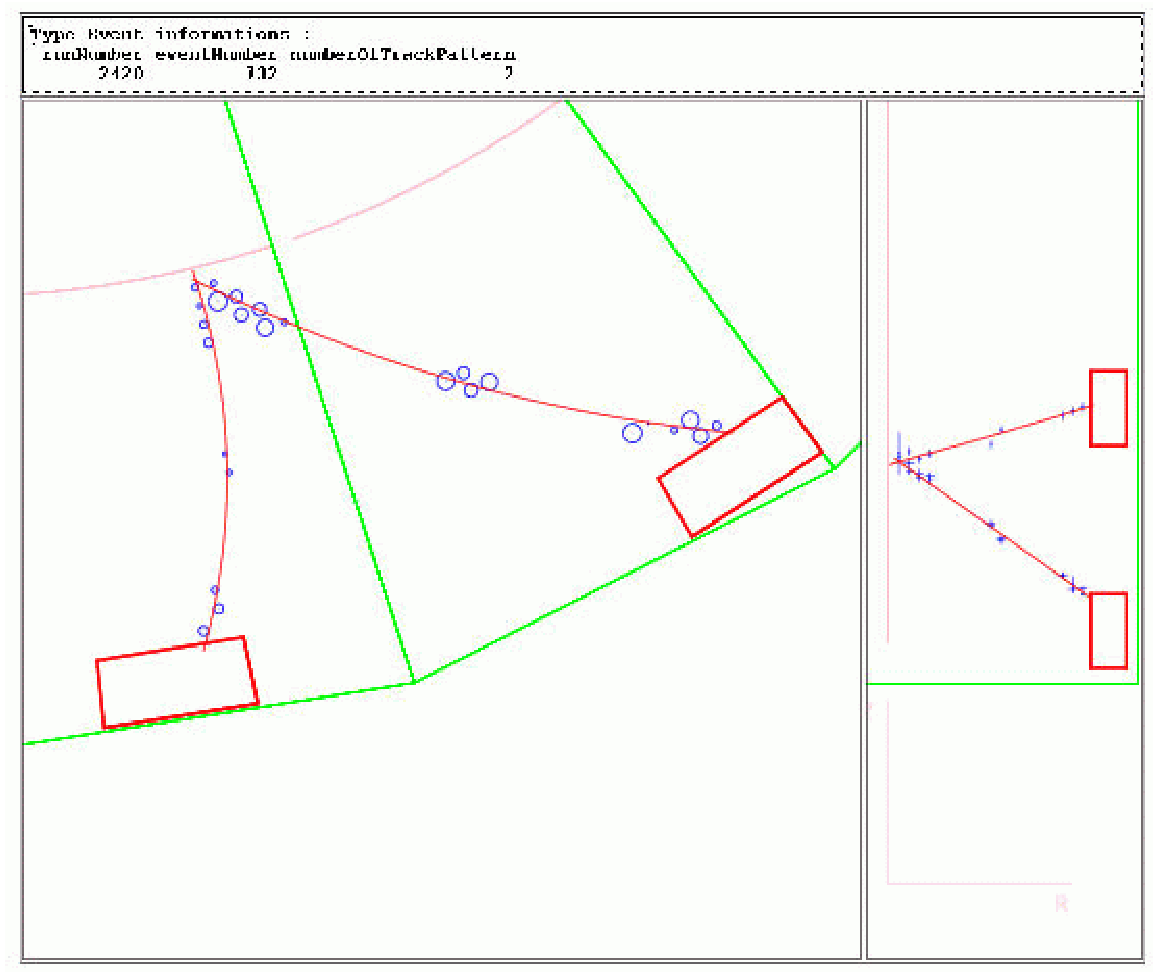}
\vspace*{0.5cm}
\caption{\label{paire} An $e^+e^-$ background event produced in a molybdenum foil (sector 15). Note the tracks of opposite curvature.}
\end{center}
\end{figure}

\begin{figure}[h]
\begin{center}
\includegraphics[width=9.5cm,scale=1]{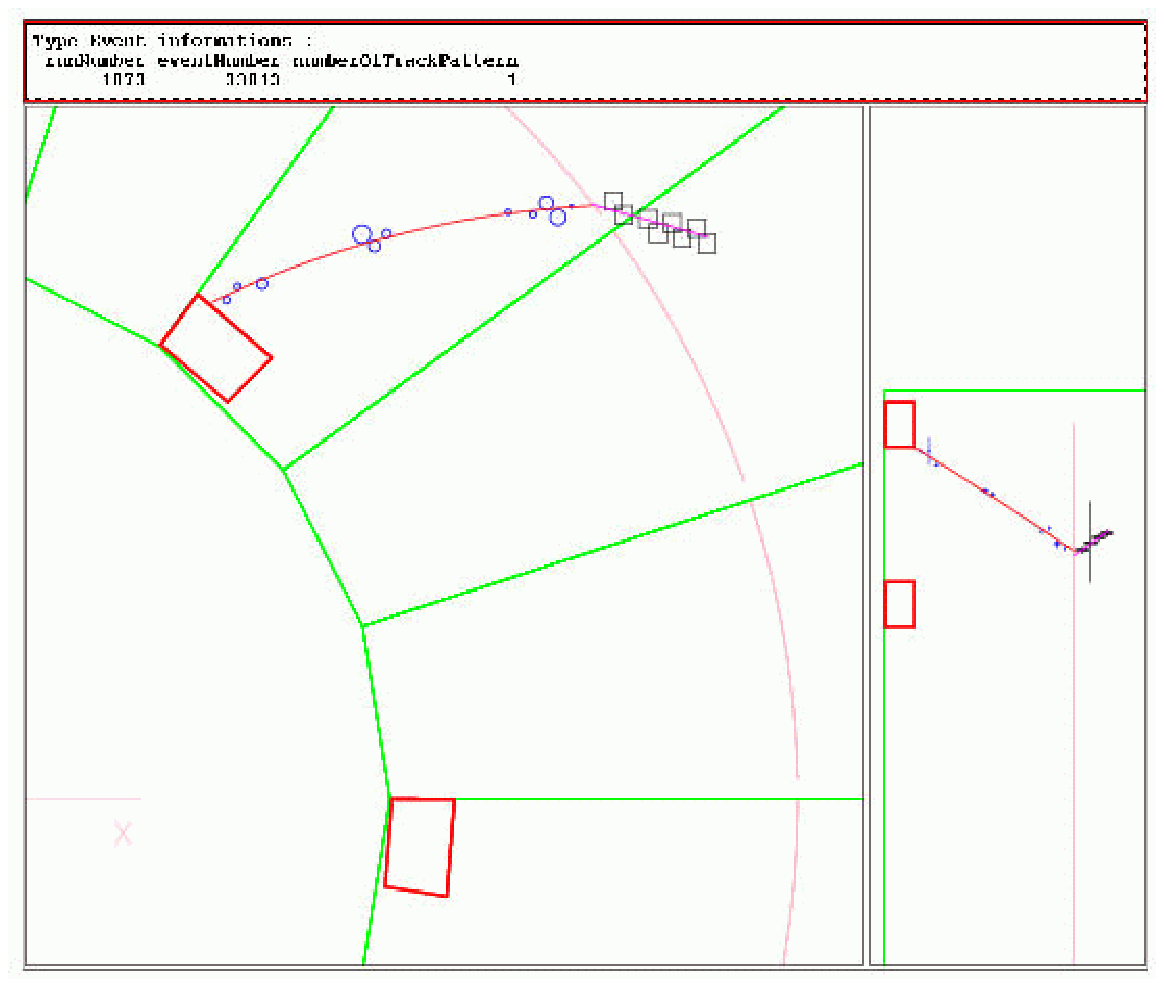}
\vspace*{0.5cm}
\caption{\label{e_alpha} Decay of some internal contamination producing a single electron event coming from a molybdenum source foil (sector 02) followed by a delayed alpha-particle, which is the short straight track represented by open squares. Note the presence of one gamma-ray.}
\end{center}
\end{figure}

\begin{figure}[h]
\begin{center}
\includegraphics[width=9.5cm,scale=1]{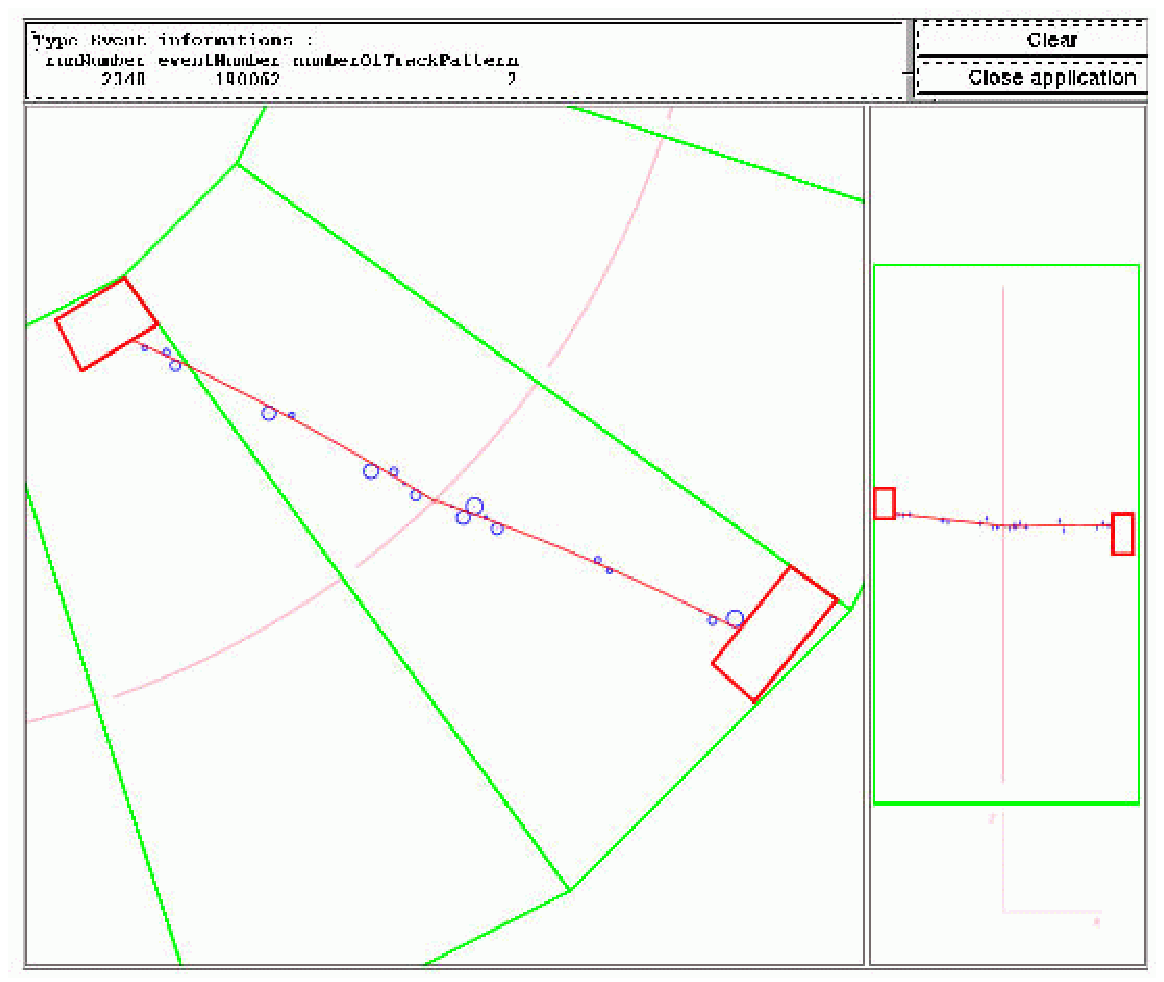}
\vspace*{0.5cm}
\caption{\label{oce} Example of an external source of background: it is a view of high energy crossing electron, which is most likely produced by a $\gamma$-ray emitted after a neutron capture in the copper frame. This type of events is used to establish reconstruction formulae for transverse and longitudinal positions in the tracking detector. It can be distinguished from back-to-back source foil events by time-of-flight measurements (see Section~\ref{cluster_def}).}
\end{center}
\end{figure}

Fig.~\ref{2e} shows a typical two electron event coming from a molybdenum source foil ($\beta\beta 2\nu$ process). In Figs.~\ref{e_gamma}, \ref{paire}, \ref{e_alpha} and~\ref{oce} one can see events characteristic of some contamination associated with the detector: $e^- \gamma\gamma\gamma$ event, $e^+e^-$ pair production, $e^- \alpha$ event and a high energy crossing electron.

\vspace*{0.5cm}

\clearpage

\vspace*{1.5cm}

\clearpage

The proportion of different types of events in a $\beta\beta$ run is given in Table~\ref{proportion_events}. Note that $\beta\beta$-like events represent 0.15\% of the registered events, which means one $\beta\beta$-like event occurs every 1.5~minute. Events which are not recorded come from electron backscatterring at the scintillators or other mechanism which drive electrons back to the source foil.

\begin{table}[h]
\centering
\begin{tabular}{|l|c|} \hline \hline
\multicolumn{1}{|c|}{Type of event} & \multicolumn{1}{|c|}{Proportion (in \%)} \\
\hline
$1e^-,$ no $\gamma$-ray & 7.7 \\
$1e^-,~N\gamma$-rays, $N\ge 1$ & 2.0 \\
$1e^+,$ no $\gamma$-ray & 1.4 \\
$1e^+,~N\gamma$-rays, $N\ge 1$ & 0.66 \\
$e^+e^-$ pairs & 1.5 \\
One-crossing-electron & 1.1 \\
{\bf Two-electron events, no~$\gamma$-ray} & {\bf 0.15} \\
\hline
\end{tabular}
\vspace*{0.5cm}
\caption{\label{proportion_events}For normal $\beta \beta$ runs the proportion of various events are given here. Note that both one-crossing-electron events and $e^+e^-$ events created in the source foils are reconstructed with a single track.}
\end{table}

\subsection{Tracking detector performance\label{tracking_def}}

\subsubsection{Final operating conditions of the tracking detector}

Since February 2003, with the new gas mixture, the longitudinal propagation velocity is 5.2~cm/$\mu$s corresponding to a full propagation time $<t_{LC} + t_{HC}> \sim 50~\mu$s and 90\% of Geiger cells have a longitudinal propagation efficiency greater than 95\%. 
The single counting rate per cell is $\sim 0.2$~Hz. 

There were eventually only 30 channels of the 6180 ($< 0.5\%$) with  missing anode signals, of which there were four cells disconnected due to anode wires in contact with ground wires, eight~cells due to interconnection problems, and 18~hot cells\footnote{self-discharging Geiger cell} ($\sim 0.3\%$).
The number of Geiger cells with at least one missing cathode signal were 160 (2.6\%), due again to interconnection problems.

\subsubsection{Geiger TDC analysis}

As discussed in Section~\ref{elec} there are four variables associated with the Geiger TDCs. They are $tdc_{LC}$ and $tdc_{HC}$ for the two cathode ring times, $tdc_A$ for the anode time and $tdc_{\alpha}$ for the delayed time (see Eq.~\ref{time_LC} to Eq.~\ref{alpha_time}). The triggered Geiger cells are classified into different types according to the values of these four TDC signals.
 There are ``in-time hits'' coming from electron or positron tracks. Also ``delayed hits'' are delayed triggers of cells used to study $\alpha$-particle events. Finally there are ``refired cells'' which are active because of cross-talk (cells fired by
a neighboring in-time cell) and ``noisy cells''.

\subsubsection{Track reconstruction}
\label{track}
$\bullet$ {\bf Principle}

Geiger events are first tagged with one of the four previously defined criteria. The refired and noisy cells are rejected, with about one or two refired cells per event. The number of noisy cells is typically negligible. Then, two different pattern recognition and track fit procedures are carried out, one for in-time hits and the other for delayed hits. 

For in-time hits the pattern recognition and tracking are carried out using a cellular automaton algorithm, which was previously used for NEMO~2 tracking~\cite{NEMO2_Cellular}. The NEMO~3 algorithm uses a set of consecutive segments which connect pairs of active cells in neighboring layers. A candidate track is defined and characterized by the number of segments (length of the track) and the sum of the angles between the segments. In NEMO~3, the longest track is favored. The curvature of the track depends on the magnetic field. 
A new track search is then begun with the other segments until there are not enough segments remaining to construct a track.

Concerning the track fitting procedure, there are two successive fits in order to solve the left/right ambiguities in the transverse plane.\\
Each reconstructed track  is extrapolated to the source foil and a vertex position is calculated whose transverse and longitudinal coordinates are $R\phi$ and $Z$, with an origin $Z=0$ at the vertical center position of the source foils. $\phi$ is the polar angle, using cylindrical coordinates, with an origin at sector~0; $R=155.5$~cm is the transversal distance between the detector's center and the source foils; for sector 00, the $R\phi$ value for the calibration source position is $(148.6 \pm 0.1)$~cm  and there is a step of 18 degrees to obtain $R\phi$ for each subsequent sector.  In the similar fashion, the track is projected onto an associated scintillator surface and the calculated coordinates of the scintillator are $R_S\phi_S$ and $Z_S$, where $R_S$ is the transversal distance between the detector's center and the entrance surface of the scintillator.\\ 
For delayed hits the treatment in the longitudinal plane is the same as that for in-time hits. The treatment in the transverse plane is slightly modified since there are only relative drift time measurements. Delayed hits can form a track only if the differences in anode times are shorter than 1.5~$\mu$s, which is approximately the maximum drift time in a Geiger cell. For each hit, the drift distance is computed from the $tdc_{\alpha}$ for the cell, assuming the delayed hit with the higher $tdc_{\alpha}$ value corresponds to a particle passing through the anode wire of the cell. 

$\bullet$ {\bf Association between tracks and energy deposited in the calorimeters}

An electron or positron event needs track associated with a scintillator. This is accomplished if there is at least one hit in the two Geiger cell layers nearest to the scintillators which belongs to the track. Additionaly, geometrical cuts are applied which extrapolate the position of the track to the surface of the scintillator ($R_S\phi_S$ and $Z_S$) which has to be located less than 3~cm away from the edges of the scintillator.

\subsubsection{Reconstruction of the particle position in the cell\label{perform_geig}}

Studies with a laser and a nine cell prototype of the tracking detector were carried out to establish formulae giving transverse ($r_{\perp}$) and longitudinal ($z$) positions in a cell from anode drift times and both cathode times~\cite{Karim_these}. These formulae have been improved with NEMO~3~\cite{Anneisa_CDF}, using data taken with high energy crossing electrons ($> 4.5$~MeV). These electrons were created by an intense Am-Be source producing fast neutrons with an activity of around $2.2 \times 10^5$~n/s. Fast neutrons are thermalized in the scintillators; then high energy $\gamma$-rays are created by the capture of the neutrons in the copper walls, producing one-crossing-electron events  by the Compton effect. With high energy crossing electrons and with an energy deposited $> 3$~MeV in the second scintillator, the multiple scattering is found to be negligible and the track is well defined.

The track resulting from the fit is assumed to be identical to the real track. Thus, the transverse and longitudinal reconstruction formulae are obtained by comparing the distance to the identified track with the drift anode time and the two cathode times.

$\bullet$ {\bf Transverse position in the cell}

\begin{figure}[h]
\vspace*{1cm}
\begin{center}
\includegraphics[width=11.5cm,scale=1]{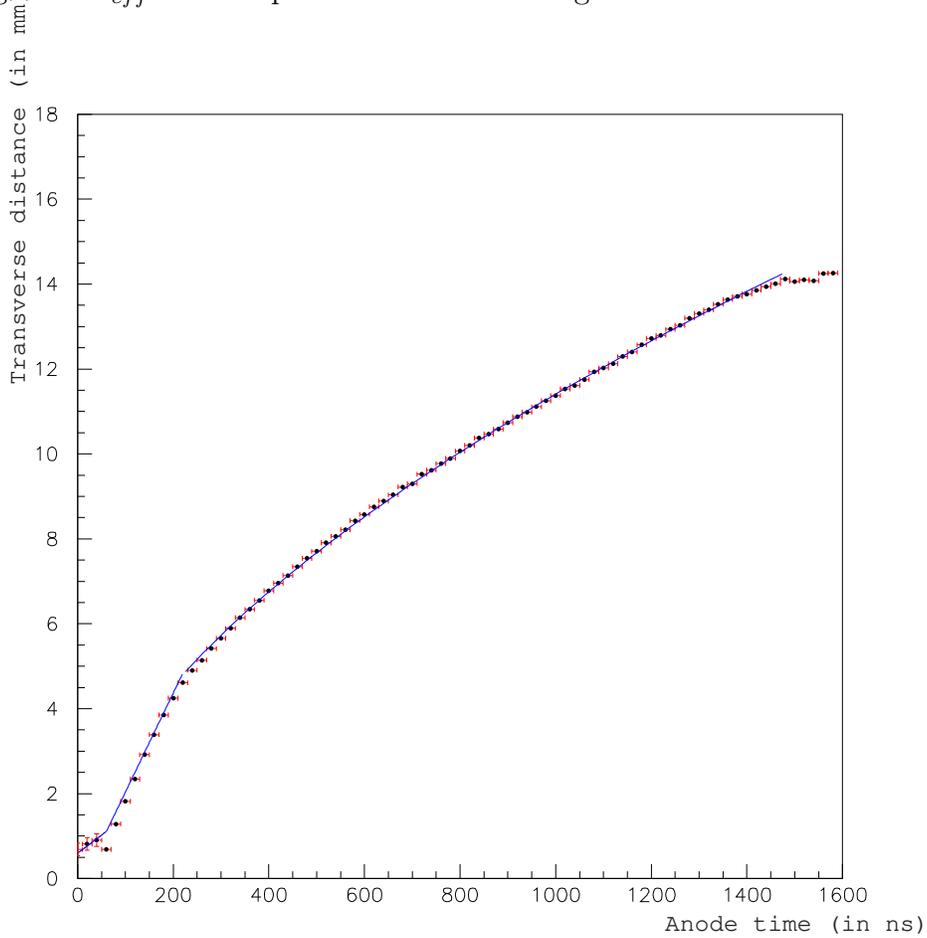}
\caption{\label{rphi_vs_ta} Transverse distance to a constructed track $r_{\perp}$ (in mm) as a function of anode time $t_A$ (in ns) obtained with high energy crossing events ($e_{cross} > 4.5$~MeV).}
\end{center}
\vspace*{0.5cm}
\end{figure}

Fig.~\ref{rphi_vs_ta} shows the transverse distance $r_{\perp}$ to the track as a function of the anode time $t_A$. Near the wire (between 1 and 4.5~mm), the drift velocity corresponds to the saturation regime and is 2.3~cm/$\mu$s. Far from the anode wire, the drift speed is proportional to the electric field. Thus, for $220 < t_A < 1480$~ns, the electric field is proportional to $1/r$, which leads to a transversal distance proportional to $\sqrt{t_A}$.

$\bullet$ {\bf Longitudinal position in the cell}

According to the two measured cathode times $t_{LC}$ and $t_{HC}$, and using the origin of the $z$ axis fixed at the vertical center of the cells, the longitudinal position $z$ is given, in mm, by:
$$
z\ = \ \frac{L_{eff}}{2} \ \frac{t_{HC} - t_{LC}}{t_{LC} + t_{HC}}  \ \left[1 \ - \ 0.505 \ 10 ^{-4}\
\frac{L_{eff}}{2} \ \left(1\ - \ \left|\frac{t_{HC} - t_{LC}}{t_{LC} + t_{HC}}\right|\right)\right] \
$$
where $L_{eff} = 2607$~mm is the effective length of the cell, which is lower than the full cell length of 2700~mm that includes the two rings (each is 30~mm long). This effective length is a consequence of the fact that cathode signals exceed their threshold a few centimeters before the plasma reaches the cathode ring. The $L_{eff}$ value depends on the HV and gas mixture.

The first order term corresponds to a constant propagation speed for the plasma. In this case, the longitudinal position is simply the propagation time of the plasma to the nearest ring, divided by the propagation velocity.

The second order term is a correction, which takes into account the reduced propagation velocity due to a decrease in the high voltage during the plasma propagation (this factor also depends on the HV and gas mixture). The correction has a maximum value of around 20~mm in two positions corresponding respectively to 25\% and 75\% of the cell length.

If one of the two cathode times is missing, the relation remains the same, but only for the TDC with a non-zero value. The other cathode time is obtained using an average plasma propagation velocity. The $z$ position in the Geiger cell is then reconstructed using the same formula as above. The propagation velocity has to be calculated and stored in the database for each Geiger cell. 
If both cathode times are missing, no reconstruction of the $z$ position of the Geiger cell hit is possible. 

$\bullet$ {\bf Transverse and longitudinal resolution of the Geiger cells~\cite{Anneisa_CDF}}

\begin{figure}[h]
\begin{center}
\includegraphics[width=12cm,scale=1]{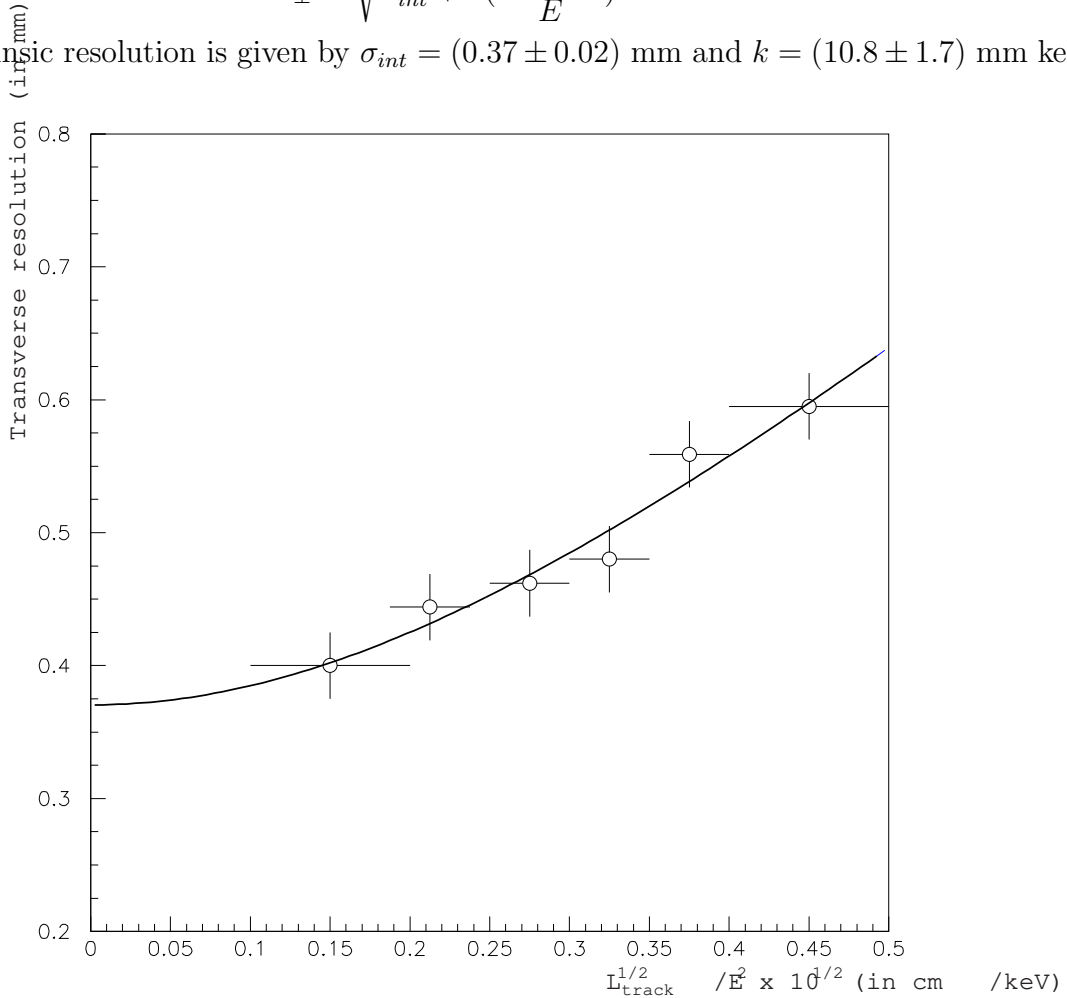}
\caption{\label{sigma_vs_LE} Distribution of transverse resolution per cell $\sigma_{\perp}$ (in cm) as a function of the variable $\sqrt{L_{track}}/E$. The associated curve is the result of the fit given by 
$\sigma_{\perp}\, = \,\sqrt{\, \sigma_{int}^2\,  +\,  k\, (\, \sqrt{L_{track}}/E)^2\, }$.}
\end{center}
\end{figure}

The residual distribution in the transverse direction is obtained by plotting the distance between the accepted track and the reconstructed position of the Geiger cell hit, using high energy crossing electrons. The full-width at half maximum $(FWHM)$ of the spectrum  is used to compute the transverse resolution of the cells as $\sigma_{\perp} = FWHM/2.35$. 
The average value of this resolution is: 
$$\sigma_{\perp} = 0.5~{\rm mm}$$
Note that $\sigma_{\perp}/R \sim 3\%$, where $R$ is the radius of the cell (15~mm). In order to estimate the contribution of multiple scattering, a study of the transverse resolution was carried out as a function of the variable $\sqrt{L_{track}}/E$, where $L_{track}$ is the track length and $E$ is the initial energy of the electron in keV (see Fig.~\ref{sigma_vs_LE}). The associated relation, which shows the resolution dependence with multiple scattering, is:
$$
\sigma_{\perp} = \sqrt{\sigma^2_{int} + k(\frac{\sqrt{L_{track}}}{E})^2}
$$
where the intrinsic resolution is given by $\sigma_{int} = (0.37 \pm 0.02)$~mm and $k = (10.8\pm1.7)$~mm~keV$^2$ is a constant.

\begin{figure}[h]
\vspace*{1cm}
\begin{center}
\includegraphics[width=12cm,scale=1]{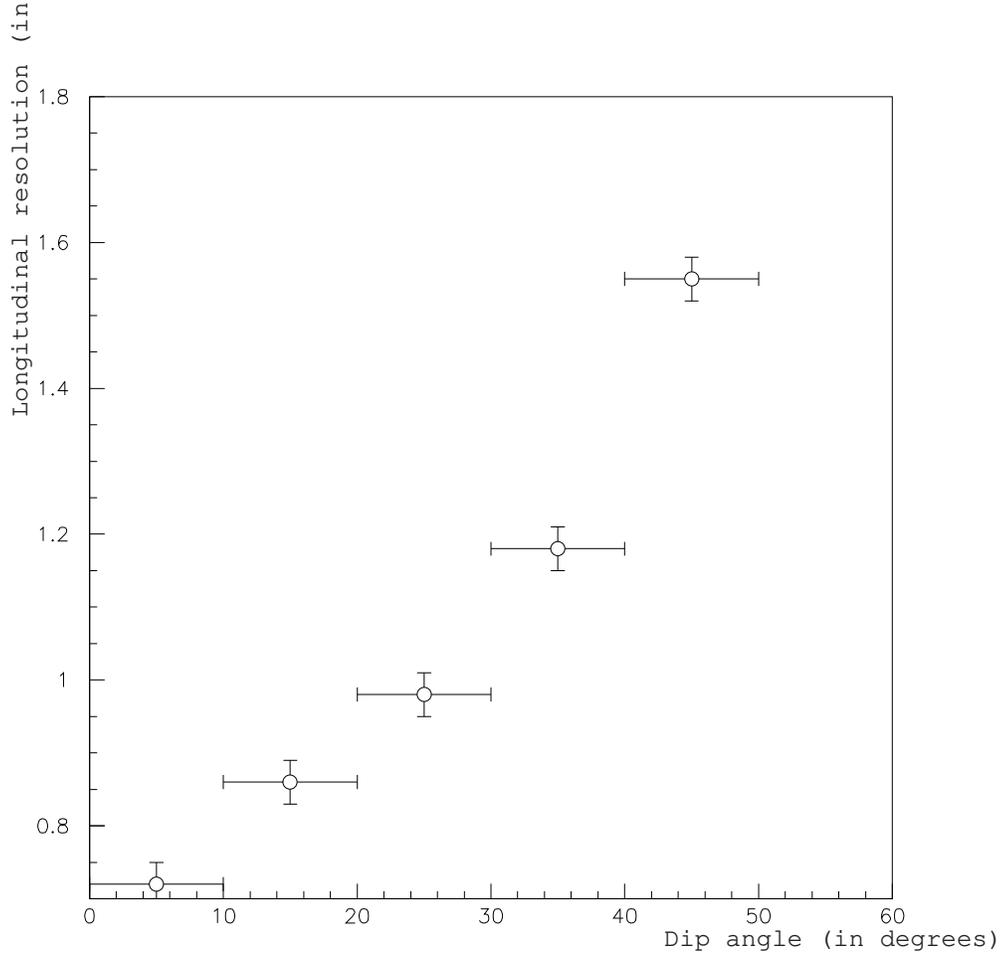}
\vspace*{1.cm}
\caption{\label{sig_para_vs_theta} Distribution of longitudinal resolution in a cell $\sigma_{//}$ (in cm) as a function of the dip angle $\theta$ (in degrees).}
\end{center}
\vspace*{1cm}
\end{figure}

\begin{figure}[h]
\begin{center}
\includegraphics[width=12cm,scale=1]{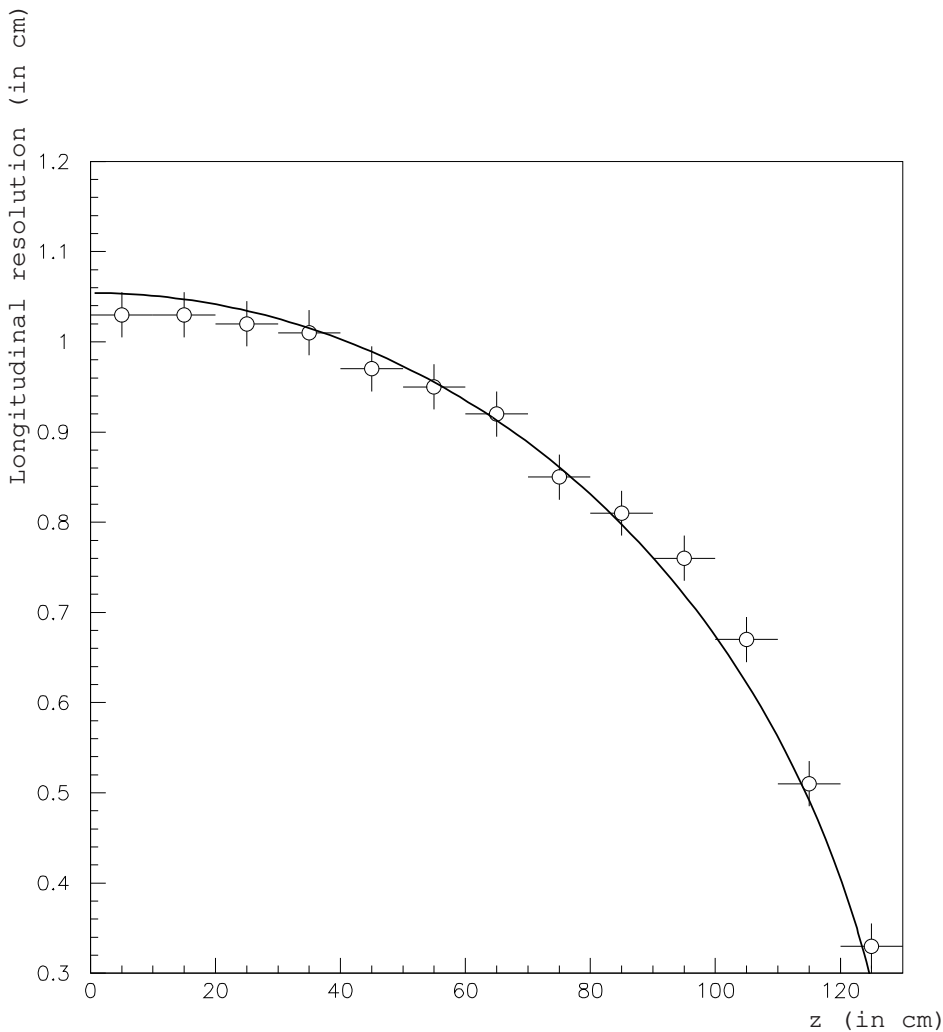}
\caption{\label{sig_para_vs_z} Distribution of the longitudinal resolution in a cell $\sigma_{//}$ (in cm) as a function of longitudinal position $z$ (in cm). The associated curve is the result of the fit given by $\sigma_{//} = \sigma_{0}\, \sqrt{\, 1\, -\, (2z/L_{eff}\,)^2}$.}
\end{center}
\end{figure}

The longitudinal resolution of the cells is obtained by the same method applied in the longitudinal plane, here the average value is
$$\sigma_{//} = 0.8~{\rm cm}$$
Note that $\sigma_{//}/L_{cell} \sim 3\%$, where $L_{cell}$ is the length of the cell. Fig.~\ref{sig_para_vs_theta} shows the influence of the dip angle $\theta$ on the longitudinal resolution. Here $\sigma_{//}$ is multiplied by a factor of two for an electron crossing the wire at an angle $\theta =45^o$ compared to a track perpendicular to the wire. This multiplication is due to the longitudinal spread of the primary ionization electrons along the anode wire. 
Fig.~\ref{sig_para_vs_z} shows the longitudinal resolution $\sigma_{//}$ as a function of the longitudinal position $z$. The associated relation is:
$$
\sigma_{//} = \sigma_0 \sqrt{1-{(\frac{z}{L_{eff}/2})}^2}
$$
where $\sigma_0 = (1.050 \pm 0.008)$~cm is the maximum resolution at the center of the detector; $L_{eff}$ is the effective length as defined above. This form is characteristic of the statistical fluctuations in the number of avalanches occuring during the longitudinal propagation.

\clearpage

\subsubsection{Misidentification of electrons and positrons~\cite{Anneisa_CDF}}

\begin{figure}[h]
\begin{center}
\includegraphics[width=12cm,scale=1]{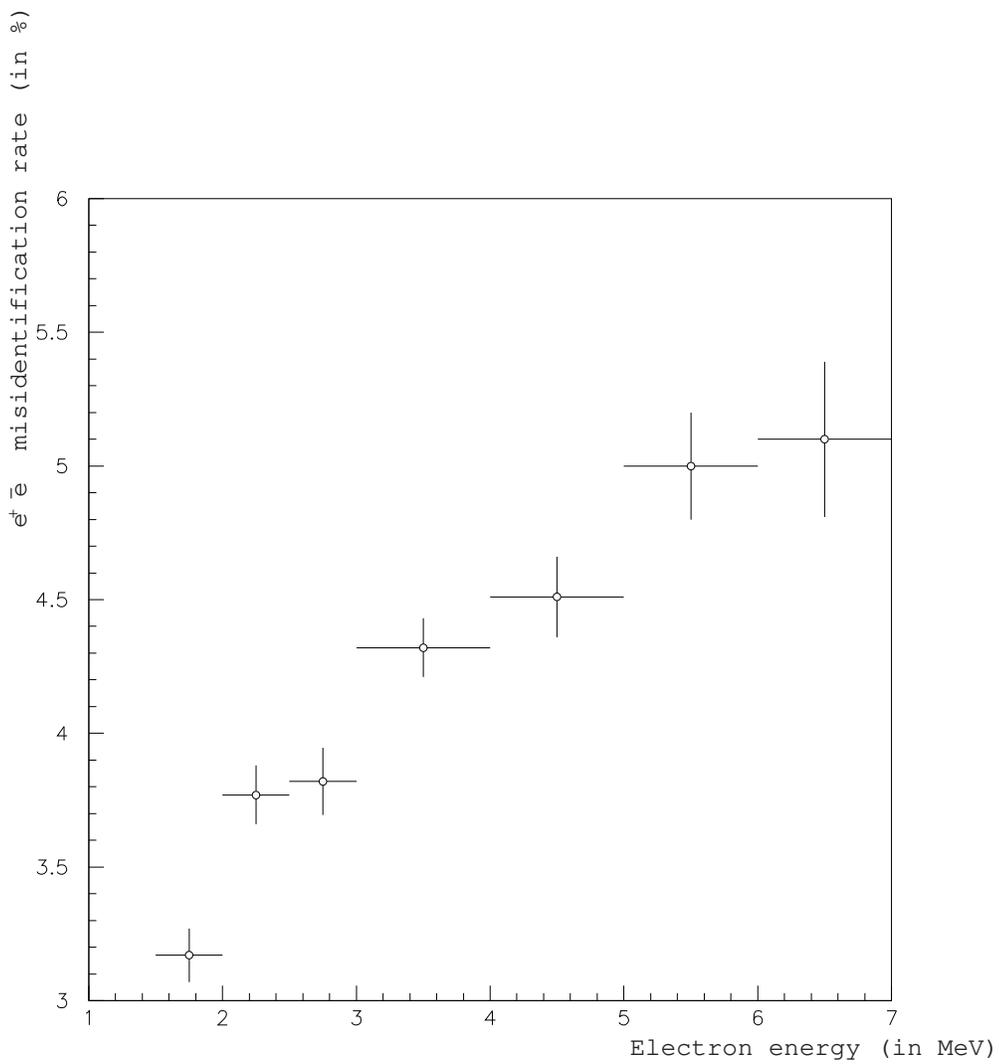}
\caption{\label{bad_e+e-} Distribution of $e^+e^-$ misidentification rate as a function of electron energy.}
\end{center}
\end{figure}

There is a possible misidentification of electrons and positrons in tests with high energy ($\ge 3$~MeV) crossing electrons which are produced during neutron source studies. These crossing events are dominated ($> 99$\%) by electrons and are reconstructed as two half-sector electron-type events with successive negative curvature. The first half from an external scintillator to the source foil on one side of the foil, the second one from the source foil to an internal scintillator, both with a common vertex on the foil. Thus it is possible to check the probability of mistaking a positive curvature positron event against an electron event after the foil. The associated distribution of the rate of error as a function of the electron energy is presented in Fig.~\ref{bad_e+e-}. The $e^+e^-$ misidentification is around 3\% at 1~MeV using extrapolated results.

\subsubsection{The vertex reconstruction~\cite{Anneisa_CDF}}

$\bullet$ {\bf Principle}

The quality of vertex reconstruction has been analyzed by taking data with 60 $^{207}$Bi sources placed in the three positions (T for Top with $Z = 90$~cm, C for Center with $Z = 0$~cm, and B for Bottom with $Z = -90$~cm) on each sector inside the calibration tubes. The $R\phi$ and $Z$ positions of these sources (see Section~\ref{track}) are known with an accuracy of 1~mm.

\begin{figure}[h]
\begin{center}
\includegraphics[width=12cm,scale=1]{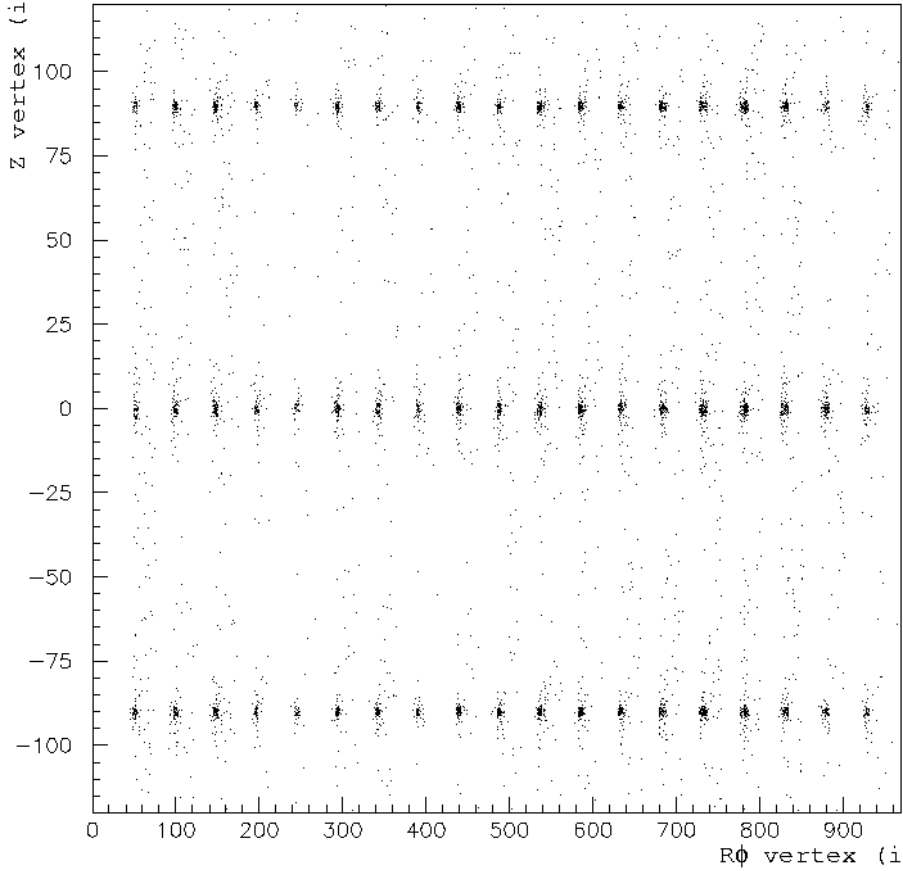}
\caption{\label{mapRZ} Distribution of reconstructed vertices for the 60 positions of the calibration $^{207}$Bi sources, in the one-electron channel.}
\end{center}
\vspace*{0.5cm}
\end{figure}

Conversion electron events coming from these sources are selected, with energies of 482~keV or 976~keV. The distribution of the 60 reconstructed vertices are presented in Fig.~\ref{mapRZ}. With this sample of data the differences between reconstructed and expected vertex positions can be estimated. The difference in the transverse plane is defined as  $\Delta R\phi = R\phi_{\rm rec} - R\phi_{\rm exp}$ and in the longitudinal plane as $\Delta Z = Z_{\rm rec} - Z_{\rm exp}$.

For transverse reconstruction of the vertex, the differences are the same on average for the three source positions (T, C and B):
$$\Delta R\phi = 1.6~{\rm mm}$$
These $\Delta R\phi$ are compatible with the accuracy of $R\phi$ positions and with the tranverse resolution $\sigma_{R\phi}$ of the vertex.

The average differences in the longitudinal reconstuction for the three positions are the following:
$$\Delta Z (T) = -0.325\pm 0.003~{\rm cm}$$
$$\Delta Z (C) = -0.164\pm 0.007~{\rm cm}$$
$$\Delta Z (B) = -0.052\pm 0.003~{\rm cm}$$
Note the maximum value of $\Delta Z$ is lower than the longitudinal vertex resolution. 
The longitudinal position of the vertex is independent of the angle.
However, an asymmetry for the three positions is observed, which corresponds to an error of around 20~ns on the cathode time measurement. It can be explained by the fact that the same discrimination threshold was used for both cathode signals although the high cathode signal is lower.

$\bullet$ {\bf Vertex resolution}

The {\bf transverse resolution} $\sigma_{R\phi}$ depends on the energy of the track. Using the two conversion electron energy values ($\sim 0.5$~MeV and $\sim 1$~MeV), it is possible to determine average values for these resolutions: $$\sigma_{R\phi} ({\rm 0.5~MeV}) = 0.3~{\rm cm}~~~~ {\rm and} ~~~~\sigma_{R\phi} ({\rm 1~MeV}) = 0.2~{\rm cm}$$

The {\bf longitudinal resolution} $\sigma_{Z}$ depends on both energy and position. Using $1$~MeV electrons, the same value of $\sigma_{Z} = 0.7$~cm was obtained for both T and B positions and $\sigma_{Z} = 0.9$~cm was obtained for the C position. For $0.5$~MeV electrons, the longitudinal resolution is $\sigma_{Z} = 1.1$~cm on average for the T and B positions and $1.3$~cm  for the C position.

\begin{figure}[h]
\begin{center}
\includegraphics[width=11.5cm,scale=1]{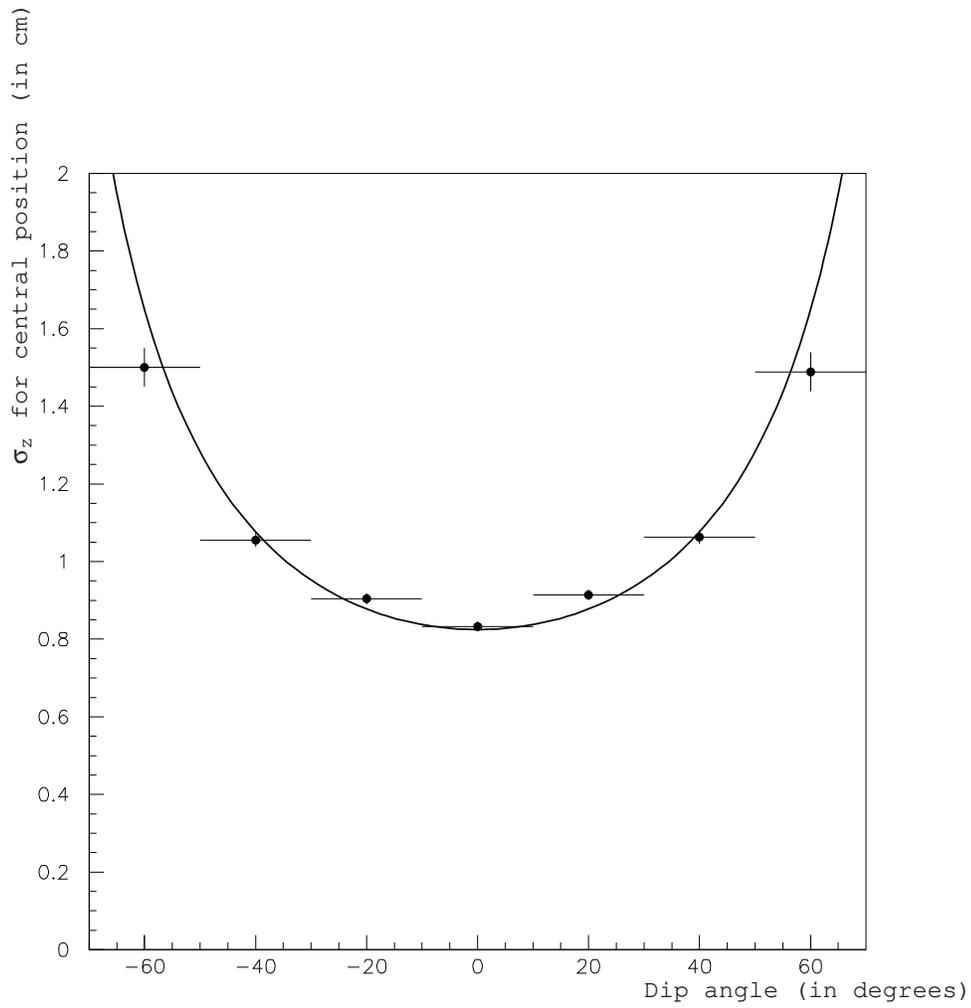}
\caption{\label{sig_vs_costheta} Distribution of the vertex longitudinal resolution  $\sigma_Z$ (in cm) as a function of the dip angle $\theta$ (in degrees), with a $\sigma'_0/cos \theta$ adjustment (obtained with $^{207}$Bi sources located in the central position).}
\end{center}
\vspace*{0.5cm}
\end{figure}

There is another dependence on the dip angle $\theta$ (see Fig.~\ref{sig_vs_costheta}), which obeys the relation:
$$\sigma_{Z} = \frac{\sigma'_0}{\cos \theta}$$
where, on average, $\sigma'_0 = (0.632\pm 0.004)$~cm for the T position, $\sigma'_0 = (0.825 \pm 0.006)$~cm for the C position and $\sigma'_0 = (0.615 \pm 0.004)$~cm for the B position.
\clearpage

$\bullet$ {\bf Study with two electrons events}

The $\beta\beta$ analysis reconstructs events with two tracks coming from the same vertex. Thus, it is important to study the vertex resolution of the two electron channel in order to check the measured transverse and longitudinal dispersions. These dispersions, $\delta R\phi$ and $\delta Z$, are defined as the distance between the vertices associated with the two reconstructed tracks.

Using events coming from the 60 $^{207}$Bi sources with two simultaneous electrons (intensity of $\sim 2\%$), one builds $\delta R\phi$ and $\delta Z$ distributions, which produce the resolutions: $$\sigma (\delta R\phi) = 0.6~{\rm cm}~~~~~~~{\rm and}~~~~~~~\sigma (\delta Z) = 1.0~{\rm cm}$$
It is also true that $\sigma (\delta R\phi) = 0.1$~cm, if one constrains the two tracks to have a common vertex. These resolutions allow one to make a distinction between two strips in a source foil in a given sector, which is crucial for sectors composed of different sources.

\subsection{Operating conditions of the calorimeter\label{cluster_def}}
\vspace*{-0.5cm}
\subsubsection{Working performance of the calorimeter}

The operating voltages of the PMTs are 1800~V for the 3'' and 1350~V for 
the 5'' PMTs. The single counting rate is $\sim 0.2$~Hz/PMT above a 48~mV threshold ($\sim 150$~keV). In total, 98\% of
the PMTs are functioning to design specifications.

\subsubsection{Energy resolution of the counters}

$\bullet$ {\bf Energy calibration results}

\begin{figure}[h]
\begin{center}
\includegraphics[width=12cm,scale=1]{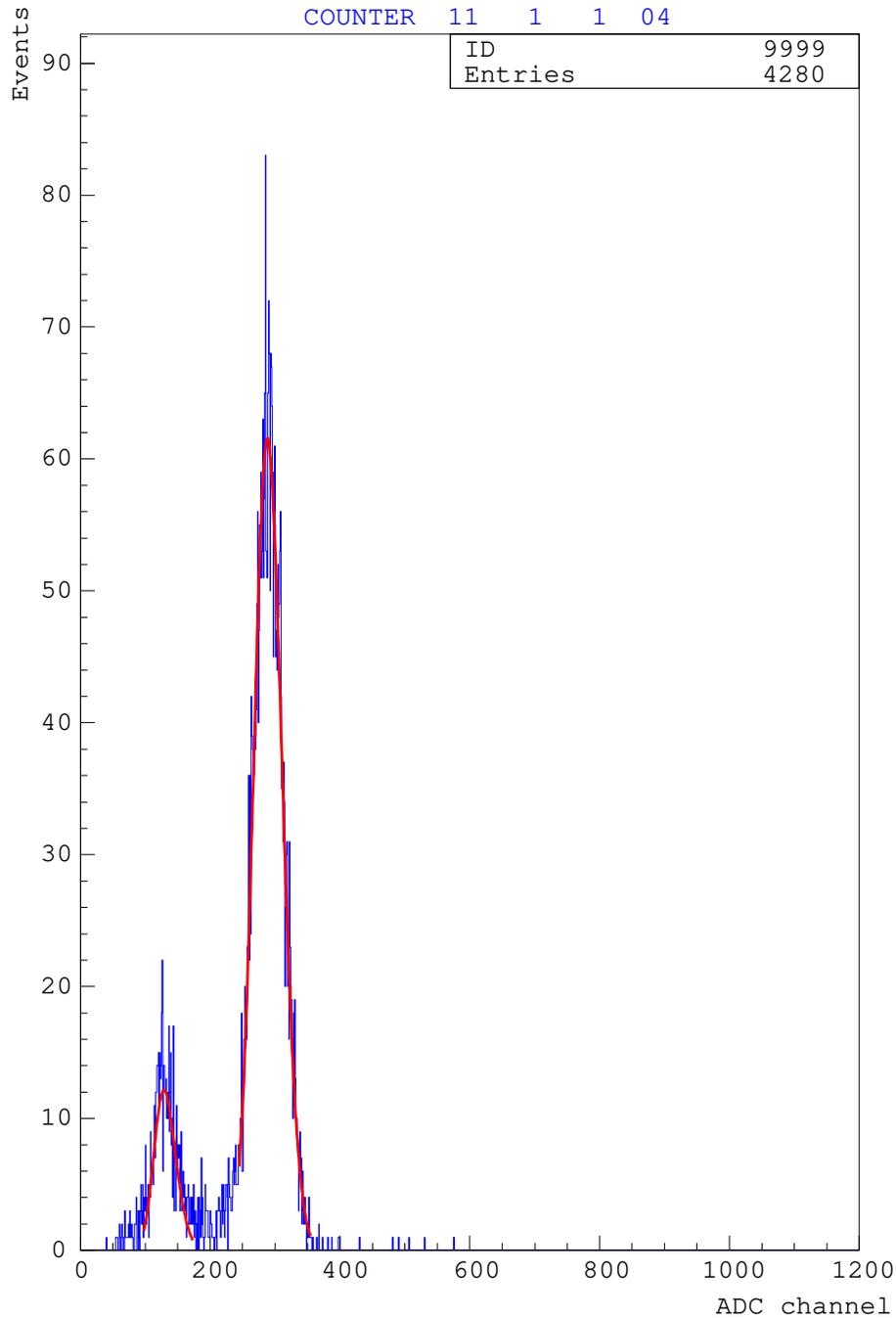}
\vspace*{0.6cm}
\caption{\label{fit_207bi} Spectral fit of the 482~keV and 976~keV $\gamma$-rays coming from $^{207}$Bi decays for one counter ($\sim 3$~keV/channel).}
\end{center}
\end{figure}

\begin{figure}[h]
\begin{center}
\includegraphics[width=12.5cm,scale=1]{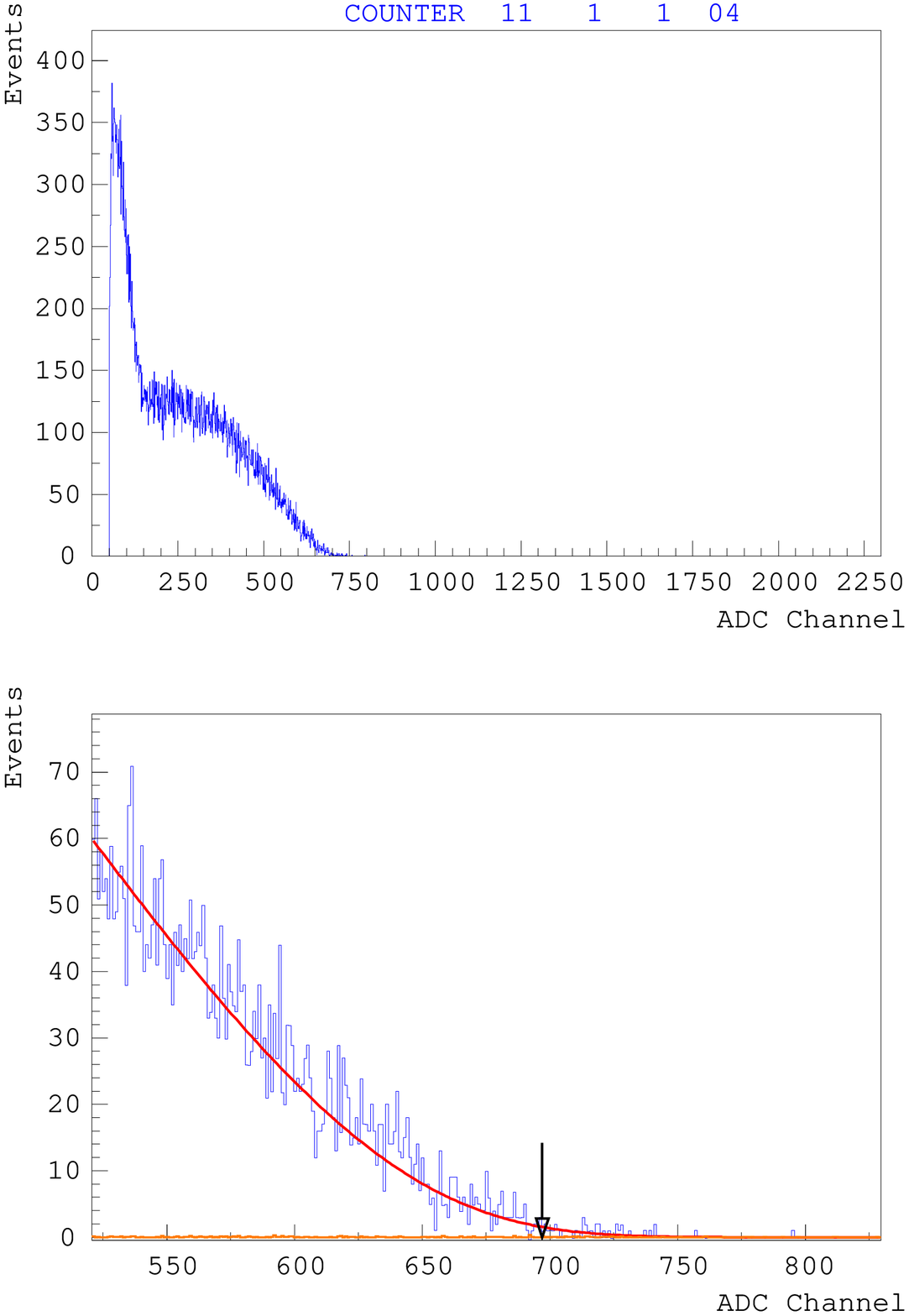}
\vspace*{0.3cm}
\caption{\label{fit_90sr} An example of beta-decay end-point adjustement (2.283~MeV) in the energy spectrum associated with $^{90}$Sr calibration sources ($\sim 3$~keV/channel). The top of the figure shows the full spectrum and the bottom shows the fit to the high-energy tail of the spectrum, which is made with a function describing the shape of a single $\beta$ spectrum of $^{90}$Y, convolved with the energy resolution function $\sigma(E)$ and taking into account the mean energy loss of the electrons.}
\end{center}
\end{figure}

As explained in Section~\ref{calib}, special calibration runs are used to obtain the three points on the energy versus ADC channel line. Two results of multi-gaussian fits (K, L, M electrons) to the peak positions in the $^{207}$Bi data  (see Fig.~\ref{fit_207bi}), and one from the beta end-point energy from $^{90}$Sr data (see Fig.~\ref{fit_90sr}). 

\clearpage

\begin{figure}[h]
\begin{center}
\includegraphics[width=13cm,scale=1]{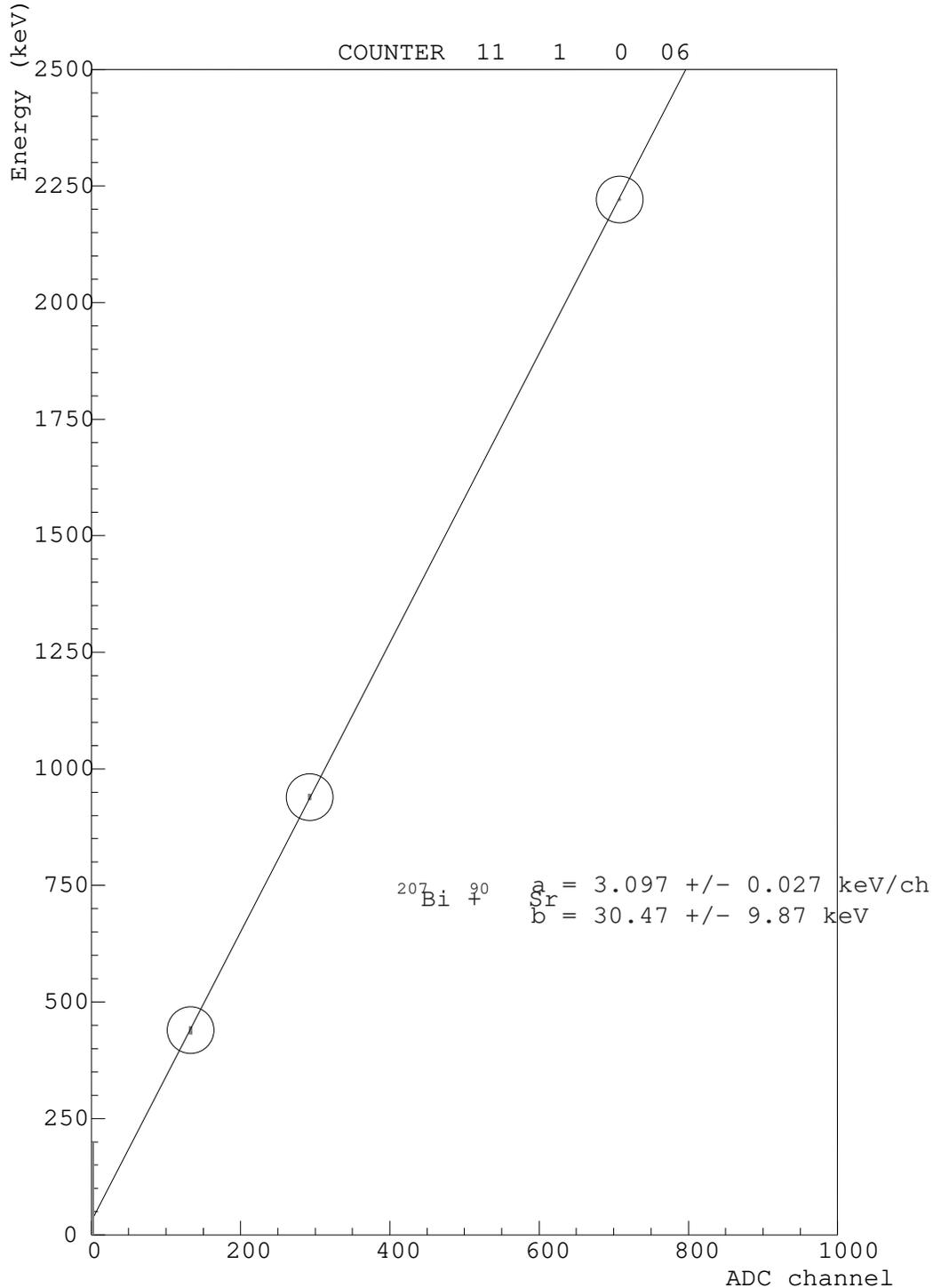}
\vspace*{0.3cm}
\caption{\label{calib_constant} An example of the energy calibration using 3 points coming from $^{207}$Bi and $^{90}$Sr data. The calibration parameters ($a$ and $b$) of the energy-channel relation are obtained this way for each counter ($\sim 3$~keV/channel).}
\end{center}
\end{figure}

The calibration lines obtained from the two $^{207}$Bi peaks as well as the fit combining $^{207}$Bi and $^{90}$Sr results do not necessarily intersect the origin of the axes. This effect was previously observed with data obtained with an electron spectrometer and in other experiments.  A typical example is given in Fig.~\ref{calib_constant}.

It was first discovered with calibration data that the response of a counter depends upon the entrance point of the electron, with a weak dependence of 1 to 2\% for counters equipped with 3" PMTs and a stronger dependence of up to 10\% for those equipped with 5" PMTs. This effect has a non-negligible consequence on the energy resolution, thus one has to correct it for the seven different types of counters, L1 to L4 for petal scintillators and EE, EC and IN for wall scintillators (see Table~\ref{resol_counters}). One can see that these resolutions depend primarily on the type of PMT associated with the scintillator which on average are 6.1\% for the 5" PMTs and 7.3\% for the 3" PMTs.

\begin{table}[h]
\centering
\begin{tabular}{|c|c|c|} \hline
\multicolumn{1}{|c|}{Block type} & \multicolumn{1}{|c|}{Associated PMT} & \multicolumn{1}{|c||}{Corrected resolution $\frac{\sigma_E}{E}$ at 1~MeV (in \%)} \\ \hline
IN & 3" & $7.3 \pm 0.1$ \\ \hline
EC & 5" & $6.0 \pm 0.1$ \\ \hline
EE & 5" & $6.0 \pm 0.1$ \\ \hline
L1 & 3" & $7.1 \pm 0.2$ \\ \hline
L2 & 3" & $7.1 \pm 0.2$ \\ \hline
L3 & 3" & $7.5 \pm 0.2$ \\ \hline
L4 & 5" & $6.3 \pm 0.2$ \\ \hline
\end{tabular}
\vspace*{0.2cm}
\caption{\label{resol_counters}Energy resolutions corrected from entrance position on scintillators.}
\end{table}

\subsubsection{Timing resolution of the counters}

$\bullet$ {\bf Timing corrections}

As explained in Section~\ref{calib}, the relative timing offset
$\varepsilon^{(i)}$ for each counter $i$ is determined using different
calibration sources, in order to provide the time alignment of the
calorimeter. 

\begin{figure}[h]
\vspace*{0.5cm}
\begin{center}
\includegraphics[width=12cm,scale=1]{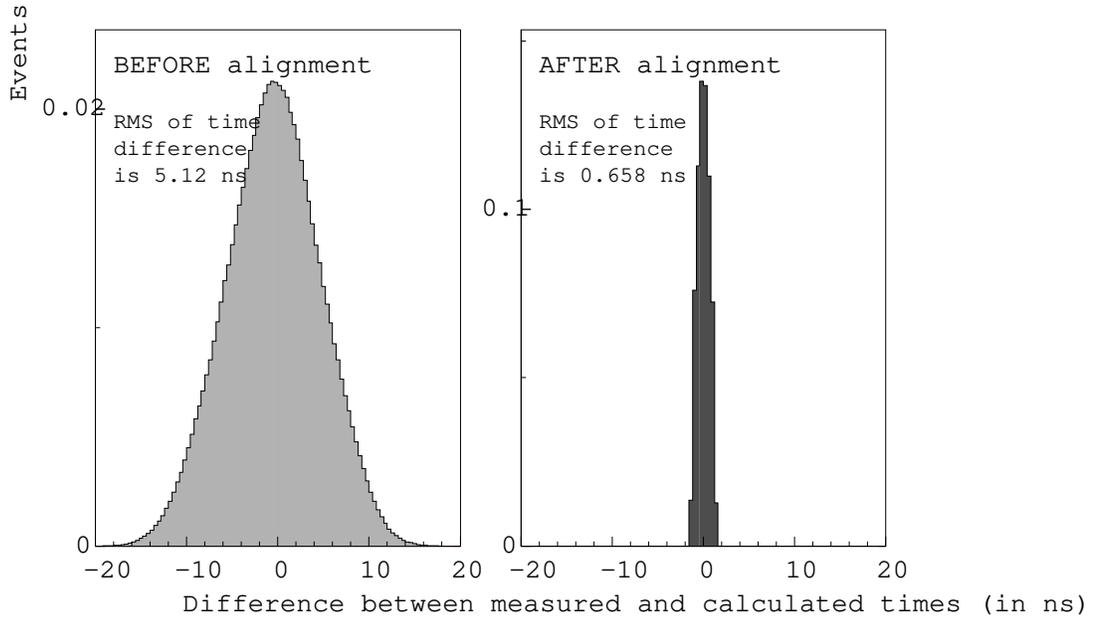}
\caption{\label{time_align} Distribution of the arrival time difference in the counters before and after alignment with $^{60}$Co calibration sources. The distribution after alignment has an RMS of $\sim 660$~ps (obtained with two-gamma events).}
\end{center}
\end{figure}

The two-gamma events from $^{60}$Co sources are used for time alignment with $\gamma$-rays. The distribution of the arrival time difference in the counters before and after alignment  is shown in Fig.~\ref{time_align}. The distribution after alignment has a resulting RMS of $\sim 660$~ps for gamma events.

\begin{figure}[h]
\begin{center}
\includegraphics[width=12cm,scale=1]{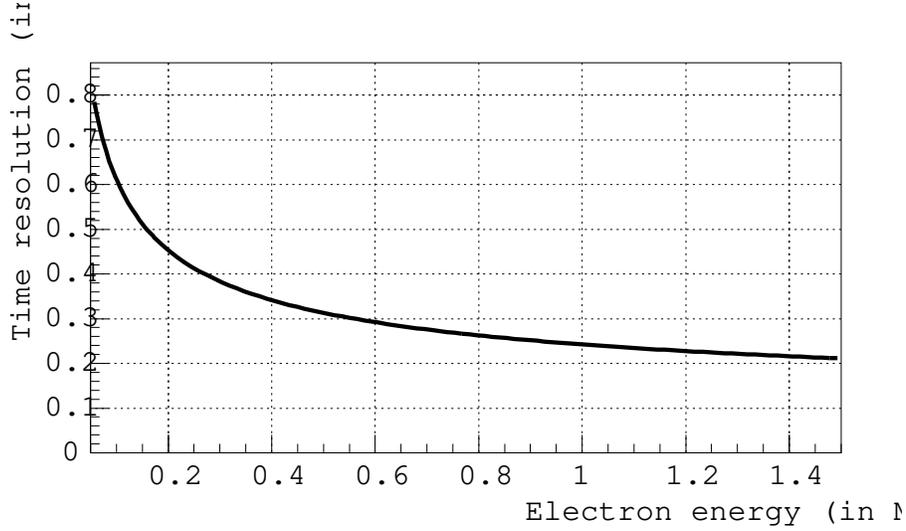}
\caption{\label{time_resolution} Time resolution (in ns) as a function of the electron energy (in MeV), obtained with two-electron events.}
\end{center}
\end{figure}

The two-electron events from $^{207}$Bi sources are used for time alignment with electrons and to obtain the time resolution as a function of electron energy as shown in Fig.~\ref{time_resolution}. This resolution is around 250~ps for 1~MeV electrons.
\clearpage

\begin{figure}[h]
\begin{center}
\includegraphics[width=12cm,scale=1]{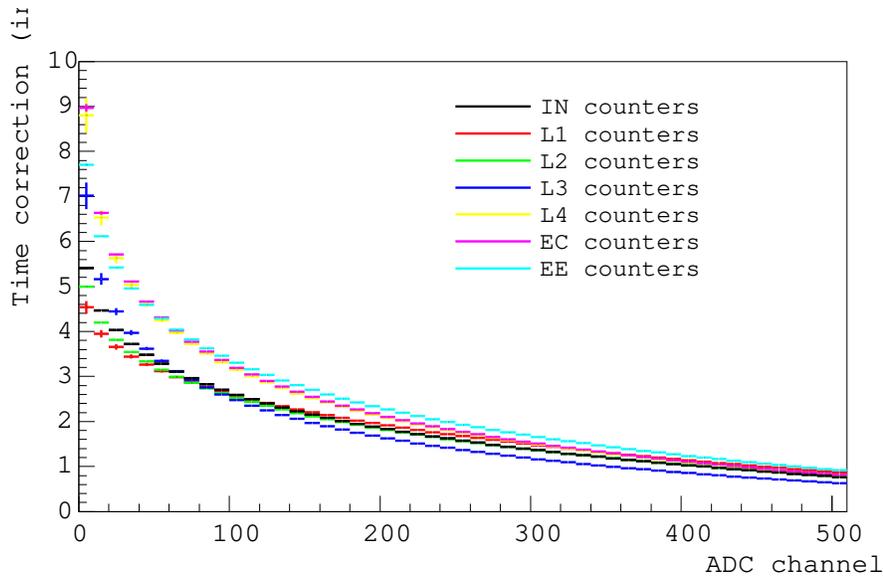}
\caption{\label{tdc_vs_adc} Time correction (in ns) as a function of the ADC channel for different types of counters ($\sim 3$~keV/channel).}
\end{center}
\vspace*{0.5cm}
\end{figure}

The energy dependence of the timing signal is adjusted by a four parameter formula, as shown in Eq.~\ref{time_vs_energy}. The first method to find this dependence is to use a ``complete'' laser run.
A second  is to use the two-electron events from $^{207}$Bi sources. The results of this second method for the different types of scintillators are  plotted in Fig.~\ref{tdc_vs_adc}.

$\bullet$ {\bf The time-of-flight (TOF) selection criterion}

To distinguish between one-crossing-electron events from external background and two-electron events from the source foil, the same 
TOF rejection criterion is applied that was used in the NEMO~2 experiment~\cite{NEMO2-NIM}. It is based on the comparison of the measured TOF ($\Delta T_{meas}$) with the calculated ones. A  single electron crosses the detector in a time $(\Delta T_{cal})_{ext}$  and two electrons emitted from the source foil have a TOF equal to $(\Delta T_{cal})_{int}$.
\clearpage

\begin{figure}[h]
\begin{center}
\includegraphics[width=12cm,scale=1]{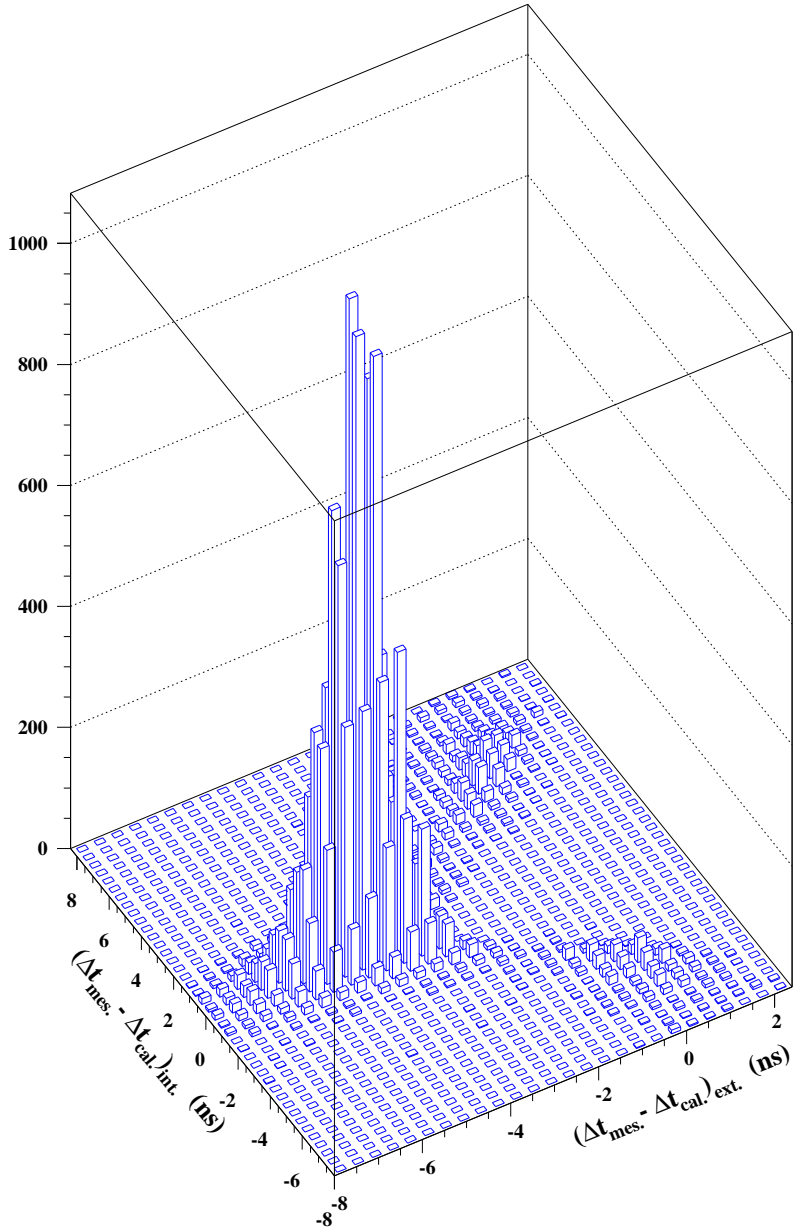}
\caption{\label{TOF} $(\Delta T_{meas} -\Delta T_{cal})_{int}$ vs $(\Delta T_{meas} -\Delta T_{cal})_{ext}$ for two-track events from $^{100}$Mo source foils. See text for more details.}
\end{center}
\end{figure}

Fig.~\ref{TOF} represents $(\Delta T_{meas} -\Delta T_{cal})_{int}$ vs $(\Delta T_{meas} -\Delta T_{cal})_{ext}$ for two track events in $^{100}$Mo foils. The plot shows well-separated bumps. The two-electron events coming from the source foil are centered around $(\Delta T_{meas} -\Delta T_{cal})_{int} \sim 0$~ns. The one-crossing-electron events have $(\Delta T_{meas} -\Delta T_{cal})_{ext} \sim 0$~ns.

\subsubsection{Counter stability}

For each counter, there are stability corrections which are calculated using the absolute calibration obtained with sources as a reference and laser spectra. These corrections take into account any variation of the laser, which is monitored by reference PMTs. This variation measurement is determined by comparing the laser peak position  and the 976~keV peak position from the $^{207}$Bi for the six reference counters (see Section~\ref{laser_description}).

\begin{figure}[h]
\begin{center}
\includegraphics{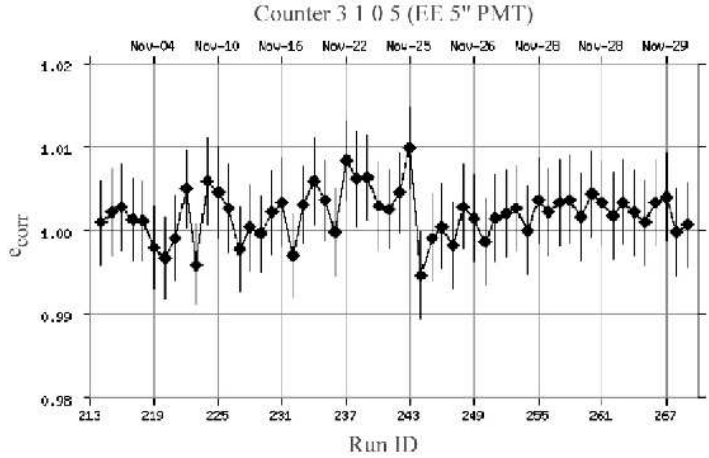}
\caption{\label{ecorr_stab} Sample of the long term (one month) stability for the energy correction parameter $e_{corr}$.}
\end{center}
\end{figure}

An example of a long term (one month) survey of the energy correction parameter ($e_{corr}$) for one counter is shown in Fig.~\ref{ecorr_stab}.
This stable behaviour with gain variations of less than 2\% is consistent with 90\% of the counters with 5'' PMTs and 96\% with 3'' PMTs. These corrections are stored in the database and used with $\beta\beta$ runs to calculate particle energies.

\section{Conclusion}

The NEMO~3 experiment is based on the direct detection of the two electrons emitted from double beta decay isotopes, with the detector and the source of the double beta decay being independent. This allows the collaboration to study seven $\beta\beta$ isotopes simultaneously. The isotopes which are distributed among 20 sectors are $^{100}$Mo, $^{82}$Se, $^{130}$Te, $^{116}$Cd, $^{150}$Nd, $^{96}$Zr and $^{48}$Ca. A calorimeter made of large blocks of scintillators coupled to very low radioactivity PMTs permits one to measure the energies of electrons, positrons, gamma-rays and also their time-of-flight, which are used to reject events from external backgrounds. In addition of a tracking volume with delayed tracking electronics for identification of alpha particles coming from $^{214}$Bi decay, a 25 Gauss magnetic field allows three-dimensional track reconstruction of charged particles. Thus the NEMO~3 detector is able to identify electrons, positrons, $\gamma$-rays and $\alpha$-particles and to detect multi-particle events in the energy domain of natural radioactivity. Using registered events in the $e^-\gamma$, $e^-\gamma\gamma$, $e^-\gamma\gamma \gamma$ and $e^- \alpha\gamma$ channels for backgrounds studies, the NEMO~3 detector is able to characterize and measure its own background, which can be subtracted from the two-electron signal.

The main objective of the NEMO~3 experiment is to search for neutrinoless double beta decay. To avoid the high energy region of natural radioactivity the $\beta\beta 0 \nu$ isotopes are selected for their high $Q_{\beta \beta}$ value. Every attempt has been made to minimize internal backgrounds in the $\beta \beta 0\nu$ sources by purification of the enriched samples as well to suppress external backgrounds using shields and by carefully selecting all the detector materials. The tail of the $\beta \beta 2 \nu$ decay distribution troublesomely overlaps the $\beta \beta 0 \nu$ distribution as a function of the energy resolution. It is the unavoidable background for neutrinoless double beta decay studies.

\begin{figure}[h]
\begin{center}
\includegraphics[width=13cm,scale=1]{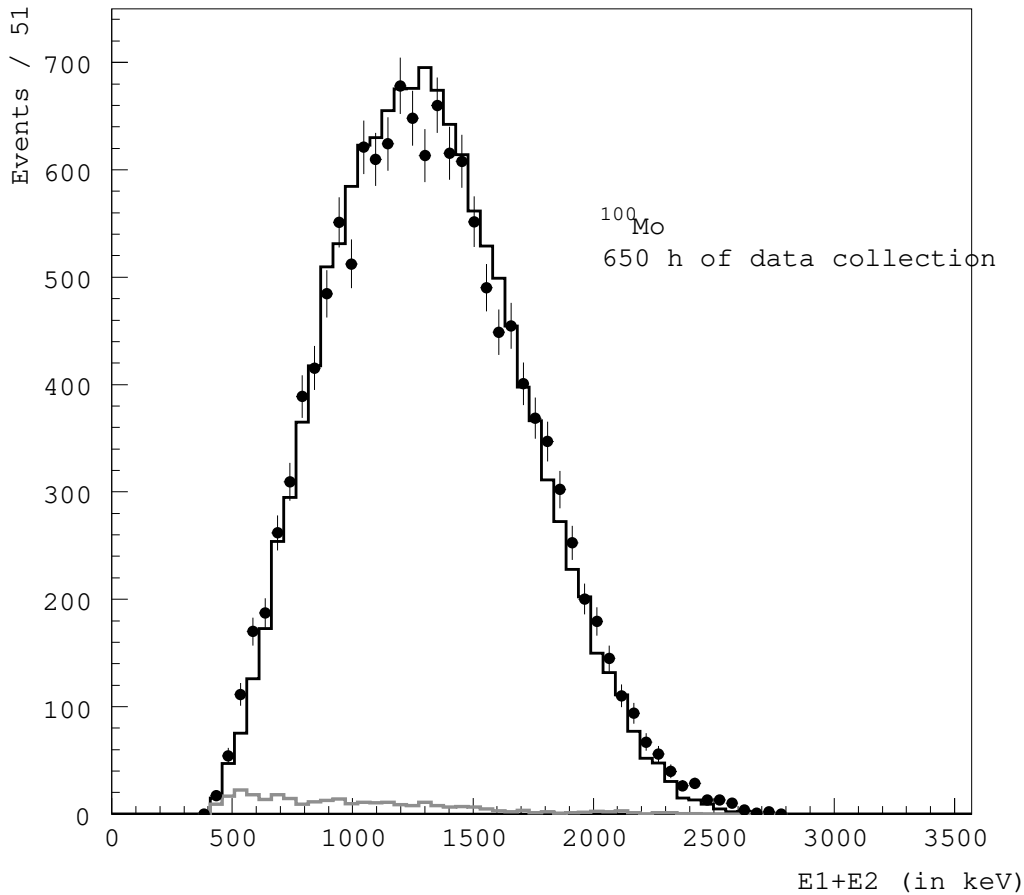}
\vspace*{0.5cm}
\caption{\label{2nu} Distribution of the experimental total energy sum measured with molybdenum sources (with background substracted) compared to $\beta\beta 2\nu$ Monte-Carlo data. It corresponds to 650 h of data collection in stable conditions, between mid-February and the end of March 2003. The number of events is $\sim 14000$, with a signal-to-background ratio of 40 to 1.}
\end{center}
\end{figure}

\begin{figure}[h]
\begin{center}
\includegraphics[width=13cm,scale=1]{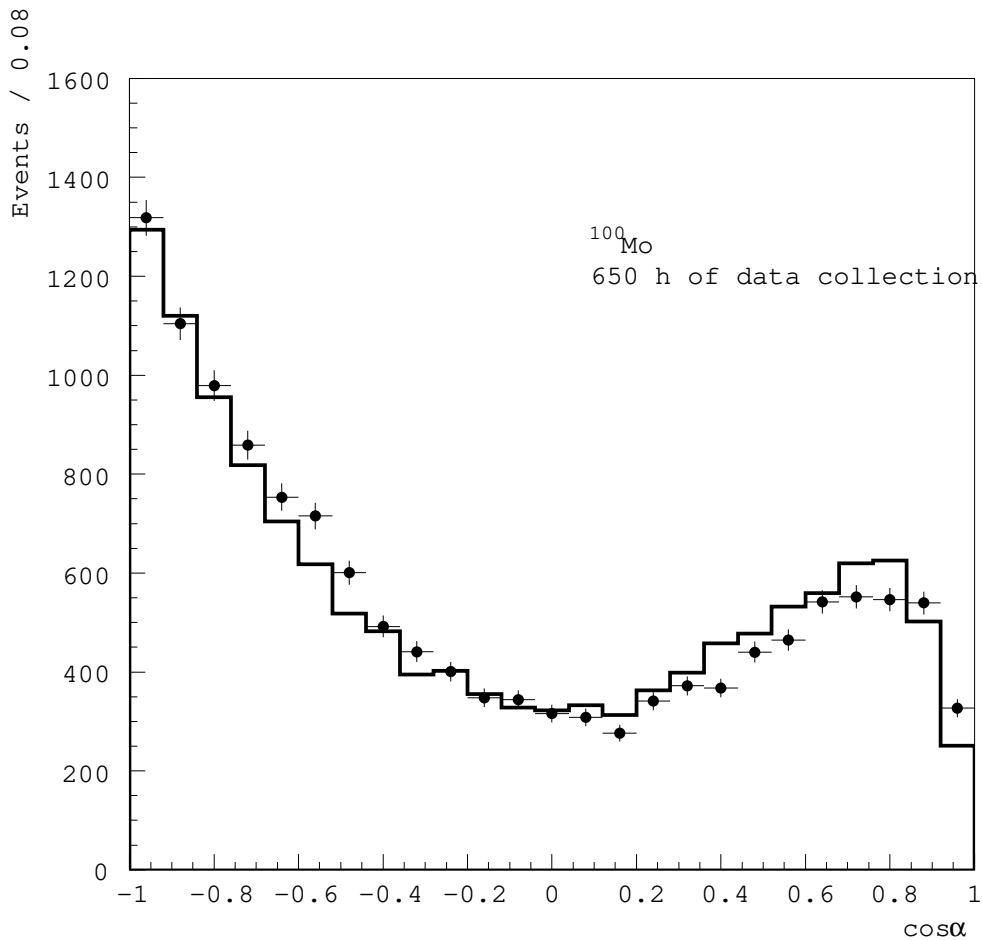}
\vspace*{0.5cm}
\caption{\label{2nu_ang} Experimental angular distribution for molybdenum sources compared to  $\beta\beta 2\nu$ Monte-Carlo calculations. The distribution corresponds to the data in Fig.~\ref{2nu}.}
\end{center}
\vspace*{0.5cm}
\end{figure}

The NEMO~3 detector has been running in the Fr\'ejus Underground Laboratory in nearly optimal conditions since mid-February 2003. Fig.~\ref{2nu} and Fig.~\ref{2nu_ang} reflect the  performance of the detector. The first shows the distribution of the summed two electron energies of $\beta \beta 2 \nu$ events measured for the molybdenum sources (background substracted). The data sample corresponds to 650 hours of data for the runs from  mid-February to the end of March 2003. The number of events is 13750, with a signal-to-background ratio of 40:1. The second figure presents the experimental angular distribution for the two emitted electrons in molybdenum sources. The same data for the summed energy and angular distribution are shown with the $\beta\beta 2\nu$ Monte-Carlo calculation. The high statistics of the data will allow for detailed checks of the models used for the Monte-Carlo calculation.

\section*{Acknowledgements}
The authors would like to thank the Fr\'ejus Underground Laboratory staff for their technical assistance in building and running the experiment.

The portion of this work conducted in Russia for source development was supported by INTAS grant number 00-00362.

\end{document}